\documentclass[11pt,a4paper]{article}

\usepackage{jheppub}

\def\s{\sigma}

\def\la{\lambda}
\def\m{\mu}
\newcommand{\C}{ C_0 }
\def\ka{ \kappa}

\newcommand{\ir}{{\mathrm{IR}}}
    
\title{Holographic RG Flows from Quasi-Topological Gravity} 
\author[a]{G.M. Sotkov}
\author[a]{and U. Camara dS} 
 \affiliation[a]{Universidade Federal do Esp\'itiro Santo,\\ 
Esp\'irito Santo, Vit\'oria, Brazil} 
\emailAdd{gsotkov@gmail.com} 
\emailAdd{ulyssescamara@gmail.com} 
\abstract{We investigate the holographic Renormalization Group (RG) flows and the critical phenomena that take place in the $QFT$'s dual to the d-dimensional cubic Quasi-Topological Gravity coupled to scalar matter. The knowledge of the corresponding flat Domain Walls (DWs) solutions,  allows us to derive the explicit form of the $QFT$'s beta -functions, as well as of the trace anomalies $a(l)$ and $c(l)$, in terms of the matter superpotential. As a consequence we are able to determine the complete set of $CFT$ data characterizing the universality classes of the UV and IR critical points and to follow the particular RG evolution of this data. We further analyse the dependence of the critical properties of such dual $QFT$'s  on the values of the Lovelock couplings and on the shape of the superpotential. For odd values of $d$,  the explicit form of the ``$a$ and $c$ - central charges'' as functions of the running coupling constant, enable us to establish the conditions under which the $a/c$ -Theorems for their decreasing are valid. The restrictions imposed on the massless Holographic RG flows by the requirements of the positivity of the  energy fluxes are derived.  The particular case  of quartic Higgs-like superpotential is studied in detail. It provides an example of unitary dual $QFT$'s having few $c\ne a$ critical points representing second or infinite order phase transitions. Depending on the range of the values of the coupling constant they exhibit  massive and massless phases, described by a chain of distinct DWs solutions sharing common boundaries.} 
\begin{document}
\maketitle 
\newcommand{\sect}[1]{\setcounter{equation}{0}\section{#1}}
\renewcommand{\theequation}{\thesection.\arabic{equation}}

\setcounter{equation}{0}
\section{Introduction} 
According to the  gauge/gravity  correspondence rules, the holographic description of certain strongly coupled $QFT_{d-1}$ at infinite volume and zero temperature is based on the particular flat Domain Walls (DW's) solutions of their dual d-dimensional Gravity models \cite{witt,VVB,rg,girardelo}. The problem addressed in the present paper concerns the systematic application  of the Gauss-Bonnet (GB) and cubic Quasi-Topological Gravity (QTG) DW's \cite{nmg,lovedw} to the investigation of the Renormalization Group properties and of the phase structure of their $QFT_{d-1}$'s dual. It represents an off-critical extension of the recent holographic studies \cite{MPS,2,My_thol,espanha,bala} of a family of $a\neq c$  $CFT_{d-1}$'s with non-vanishing energy flux parameter $t_4$. Special attention is devoted to the unitarity and causality of the massless RG flows described by these DW's solutions. The explicit proof of the corresponding holographic ``$a$ and $c$ - Theorems''\cite{x,cardyth,rg}, together with the implementation of the energy fluxes positivity requirements \cite{maldahof,hofman,BM} on the UV- and IR- $CFT$'s data, allows us to select the physically consistent ``chains'' of DW's and to establish a set of conditions on the parameters of the matter superpotential $W(\s)$. The aim of the present research is to provide more arguments confirming the conjecture that by holographic calculations based on $d=5$ GB and cubic QT Gravity models \cite{My_qtop,c_th} one can reproduce the strong coupling limits of certain unitary (non-supersymmetric) $QFT_4$'s, whose massive and massless phases exhibit ``phenomenologically reasonable'' off-critical features.

\textit{ Domain walls.} The flat static domain walls of $AdS_d$-type represent a special class of $SO(d-2,1)\rtimes T_{d-1}$ symmetric solutions \cite{lovedw,nmg,6} of $d$-dimensional Einstein (super)gravity and its ``higher derivatives'' generalizations \cite{My_qtop,c_th,love} of negative cosmological constant $\Lambda<0$ coupled to an appropriate scalar matter of potential $V(\s)$. They are usually defined by the following ansatz:
\begin{eqnarray}
ds^2_d&=&g_{\mu\nu}(x_{\rho})dx^{\mu}dx^{\nu}=dy^2+e^{2A(y)}\eta_{ij}dx^idx^j ,\quad\quad\mu,\nu=0,1,2..., d - 1    \nonumber\\ 
\eta_{ij}&=&\big(-,+,...,+\big),\quad\quad\quad i,j=0,1...,d-2,\quad\quad\quad\sigma(x_i,y)=\sigma(y) \label{dw}
\end{eqnarray} 
together with the standard DW's boundary conditions (b.c.): 
$$e^{A}(y\rightarrow\pm\infty) \approx e^{\frac{y}{L_{\pm}}},\quad\quad\quad\sigma(\pm\infty)=\sigma^*_{\pm}, \quad\quad\quad  V'(\s^*_{\pm})=0.$$
Such DW's describe asymptotically $AdS_{d}$ space-times\footnote{denoted as $(a)AdS_{d}$ in what follows}, whose main feature is that the matter energy is concentrated around certain $(d-2)$-dimensional subspaces dividing the $(a)AdS_{d}$ space-time in two parts of different $AdS_d$ vacua \cite{lovedw,6,5} related to the extrema of the matter potential. 

\textit{Off-critical Holography.} The increasing interest in the construction of exact GB and QTG DW's solutions is mainly motivated by their role  in the description of the holographic Renormalization Group  flows  \cite{rg,My_thol,town2} between specific conformal field theories $CFT_{d-1}^{UV/IR}$  dual to two consecutive $AdS_d$-vacua of different \emph{small} cosmological constants $\Lambda_{UV/IR}=-(d-1)(d-2)/2L^2_{UV/IR}$. The \emph{off-critical} holography thus offers a \emph{non-perturbative} information about certain \emph{non-conformal} (sypersymmetric) $QFT_{d-1}$'s of few ``critical UV/IR-points'', that are conjectured to be dual to the considered GB and cubic Quasi-Topological Gravity-matter models \cite{My_qtop,c_th}.
It is convenient to realize them as appropriate $CFT_{d-1}^{UV}$'s perturbed by certain relevant operators $\Phi_{\sigma}(x^i)$, which break the conformal $SO(d-1,2)$ symmetry to its Poincare subgroup $SO(d-2,1)\rtimes T_{d-1}$ \cite{VVB,rg}. The actions of such $QFT$'s can be written in the form of perturbed $CFT$'s \cite{x,cardy}: 
\begin{eqnarray}
S_{QFT_{d-1}}(\s)=S_{CFT_{d-1}}^{UV}+(\s(L_{rg})-\s_{UV})\int d^{d-1}x\Phi_{\s}(x^i)\label{pert}
\end{eqnarray}
Then the ``running'' coupling constant $\sigma(L_{rg})$ of  the dual  $QFT_{d-1}$ is identified  with the scalar field  $\sigma(y)$ and the RG ``energy scale'' $L_{rg}^2$ is simply related to the scale factor $e^{2A(\s)}$ of certain d-dimensional extended EH gravity DW's: $A(\s)=-l=ln(l_{pl}/L_{rg}(\s))$ \cite{VVB,rg,akhmed}.

\textit{Holographic $a\neq c$ Theorems.} Given a pair of dual $(a)AdS_d /QFT_{d-1}$ theories, by applying the well known ``holographic rules'' \cite{witt,VVB,rg}, one can reproduce  important strong coupling ($i.e.$ small $\Lambda^{eff}$) features of the \emph{quantum} $QFT_{d-1}$ as for example the RG evolution of all the UV conformal data - the $a(l)$ and $c_n(l)$ - central functions, the anomalous dimensions $\Delta(\s)=d-1-s(\s)$ etc. - from the corresponding \emph{classical} DW's solutions. For odd values of $d$, the knowledge of the explicit form of the ``$a$ and $c$ - central charges'' as functions of the RG scale $l$, allows to further investigate the conditions under which the Zamolodchikov-Cardy's $a/c$-Theorems \cite{x,cardyth} for their decreasing are valid\footnote{see for example ref. \cite{My_thol} for recent discussion.}. It is worthwhile to mention that the DW's of the Einstein (super)gravity of $\Lambda<0$ coupled to matter turns out to lead to very restricted $c=a$ class of ${\cal N}=4$ suppersymmetric $CFT$'s and $QFT$'s, while the $CFT$'s dual to the extended GB and Quasi-Topological gravity-matter models are known to describe a family of  (non-)sypersymmetric field theories, having (at least) two distinct central charges $a\neq c$. 

One of the important results established in the present paper is the following  holographic $a\ne c$ Theorem: \emph{in all the $QFT_4$'s duals to the $d=5$ GB and cubic QT Gravity of negative matter superpotential $W(\s)<0$, having at least two extrema, the $a(l)$ and $c(l)$ -central functions are both positive and decreasing during the massless RG flow}. The conditions of its validity turns out to impose a specific lower bound $L_{min}$ on the values of the UV and IR scales\footnote{determined by the central charges $a_{UV/IR}$ and $c_{UV/IR}$, or asymptotically by the number of colors $(N^2_c)_{UV}\approx (L_{UV}/l_{pl})^3$ of the corresponding  $SU(N_c)$ $CFT_4^{UV/IR}$.} $L_{UV/IR}$ of the $CFT_4^{UV/IR}$: $L_{UV}>L_{IR}>L_{min}$. The exact values of these minimal scales are obtained from the requirements of \emph{positivity of the energy fluxes} \cite{MPS,My_thol,maldahof} in the corresponding dual $CFT_4^{UV/IR}$'s.

\textit{DW's chains and QFT's phase transitions}. The concept and the explicit constructions of \emph{chains} of standard and singular (involving naked singularities) DW's of common boundaries \cite{lovedw,nmg,holo}  are our basic tools in the analysis of the critical phenomena that occur in the considered dual $QFT$'s. The close relationship between the QFT's phase transitions and the existence of chains of distinct gravitational DW's, suggests an  interpretation of such couples of DW's as representing geometric phase transitions in the space of all the DW's solutions of the GB and cubic QTG models. We further investigate the dependence of the critical properties of these dual $QFT$'s  on the values of the Lovelock couplings, on the shape of the superpotential and on the scalar field $\s$ boundary conditions compatible with the unitarity and causality requirements. 

\textit{$QCD_4$ and $sQGP$-hydrodynamics applications?} The relevance of certain $a\neq c$ four-dimensional $CFT$ models dual to $d=5$ \emph{extended} EH gravity, in the description of the hydrodynamics of  strong coupled quark-gluon plasma (sQGP) \cite{hydro,shydro,hidro}\footnote{in particular the evidences for possible violations of the ``universal lower bound'' $\eta/s = 1/4\pi$ of the ratio of shear viscosity to entropy density for such ``conformal fluids''.}, motivates the recent intensive  investigations  of the effects of ``higher curvature'' gravitational interactions on the intrinsic unitarity and causality properties of their dual $CFT_4$'s \cite{2,
espanha,bala,maldahof}. At the energy scales of experimental interest, the realistic sQGO models, based on the (supersymmetric) $QCD_4$  are however known to be \emph{non-conformal} $QFT$'s. Therefore in this holographic $(a)AdS_5/QFT_4$ context, the problem of construction of appropriate classical solutions of the extended $d=5$ GB and Quasi-Topological gravity-matter models as for example DW's and of their finite temperature analogues called thermal gas solutions \cite{ven-kir}, becomes of real practical importance \cite{bviscos}. 

One relevant result concerning the RG behaviour of the ratio $\eta/s$ is provided by the $\frac{a}{c}$ -Theorem, established in Sect.6 in the case of (zero temperature) massless RG flows in the $QFT_4$'s dual to $d=5$ GB gravity-matter model. It suggests that for positive GB couplings $\la$, the corresponding UV- and IR- values of $\eta/s$ satisfy the following inequality: $\left(\eta/s\right)_{UV}\geq \left(\eta/s\right)_{IR}$, while for negative $\la$ we find an indication that  $\eta/s$ should be increasing. The complete description of the off-critical behaviour of $\eta/s$ as a function of the RG scale (or of the coupling constant $\s$) in the massive and massless phases of the dual $QFT_4$, requires further investigations.

\textit{The layout of paper.} In Sect.2.1. we review some basic concepts and results concerning the cubic QT Gravity vacua structure, its  DW's solutions and the superpotential method. Sects.2.2. and 2.3. contain a brief summary of few important $CFT_{d-1}$'s and renormalization group topics, involving the definitions of the scaling dimensions, central charges, OPE's and  their relations to the holographic $\beta(\s)$- function. Our proofs of the holographic ``$a$ and $c$ - Theorems'' for the $QFT_{d-1}$ dual to GB and cubic QT Gravity coupled to scalar matter, together with the specific restrictions on the values of the UV and IR scales $L_{UV/IR}$ and on the Lovelock's gravitational couplings $\la$ and $\mu$  necessary for their validity are presented in Sects.3,4,5 and App. \ref{apexA}. We derive in Sect.6 the additional ``stronger'' requirements imposed on the physical scales $L_{UV/IR}$ by the positivity of the energy fluxes at the $CFT_{d-1}$'s, representing the UV and IR critical points. Sect.7  is devoted to the detailed study of the critical phenomena in a specific family of $QFT$'s, corresponding to a particular choice of \emph{ quartic Higgs-like matter superpotential}. They provide examples of dual $QFT_{d-1}$'s having few $c\ne a$ critical points of second and infinite order phase transitions and specific  massive and massless phases. Few important results concerning the explicit form and the properties of the GB domain walls for negative values of $\la$, that are important in the description of the massless-to-massive phase transitions in the dual $QFT$'s (see Sect.7.3.), are derived in our App.\ref{apexB}.


\setcounter{equation}{0}
\section{CFT-data from QT Gravity DWs}

We are interested in the off-critical properties of the $QFT_{d-1}$ dual to the  cubic QT Gravity \cite{ My_qtop,c_th} coupled to scalar matter:
\begin{eqnarray}
S^{{\mathrm{eff}}}_{GBL}= \frac{1}{\kappa^2}\int \! d^d x \, \sqrt{-g} \, \Bigg[ R + \frac{ \lambda L^2}{(d-3)(d-4)}\big(R_{abce}R^{abce}-4R_{ab}R^{ab}+R^2\big)\nonumber\\
- \frac{8\mu (2d-3)L^4}{(d-6)(d-3)(3d^2-15d+16)}{\mathcal{Z}}_d  - \kappa^2\left(\frac{1}{2} g^{ab}\partial_{a} \sigma \, \partial_{b} \sigma + V(\sigma)\right) \Bigg] ,\label{qtop}
\end{eqnarray}
where $\kappa^2=\frac{8\pi^{(d-1)/2}}{\Gamma((d-1)/2)} G_d$ is defining the d-dimensional Plank scale, the $L^2$ is representing the new scale, related to the bare cosmological constant\footnote{The negative bare cosmological constant $\Lambda$ is implicitly defined by the vacuum value of the matter potential $\kappa^2 V(\sigma^*)=2\Lambda=-(d-1)(d-2)/L^2$ and should be distinguished from the effective cosmological constant $\Lambda^{{\mathrm{eff}}}$.}, specific for the ``higher order'' gravitational models \cite{deser}, while  $\la$ and $\m$ denote the appropriately normalized dimensionless Gauss-Bonnet and Lovelock ``gravitational'' couplings. The cubic  ${\mathcal{Z}}_d $-invariant \cite{My_qtop} has the specific form:  
\begin{eqnarray}
&&{\mathcal{Z}}_d = R_a{}^c{}_b{}^d \, R_c{}^e{}_d{}^f \, R_e{}^a{}_f{}^b + \frac{1}{(2d-3)(d-4)} \Big[ \frac{3(3d-8)}{8} R_{abcd} R^{abcd} R - 3(d-2) R_{abcd} R^{abc}{}_e R^{de} + \nonumber\\
&&+ 3d \, R_{abcd} R^{ac} R^{bd} + 6 (d-2) R_a{}^b R_b{}^c R_c{}^a - \frac{3 (3d - 4)}{2} R_a{}^b R_b{}^a R + \frac{3d}{8} R^3 \Big], \nonumber
\end{eqnarray}
ensuring that the equations of motion for the black  hole solutions are  of second order. In general this action leads to fourth order equations \cite{My_qtop,c_th}. For conformally flat metrics  it reduces to the action of  d-dimensional extension of the cubic NMG model \cite{sinha}, derived in ref. \cite{lovedw} by requiring that the corresponding DW's (\ref{dw}) obey \emph{second} order equations as well.

In order to make self-consistent our investigation of the Holographic RG flows derived from certain QT Gravity domain walls, we review in this section the basic concepts concerning the QTG vacua and DW's properties and the most important ingredients in the description of the dual $CFT_{d-1}$'s - as scaling dimensions, central charges and  OPE's, together with their relations to the corresponding quantities in the Wilson and holographic RG approach to the critical phenomena. Subsection 2.1. contains a short summary of the recent results on the GB and QT Gravities  DW's \cite{lovedw}, while in subsections 2.2. and 2.3 we give a brief introduction to some selected topics in $CFT$'s, that are essential in the derivation of the holographic ``$a\neq c$ - Theorems'' and of the corresponding positive energy fluxes requirements.

\subsection{Superpotential and  Lovelock's vacua}

The family of stable flat  DWs (\ref{dw}) solutions of the cubic Quasi-Topological Gravity (\ref{qtop}) can be obtained by solving the following $I^{st}$ order system of equations\footnote{we have denoted $W'(\sigma) = \frac{dW}{d\sigma}$ and $\dot{\sigma}=\frac{d \sigma}{dy}$}:
\begin{equation}
\dot\sigma =\frac{2}{\kappa} W' \C(W) ,\quad \dot{A} = - \frac{\kappa}{d-2} W(\sigma),\quad \C(W)= 1 - 2\la \frac{\kappa^2L^2 W^2}{(d-2)^2} -3 \m\frac{\kappa^4L^4 W^4}{(d-2)^4},\label{sys}
\end{equation}
derived by an appropriate extension \cite{lovedw,nmg} of the superpotential method \cite{6}. The relation  between the matter potential $V(\s)$ and the superpotential $W(\s)$ is given by:
\begin{eqnarray}
\kappa^2 V(\sigma) = 2 ( W' )^2 \C^2(W)- \left( \frac{d-1}{d-2} \right) \kappa^2 W^2\left(1 - \la \frac{\kappa^2L^2 W^2}{(d-2)^2} - \m\frac{\kappa^4L^4 W^4}{(d-2)^4}\right),    \label{pot}
\end{eqnarray}
Let us remind  the common property of all the ``higher curvature'' generalizations of EH gravity, namely that the effective cosmological constants related to the vacuum values $\s_k^*$ ($V'(\s_k^*)=0)$ of the superpotential: 
$$f_k \equiv L^2/L_k^2 = L^2 \dot{A}^2(\s^*_k) = \ka^2 L^2 W^2(\s^*_k) / (d-2)^2 \; $$ 
are \emph{different} from the  corresponding ``bare'' ones $$h_k \equiv  L^2/L_{0k}^2 = - L^2 \, V(\s^*_k) / (d-1)(d-2). $$ 
As  a consequence of eq. (\ref{pot}), they satisfy the following cubic ``vacua'' equation: 
$$h_k =  f_k ( 1 - \la f_k - \m f_k^2 ).$$ 
An important feature of the $AdS_d$ type of vacua ($h_k > 0$, or $V(\sigma^*_k) < 0$) of the considered model is the existence of very special kind of \emph{topological} vacua\footnote{ one can always find an appropriated range of values of the gravitational couplings $\la$ and $\mu$ (not both negative), such that  at least one of the vacua of the Lovelock gravity-matter model (\ref{qtop}) to be  of topological nature, as it is demonstrated in Sect.3 of ref. \cite{lovedw}.} defined by the real solutions of the equation $\C(f_{top}) = 0$. They represent few extrema of  the matter potential $V(\s)$ that however \emph{are not}  extrema of the superpotential.

The second kind of Lovelock's vacua ($\sigma_k^*$, $f_k$) are  determined by  all the  extrema of the superpotential $W'(\s_k^*) = 0$ with $W(\s_k^*)\neq0$. According to ref. \cite{My_qtop}, the \emph{stable physical} vacua are selected by the ``causality" requirement $\C(f_k) = 1 -2 \la f - 3 \m f^2>0$. It excludes all the $f_k$'s with $\C(f_k)<0$ since they  lead to the wrong ``ghosts-like" sign of the graviton's kinetic terms in the  ``linearized" (Gaussian) form of the Quasi-topological Gravity action (\ref{qtop}).

We next consider the case when $\la$ and $\m$ are not both negative and thus we can have at least one topological vacuum, given  by the solutions $$f_{top}^{\pm}=- \frac{1}{3\m}( \la \mp \sqrt{  \la^2 + 3 \m  } )$$ 
of the ``topological" vacua equation $C_0(f_{top}^{\pm})=0$. They determine the following particular values of the topological ``bare" cosmological constants (and of the related  $AdS_d$ scales $L_{o,top}^{\pm}$)
\begin{equation}
h_\pm = h(f_\pm) = \frac{1}{27\m^2}\left[ - \la (2 \la^2 + 9 \m) \pm 2 (\la^2 + 3 \m)^{3/2} \right]\; , \label{h top la and mu}
\end{equation}
It turns out that $h_{\pm}$  represent the extremal values of the $h(f)$: $f_+$ is a local maximum, while $f_-$ is a local minimum. By construction the $h_{top}^{\pm}$ are related to the values of the potential $V(\s_{top}^{\pm})$ at two of its extrema and therefore they make a part of the boundary data for the considered gravity-matter model (\ref{qtop}). This suggests that for each fixed value of $0<\la<1/3h_{top}^{+}$  we can consider eq. (\ref{h top la and mu}) as a \emph{quadratic} equation for $\m$, providing two different real\footnote{remember that  $h_+> h_-$ and therefore $\la<1/3h_+<1/3h_-$ assures the reality of $\m_{\pm}$ below.}  values $\m_{\pm}(\lambda)$. Notice that for all the values of $\la$ we have  $h_-(\m_-) > 0$ and $h_+(\m_\pm) > 0$, which allows us to fix the fundamental  scale  as  $L^2=L_{0+}^2(\mu_+)=L_{0-}^2(\m_-)$ (see eq. (\ref{qtop})), by normalizing the ``bare'' topological vacuum : $h_+(\m_+, \la ) = h_-(\m_-, \la) = 1 $. Our choice
consists in taking the smallest topological scale of the $\mu_+$ model, and the largest one for the $\mu_-$ model. Then the two distinct gravitational models ($\la$,$\m_{\pm}$) of equal fundamental scale, are defined by the  following particular  form of the Lovelock coupling $\mu$ as a function of $\lambda$:
\begin{equation}
\m_\pm (\la) = \frac{1}{ 27}\left( 2 - 9 \la \pm 2 ( 1 - 3 \la  )^{3/2} \right)   \; . \label{mu pm}
\end{equation}
According to the analysis of ref. \cite{lovedw}, the restrictions on the values of $f_k$ (and $h_k$) which lead to stable physical vacua in the different domains of admissible values of $\la$ and $\m$  can be summarized as follows:

(a)\textit{The $\mu_+$ model:} Taking into account its main characteristics  given by the  extrema of the  curve $h(\la, \m_+ ; f) \equiv h(f)$: 
\begin{eqnarray}
&& h_+ = h(f_+) = 1\, , \;\; h_- = h(f_-) =  \frac{4 - 45 \la + 108 \la^2 - 4 (1 - 3\la)^{3/2}}{27 \la (1 - 4 \la )^2}, \nonumber\\
&& f_+ =\frac{1}{\la} ( 1 - \sqrt{1 - 3\la} ) \, , \;\; f_- = \frac{1}{3\la(1 - 4\la)}\left( 1 - 6 \la - \sqrt{1 - 3  \la} \right)\label{h}.
\end{eqnarray}
it is easy to verify that: 

(a1) for $\la < 1/4$, the minimum  $f_- < 0$ is negative  and therefore the physical conditions $h, \C > 0$ require that  $0 < f_{phys} < f_+$ ;

(a2) in the interval $1/4 < \la < 8/27$  both  $f_{\pm} > 0$ are positive, but now the $h_-$ is negative and thus violating\footnote{Nevertheless one can have stable physical vacua (of largest scale) when 
$$f_{phys} > f_0= - \la / 2 \m_+ + \sqrt{ \left( \la^2 + 4 \m_+ \right)/4 \m_+^2},$$
 where we have denoted by $f_0$ the greatest root of $h(f) = 0$ equation.} the $AdS_d$ stability condition: $h > 0$. Therefore the $\C(f)>0$ and $h>0$ requirements are satisfied again in the  ``smallest scale'' physical region: $0 < f_{phys} < f_+$; 

(a3) for  $8/27 < \la < 1/3$  both the $f_\pm$ as well as both $h_\pm$ are positive. This case permits two qualitatively different physical regions: one is the usual Gauss-Bonnet-like region given by $0 < f_{phys} < f_+$, while the second one, $f_- < f_{phys}$, is of a new type with no upper limit on the values of the effective cosmological constants. The topological vacuum $f_-$ introduces a \emph{maximal scale}  $L_-^2 \sim f_-^{-1}$, restricting the possible values of the physical vacua scales from above.

(b)\textit{The $\mu_-$ model:} The corresponding definitions of the physical regions are now given by the extrema of  another curve  $h(\la, \m_-; f) \equiv \tilde{h}(f)$: 
\begin{eqnarray}
&& \tilde{h}_+ = \tilde{h}(\tilde{f}_+) = \frac{4 - 45 \la + 108 \la^2 + 4 (1 - 3\la)^{3/2}} { 27 \la (1 - 4 \la )^2} \, , \;\; \tilde{h}_- = \tilde{h}(\tilde{f}_-) = 1 ,  \nonumber\\
&&   \tilde{f}_+ = \frac{1 - 6 \la + \sqrt{1 - 3  \la} }{  3\la(1 - 4 \la)} \, , \;\; \tilde{f}_- = \frac{ 1 + \sqrt{1 - 3\la} }{  \la} \label{h tilde},
\end{eqnarray}
It has the important property that $\tilde{h}_+ > 0$ for all $\la > 0$. The physical region of  \emph{minimal} scale  for this models is then defined  by $0 < f_{phys} < \tilde{f}_+$, while  the region $\tilde{f}_- < f_{phys}$ represents the vacua of \emph{maximal} scales $L_{-}^2(\m_-)>L_{phys}^2(\m_-)$.

The above $AdS_d$ type of physical vacua $(\sigma_k,L_k)$  provide a set of all the admissible b.c.'s for $\s(y)$ and $e^{2A(y)}$ at $y \rightarrow\pm\infty$ for the  corresponding stable QT-Gravity DW's solutions. Their explicit form is given by the solutions of the $I^{st}$ order eqs. (\ref{sys}) that relate two neighbouring vacua (see sect.4 of ref. \cite {lovedw}, where the explicit form of such DWs was found).

\subsection{Renormalization Group and CFT-data}

2.2.1. \textit {RG-data: $\beta-$function}. 
Given the form of the superpotential $W(\s)$ and related to it $I^{st}$ order system (\ref{sys}), the scale-radial duality determines the $\beta-$function of the conjectured dual  $QFT_{d-1}$ \cite{VVB,rg,girardelo} in terms of the superpotential:
\begin{eqnarray}
\frac{d\sigma}{dl}=-\beta(\sigma)=\frac{2(d-2)}{\kappa^2}\frac{W'(\sigma)}{W(\sigma)}\C(W) ,\quad\quad\quad l=-A(\s)=ln(l_{pl}/L_{rg}(\s))
\label{rg}
\end{eqnarray}
The constant solutions $\s^*_{k}$ of the above RG equation (\ref{rg}) are defined by the zeros of the $\beta$-function and by construction coincide with the Lovelock vacua solutions of $AdS_d$ type. As is well known at such \emph{critical points} in the coupling space the corresponding $QFT_{d-1}$ becomes conformal invariant and  phase transitions of  second  or infinite order  take place \cite{polya,polboot,cardy}. 
The different \emph{non-constant} solutions $\sigma_{kj}=\sigma(l;\sigma^*_{k},\sigma^*_{j})$ are
 directly related to the flat DWs solutions of the considered Quasi-Topological Gravity-matter models (\ref{qtop}). As it is demonstrated in ref.\cite{lovedw}, the near-boundary/horizon properties of such DW's, representing specific $(a)AdS_d$ space-times, determine important characteristics of the Lovelock's vacua and of their dual $CFT_{d-1}$'s, namely the ``critical exponents'' $s_k=-\frac{d\beta}{d\s}(\s_k)$ (for  $W_{k}\neq0$ ):
\begin{eqnarray}
&s_a = 2 (d-2) \frac{W''_a}{\kappa^2W_a} \left[ 1 - \frac{2 \la L^2}{(d-2)^2} \kappa^2W_a^2 - \frac{3 \m L^4}{(d-2)^4}\kappa^4 W_a^4 \right]
    = 2 (d-2) \frac{W''_a}{\kappa^2W_a} \left(1 - \frac{L_+^2}{ L_a^2} \right)\left(1 -\frac{ L_-^2 }{ L_a^2} \right),\nonumber\\
&s_{top}^\pm = - \frac{8}{(d-2)} W_{\pm}'^2 \left[  \la L^2 + \frac{3  \m L^4}{(d-2)^2}\kappa^2 W_{\pm}^2 \right] = - \frac{4 L_\pm^2}{(d-2)} W_{\pm}'^2 \left[ 1 - \frac{L_\mp^2}{L_\pm^2} \right]\label{criti}
\end{eqnarray}
These  new parameters  characterize the near-boundary (and near-horizon) behaviours of the scalar field and of the scale factor:
\begin{eqnarray}
\sigma(y)\stackrel{y\rightarrow\infty}{\approx}\sigma_{k}^* +const.\, e^{-
s_k\frac{y}{L_k}}, \quad \quad e^{2A(\s)} \sim ( \s - \s_k)^{-\frac{2}{s_k}}, \label{asymp}
\end{eqnarray}
which can be easily checked by taking appropriate limits of eqs. (\ref{sys}) with $W(\s)$ replaced by:
\begin{eqnarray}
W(\sigma)\approx W(\sigma_k^*)+\frac{W'(\sigma_k^*)}{2}(\sigma-\sigma_k^*)^2.
\end{eqnarray}
for each one of the physical vacua  $W'(\s_k^*)=0$. Observe that the vacua $(\sigma_k,L_k,s_k)$ with $s_k < 0$ describe the horizons ($i.e.$ we have that $e^{2A} \rightarrow 0$), while for the vacuum  with  $s_k >0$ we find that $e^{2A} \rightarrow \infty$, thus representing  the $AdS_d$  boundary.

In the case of higher order $n_k>1$ zeros of the $\beta(\s_k)-$function, $i.e.$ when also all its derivatives  $\beta^{(n_k)}(\s_k)=0$ are vanishing, we find a new type of near-boundary behaviour of the matter field:
\begin{eqnarray}
\sigma(l)\stackrel{y\rightarrow \infty}{\approx}\sigma_{k}^* + {\mathrm{const.}}\,(\frac{y}{L_k} )^{\frac{1}{1-n_k}} ,\label{inford}
\end{eqnarray}
Notice that such asymptotic values of the running coupling constant in the dual  $QFT_{d-1}$ models are specific for the neighbours of a ``marginal'' critical point $\s_k^{marg}$  of vanishing critical exponent $s_k^{marg}=0$, where an infinite order phase transition takes place.

\vspace{0.5cm}
2.2.2. \textit {CFT-data: BF unitarity conditions.} 
The conditions of \emph{stability} of the physical vacuum, expressed by $\C(f_k)>0$, are known to be a part of the consistency requirements for the unitarity of the dual $CFT_{d-1}$'s \cite{lovedw,My_qtop}. 
Let us remind the additional restrictions obtained by  imposing the Breitenlohner-Freedman (BF) unitarity condition \cite{BF}:
\begin{eqnarray}
-\frac{(d-1)^2}{4 L_k^2}\le m_{\sigma}^2(\sigma_{k}^*)=V''(\sigma_{k}^*)\label{BFred}
\end{eqnarray}
It assures the stability of the gravity-matter  vacua $(\sigma_{k}^{*} \, , \, \Lambda_{\mathrm{eff}}^{k} < 0)$ with respect to the linear fluctuations of the scalar field. Taking into account (\ref{BFred}), together with the definitions (\ref{criti}) of the critical exponents  $s_k$, it is easy to derive the relation of the ``vacua masses''  $m_{\sigma}^2(\sigma_{k}^*)=V''(\sigma_{k}^*)$ of the scalar field to the anomalous dimensions $\Delta_k=d-1-s_k$ of the corresponding ``dual'' conformal fields $\Phi_{\sigma_k}(x_i)$:
\begin{eqnarray}
m_{\sigma}^{2}(\sigma_{k}^{*})=\frac{\kappa^2W_{k}^2}{(d-2)^2}s_{k}(s_{k}-d+1)=\frac{1}{L_k^2}s_{k}(s_{k}-d+1).\label{mas}
\end{eqnarray}
The above mass formulas are valid for all the Lovelock's vacua $\sigma_{A}^{*}$ -- either the ``physical'' (or ``ghost''-like), $\sigma_k^*$, or the topological, $\sigma_{top}^{\pm}$, ones. They are obtained  from  eq. (\ref{pot}) by calculating  the values of $V''(\sigma)$ at the corresponding vacuum and next by expressing them in terms of the critical exponents (\ref{criti}).

\begin{figure}[ht]
    \centering
    \includegraphics[scale=0.6]{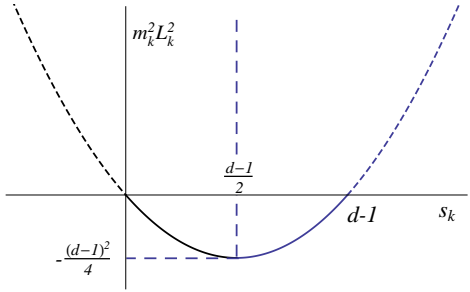}
    \begin{quotation}
    \caption[ed]{The  $m_k^2L_k^2$ curve as a function of $s_k$: l.h.s. of the parabola plotted in black corresponds to $\Delta_k=\Delta_k^+$ and its r.h.s.(in blue) represents $\Delta_k=\Delta_k^-$. On the dashed line $m^2_k>0$, while on the continuous one $m^2_k<0$.}
     \label{fig:m_2}
     \end{quotation}
\end{figure}

Notice that  the BF-condition (\ref{BFred}) is  automatically satisfied for all the values of $s_{k}$. As one can see from Fig.(\ref{fig:m_2}), for each (positive or negative) value  $m^2_k\neq -\frac{(d-1)^2}{4L^2_k}$ of the mass, there exist always two different critical exponents $s_k^{\pm}$, such that $s_k^+ + s_k^-=d-1$. They provide two different asymptotic behaviours of the scalar field $\s(y)$, as we have explained above. The scaling dimensions of the corresponding $CFT_{d-1}$ fields $\Phi_{s_k^{\pm}}(x_i)$ are given by: 
\begin{eqnarray}
\Delta_{k}^{\pm}=d-1-s_{k}^{\pm}=\frac{d-1}{2}\pm\sqrt{\frac{(d-1)^2}{4}+m^2(\sigma_{k}^*)L_{k}^2}\label{sdime}
\end{eqnarray}
They  are known to describe  different states that give rise to different quantizations of the ``boundary'' $CFT_{d-1}$. As argued in ref. \cite {witku}, only the field corresponding to $\Delta^+$ (under certain restrictions) can drive the RG flow from one UV to the IR CFT's, thus defining a  non-conformal perturbation of the $CFT_{UV}$ as in eq. (\ref{pert}). Instead, the conformal field  with $\Delta^-$ leads to spontaneous breaking of the conformal symmetry by introducing a new vacuum state and non-vanishing mean values $<\Phi_{s^-}(0)>\neq 0$.

As is well known one of the conditions for \emph{unitarity} of these  $CFT_{d-1}'s$ is the \emph{positivity} of the corresponding scaling dimensions $\Delta_{k}^{\pm}>0$, which  requires that\footnote{ the values of $s_k^{-}>d-1$  correspond to $\Delta_k^{-}<0$, thus giving rise  to non-unitary CFT's.} $s_{k}<d-1$. Since we are interested in the holographic description of the  RG flows in such perturbed CFT's, we shall  consider  the case $\Delta_k^+$ only. We further impose an important extra condition, namely that the operators $\Phi_{s^+}^{UV}(x_i)$  are \emph{relevant}: $0<\Delta_{UV}^{+}<d-1$. Then, according to eq. (\ref{sdime}) we obtain  the following restriction: 
$$\frac{d-1}{2}<\Delta_{UV}^+<d-1 \quad\quad \textrm{ or equivalently}\quad\quad  0<s_{UV}^+<\frac{d-1}{2},$$ 
which ensures the existence UV/IR- RG flows in the dual \emph{unitary} $QFT_{d-1}$. Observe that the above conditions are satisfied only for \emph{negative} values\footnote{ it is worthwhile to mention that for positive masses  $m^2_{UV}>0$ we have that $\Delta_{UV}^+>d-1$, i.e $s_{UV}^+<0$ is negative, and therefore the corresponding operators $\Phi_{\sigma_k}(x_i)$ are irrelevant.} of the scalar field masses $m^2_{UV}$:
\begin{eqnarray}
-\frac{(d-1)^2}{4}\le m_{\sigma}^2(\sigma_{UV}^*)L_{UV}^2<0. \label{news}
\end{eqnarray}
This stronger form of the BF-condition, together with the other physical restrictions on values of $AdS_d$ physical scales $L_k$, are responsible to qualitative changes in the shape of the matter potential, compared to the case when $s_{UV}>d-1$, as it is demonstrated in refs. \cite{lovedw, Myers-new} (see  also Sect.7.3 below).

\vspace{0.5cm}  
2.2.3. \textit {CFT-data: OPE's structure constants.} Let us remember that the conformal invariance \cite{cardy,valyatod} determines the exact form of all the 2- and 3-point functions of the fields $\Phi_{(\Delta,j_k)}(\vec{x})$ of conformal (scaling) dimensions $\Delta_{\Phi}$ and spins $j_k$  in a given $CFT_{d-1}$. For example, in the case of scalar fields they are given by:
\begin{eqnarray}
&&<\Phi_{\Delta}(\vec{x_1})\Phi_{\Delta}(\vec{x_2})\Phi_{\Delta}(\vec{x_3})>=C^{(3)}_{\Phi}\left(|\vec{x_{12}}||\vec{x_{13}}||\vec{x_{23}}|\right)^{-\Delta_{\Phi}},\nonumber\\
&&<\Phi_{\Delta}(\vec{x_1})\Phi_{\Delta}(\vec{x_2})>=C^{(2)}_{\Phi}|\vec{x_{12}}|^{-2\Delta_{\Phi}}\label{twopoint}
\end{eqnarray}
In order to complete the description of the properties of the relevant operator $\Phi_{s^+}^{UV}(x_i)$ we need to know  the value of its scaling dimension $\Delta_{UV}^+=d-1-s^+$, given by eq. (\ref{criti}), and the explicit form of the short-distance OPE's\footnote{notice that this form of the OPE's is oversimplified by neglecting the contributions of the irrelevant operators and of the less singular relevant operators as well.} as well:
\begin{eqnarray}
\Phi_{s^+}(1)\Phi_{s^+}(2)\approx C^{(2)}_{\Phi}|x_{12}|^{2(s^+ -d+1)}+C_{\Phi\Phi\Phi}|x_{12}|^{s^{+}-d+1}\Phi_{s^+}(2)+... \label{ope}
\end{eqnarray}
 We have introduced the OPE's \emph{structure constant} $C_{\Phi\Phi\Phi}=C^{(3)}_{\Phi}/C^{(2)}_{\Phi}$, which turns out to coincide with the ratio of the 2- and 3-point normalization's constants.

The standard perturbative QFT's calculations \cite{x,cardy,kleba} based on the ``perturbed CFT'' ($pCFT$) action (\ref{pert}), involving the exact form of the 2- and 3-point functions (\ref{twopoint}) and the most singular terms of the OPE (\ref{ope}), give the well known result for the corresponding perturbative $\beta-$function: 
\begin{eqnarray}
\beta(\sigma)\approx -s^{+}(\sigma- \sigma_{UV}^*)+ C_{\Delta}(\sigma- \sigma_{UV}^*)^2 +...,\quad\quad C_{\Delta}=\frac{\pi^{(d-1)/2}}{\Gamma((d-1)/2)} C_{\Phi\Phi\Phi}\label{betapert}
\end{eqnarray}
Therefore the knowledge of the holographic $\beta-$function (\ref{rg}) allows to determine the \emph{exact value} of the  CFT's OPE structure constants\footnote{these exact values of the structure constants have to be distinguished from  the perturbative values  $C_{\Phi\Phi\Phi}$, calculated from the limits of the corresponding bulk-to-boundary propagators of the scalar field $\s(y,x_i)$, by applying the Witten's diagrams techniques \cite{witt}.} in terms of the values of the superpotential and its derivatives at the corresponding UV vacuum $\s_{UV}$:
\begin{eqnarray}
2C_{\Phi\Phi\Phi} \frac{\pi^{(d-1)/2}}{\Gamma((d-1)/2)}= \beta^{''}(\s_{UV}^*)=\frac{2(d-2) W_{UV}^{'''}}{\kappa^2 W_{UV}}\C(W_{UV}).\label{str}
\end{eqnarray}
This  identification provides not only an important non-perturbative CFT-data concerning the OPE's of the dual fields $\Phi_{s^+}^{UV}(x_i)$, but it also imposes specific restrictions on the form of the superpotential $W(\s)$. Namely, in order to have consistent $pCFT$'s defined by the action (\ref{pert}) the requirement $W_{UV}^{'''}\C(W_{UV})\neq 0$ must be satisfied, $i.e.$ in the case of polynomial superpotentials, $W$ should be at least cubic.

\vspace{0.5cm}
2.2.4  \textit{Conserved currents OPE's}. As usually the set of conserved currents - $J_k(\vec{x})$ of dimension $\Delta_J=d-2$, $T_{ij}(\vec{x})$  of dimension $\Delta_T=d-1$, etc. -  and their correlation functions play an important role in the definition of the corresponding conserved charges, of their algebra and of the physical states of an arbitrary $QFT_{d-1}$. It is well known that, in the absence of additional (extra) symmetries, almost all the important parameters specifying each $CFT_{d-1}$ are realized in terms of the constants of the 2- and 3-point functions of their  stress-tensors $T_{ij}(x_k)$ (with $T^i_i=0$). As in the case of the scalar field\footnote{ in fact the explicit form of all the 2- and 3- point functions of fields of arbitrary Lorentz spins is well known \cite{raiko,osborn1,valyatod}, involving an increasing number of independent tensor structures determined by the conditions of conformal symmetry.} the corresponding 2-point function:
\begin{eqnarray}
<T_{ij}(x_k)T_{pq}(0)>&=&\frac{\pi^{(d-1)/2}}{\Gamma((d-1)/2)} \frac{c(d)}{x^{2(d-1)}}J_{ij,pq}(\vec x),\quad\quad\quad I_{ij}=\eta_{ij}-\frac{2 x_ix_j}{x^2}\nonumber\\
 J_{ij,pq}(\vec x)&=&\frac{1}{2}\left[I_{ip}(\vec{x})I_{jq}(\vec x)+I_{iq}(\vec{x})I_{jp}(\vec x)\right]-\frac{\eta_{ij}\eta_{pq}}{d-1}\label{TT}
\end{eqnarray}
contains only one  parameter $c(d)$, while in the 3-point function:
\begin{eqnarray}
<T_{ij}(\vec{x_1})T_{mn}(\vec{x_2})T_{pq}(\vec{x_3})>=\frac{\left({\cal A}{\cal I}^{\cal A}_{ij,mn,pq}(\vec{x_{st}})+{\cal B}{\cal I}^{\cal B}_{ij,mn,pq}(\vec{x_{st}})+{\cal C}{\cal I}^{\cal C}_{ij,mn,pq}(\vec{x_{st}})\right)}{\left(|\vec{x_{12}}||\vec{x_{13}}||\vec{x_{23}}|\right)^{d-1}}\label{TTT}
\end{eqnarray}
we need \emph{three} constants $\cal A$, $\cal B$ and $\cal C$, representing the number of the different independent (conformal covariant) tensor structures ${\cal I}^{\cal K}_{ij,mn,pq}(\vec{x_{st}})$, that also  satisfy the corresponding traceless and conservation ${\partial }^i T_{ij}(\vec{x})=0$ conditions\footnote{ we are omitting  the explicit forms of these tensor structures \cite{osborn2}, since they are irrelevant for the problems discussed in the present paper.}. The conformal Ward identities further imply that the following relation between the constants of  the $T_{ij}$'s 2- and 3-point functions to take place \cite{osborn2}:    
\begin{eqnarray}
  c(d)=\frac{(d-2)(d+1){\cal A}-2{\cal B}-4d{\cal C}}{d^2-1}. \label{ctensor}
\end{eqnarray}
 Similarly to the scalar fields OPE's (\ref{ope}), the the structure constants of the stress-tensor $T_{ij}(\vec{x})T_{mn}(0)$ OPE's turns out to be simply related to the ratios of  specific linear combinations of $\cal A$, $\cal B$ and $\cal C$ with the $c(d)$ -central charge \cite {osborn2,anselmiope}.

\subsection{Weyl anomalies and central functions}

Let us remind that the \emph{higher} $d\geq 4$ dimensional conformal $SO(d-1,2)$ symmetry is not restricting completely the $CFT_{d-1}$ dynamics \cite{cardy,valyatod}. Although it is again enough to determine  all the 2- and 3-point functions (\ref{twopoint}) and (\ref{TTT}), it is not sufficient to fix the form of the $n\geq4$-point functions, which require  the knowledge of the solutions of   the conformal bootstrap equations \cite{polboot}. The corresponding $CFT_{d-1}$ are characterized by: (1) the \emph{anomalous dimensions} $\Delta_k =d-1-s_k$ and the \emph{spins} $j_a$ of the conformal fields $\Phi_{\Delta,j_a}(x_i)$, defining certain unitary representations of the conformal group \cite{valyatod}; (2) the so called \emph{central charges} $c_n$, $a$, $b_l$ and related to them coefficients - ${\cal A}, {\cal B}, {\cal C}$ - of the stress-tensor 3-point function; and (3) the \emph{structure constants} $C_{\Phi\Phi\Phi}$ of the OPE's of  $\Phi_{k}(x_i)$'s.
 
In order to complete the list of the remaining $CFT_{d-1}$ data that can be calculated by the RG and/or other perturbative methods, we next consider the definitions of a part of the  central charges, namely $c=c_2$ and $a$. For odd values of $d$, $i.e.$ for even dimensional $CFT_{d-1}$'s with $d=2n+1$, the above central charges appear as the (coupling dependent) coefficients in the Weyl (trace) anomaly expression \cite{duff,deseranom}:
\begin{eqnarray}
<T^i_i>\sim \sum c_{n}(d,\s)I_{n}(d)- 2(-1)^{(d-1)/2} a(d,\s)E_{d-1}+ b_l(\textrm{current's  anomalies})_l, \label{anom}
\end{eqnarray}
where $I_n$ are the invariants constructed from up to $n=(d-1)/2$ Weyl tensors $W_{ijkl}$ in $(d-1)-$dimensions and its covariant derivatives  as for example $I_2(d=5)=W_{ijkl}W^{ijkl}$ and $E_{d-1}$ is the corresponding Euler invariant. The standard methods of the perturbative $QFT_{d-1}$ calculations on arbitrary curved space background \cite{birell,duff,osborn3} provide the explicit form of the gravitational part of these anomalies \footnote{related at the lowest order to the corresponding $R^2-$type counter-terms } together with the one- and two-loop values of the anomaly coefficients $c$ and $a$ for considerably large family of $CFT_4$'s. One representative example is given by the $L_{int}=\frac{g_0}{16}\phi^4$-model in four dimensions\cite{hathrell}: 
\begin{eqnarray}
 c(g_0)\approx 1-\frac{5g_0^2}{36(4\pi)^4}+...,\quad\quad\quad a(g_0)\approx 1+\frac{85}{288}\left(\frac{g_0}{(4\pi)^2}\right)^4+..\nonumber
\end{eqnarray}
The exact values of $a$ and $c$ are known for  the extended $\cal N$-supersymmetric $SU(N_c)$ gauge theories and in particular for the 
${\cal N}=4$ SUSY Yang-Mills they do coincide: $c=a=\frac{(N_c^2-1)}{4}$, see for example refs.\cite{anselmi,anselmib,osborn1,beyon}. 

\vspace{0.5 cm}
2.3.1. \textit{The c-central charge}. An equivalent definition of the central charge  $c_2(d)=c(d)$, valid for arbitrary dimensions d, is as the properly normalized coefficient of the 2-point function of the stress-tensor (\ref{TT}). One important consequence  is 
that $c(d)$ in fact determines the norms of the stress-tensor's ``states'' $T_{ij}(0)|0>$ and therefore it must be positive $c(d)>0$ as a part\footnote{together with certain restrictions on the conformal dimensions $\Delta_k\geq 0$} of the \emph{unitarity} conditions for such $CFT_{d-1}$. For each given even-dimensional $d-1=2n$ (super)conformal model 
containing  few \emph{massless} free fields: $n_s$ scalars, $n_v$ vectors and $n_f$ spinor fields it is relatively easy to calculate from eq. (\ref{TT}) the exact value of the $c$- central charge \cite{anselmirev,osborn1}:
$$c_{free}(2n)=n_s+ 2^{n-1}(2n-1)n_f+\frac{(2n)!}{2((n-1)!)^2}n_v  \neq a_{free}(2n).$$ 
Notice  that the interaction is modifying  these values as in the case $n_s=1$ and  $n_f=0=n_v$ of the above mentioned $\phi^4$ model.

We next briefly recall the holographic derivation of eq. (\ref{TT}). The starting point is the effective action of the cubic QT Gravity considered as a functional of the arbitrary ``boundary metrics''
 $h^0_{ij}(x_i)=h_{ij}(y\rightarrow \infty,x_i)$, calculated in the linear approximation $g_{\m\nu}(y,x_i)=g_{\m\nu}^{vac}(y)+\kappa h_{\m\nu}(y,x_i)$ around a given $AdS_d$ vacuum $(\s_k^*,L_k)$ \cite{My_qtop,My_thol}:
\begin{eqnarray}
S_{gbl}(h^0_{ij})=\frac{\pi^{(d-1)/2}}{\Gamma((d-1)/2)}c_{QT}(f_k)\int d^{d-1}x_1 d^{d-1}x_2 h^0_{ij}(x_1)\frac{J^{ij,pq}(\vec x_{12})}{|\vec x_{12}|^{2d}} h^0_{pq}(\vec x_2)\label{hij}
\end{eqnarray}
with $c_{QT}(f_k)=C_0(f_k)\left(\frac{L_k}{l_{pl}}\right)^{d-2}$. According to the $AdS_d/CFT_{d-1}$ correspondence \cite{witt}, in the limit of large $(\frac{L_k}{l_{pl}})^{d-2}\approx N^2_c\gg 1$, it can be considered as a generating functional of  the 2-point correlation function of the stress-tensor of the  dual $CFT_{d-1}$: 
\begin{eqnarray}
\frac{1}{2} \frac{\delta^2 S_{gbl}(h^0_{ij})}{\delta h^0_{ij}(x)\delta h^0_{pq}(y)}\equiv <T^{ij}(\vec x)T^{pq}(\vec y))>_{CFT},\nonumber
\end{eqnarray}
This identification  provides an important non-perturbative information about the  strong-coupling limit of $c(d)$-central charge:
\begin{eqnarray}
  c_{QT}(d;L_k,\la,\m)=(\frac{L_k}{l_{pl}})^{d-2}(1-2\la f_k-3\m f^2_k),\quad f_k =\frac{\ka^2 L^2 W^2_k}{(d-2)^2}=\frac{L^2}{L_k^2}. \label{cch}
\end{eqnarray}
Notice that  the new QT gravity contributions can be considered as the next-to-leading order term in the $l_{pl}/L_k$ expansion of $c_{QT}(\la,\m)$  to the (leading) EH-order\footnote {in the dual $SU(N_c)$ CFT's with $N_c\gg1$ it can be considered as representing the next order in $1/N_c$ expansion.}, $i.e.$ $c_{EH}\approx(\frac{L_k}{l_{pl}})^{d-2}$. The fact that the stability condition selecting the physical  vacua $\C(f_k)>0$ of d-dimensional extended cubic Lovelock Gravity (\ref{qtop}), discussed in Sect.2.2., is responsible for the unitarity  of the dual $CFT_{d-1}$'s, is an example of the  holographic relationship between the \emph{causality} (and stability) of the (semi)classical Gravitational models and the \emph{unitarity} of their dual $QFT_{d-1}$'s.

The above expression (\ref{cch}) for the UV and IR values of the $c$-central charge (for $L_k=L_{UV/IR}$), together with the well known form \cite{rg} of the coinciding $c_{EH}=a_{EH}$-central functions in the case of EH-$AdS_d$ gravity-matter models, suggest the following definition of the cubic QT Gravity  $c$-central function \cite{My_thol}:
\begin{eqnarray}
c(d,\s)=\frac{(d-2)^{d-2}}{(-l_{pl}\kappa W)^{d-2}}\left(1-2\la\frac{\kappa^2L^2W^2(\s)}{(d-2)^2}-3\m \frac{\kappa^4L^4W^4(\s)}{(d-2)^4}\right),  \label{cfun}
\end{eqnarray}
It is worthwhile to mention its  relation to the holographic $\beta-$function (\ref{rg}):
\begin{eqnarray}
\beta(d,\s)=-2(d-2)\frac{W'}{\kappa^2W}\C(\s)= -c(d,\s)\frac{dg^{-1}(\s)}{d\s},\quad g(\s)=\frac{\kappa^2 (d-2)^{d-2}}{2(-l_{pl}\kappa W)^{d-2}},\label{cbeta}
\end{eqnarray}
where the function $g(\s)$ represents the one-dimensional analogue of the Zamolodchikov's coupling space metrics \cite{x,anselmi,freedtachy} and it turns out to coincide with the constant of $\Phi_{\s}$ 2-point function $g=C^{(2)}_{\Phi}$.

An important comment about the relations between the UV and IR values of the $c(d)$-central charge is now in order. In \emph{two dimensional} $CFT$'s ($i.e.$for $d=3$) it represents the central charge of the Virasoro algebra \cite{bpz} and according to the Zamolodchikov's $c$-theorem \cite{x} for a large class of unitary perturbed $CFT_2$'s, admitting peturbative UV and IR critical points, we have $c_{UV}>c_{IR}$ and the corresponding $c(d=3,\s)$-function is \emph{monotonically decreasing} during the massless RG flow. In higher dimensions, according to the Cardy's conjecture\cite{cardyth}, a similar role is played by the $a(d>3)$-central charge, which  satisfies  a d-dimensional  ``$a$-theorem'': $a_{UV}\geq a_{IR}\geq0$. On the other hand, for  the corresponding $c(d>3)$-central function there exist representative $QFT$'s examples demonstrating that one can observe  three different behaviours  along the RG flow: \emph{decreasing $\frac{dc}{dl}<0 $, increasing $\frac{dc}{dl}>0 $ or  non-monotonic $c(l)$-function}\footnote{see for example refs. \cite{anselmi,anselmirev} for the systematic discussion of such models}. Although in the few known cases such behaviour is related to the particular restrictions on the values of the ratio $N_f/N_c$ of numbers of the flavors and colors in the corresponding $SU(N_c)$ $QCD_4$ (see for example ref. \cite{anselmirev}), a clear field-theoretical explanation of these different RG evolutions of the $c-$central charge is still \emph{missing}. In order to find the corresponding \emph{holographic} explanation, the RG flows in certain (super)conformal models with \emph{non-equal central charges $c\neq a$} have to be studied. Their holographic description is known to require the explicit form of certain domain walls in the GB or/and cubic QTG extensions of the EH-gravity \cite{2,MPS,lovedw}. This problem is addressed in Sect.3. below in the frameworks of our proof of the cubic Quasi-Topological Gravity holographic ``$a/c$-Theorems''.

\vspace{0.5 cm}
2.3.2. \textit{The $a$-central charge}. The coefficient in front of the Euler invariant $E_{d-1}$ in the eq. (\ref{anom}), known as  $a$-central charge, possesses few others  holographic and purely $QFT_{d-1}$'s definitions and physical interpretations. Its $CFT$'s origin is as an independent parameter that, together with the c-central charge (\ref{ctensor}) and the energy fluxes parameter $t_4$,\footnote{Notice  that $t_4$ is zero for arbitrary ${\cal N}=1,2,4$ SUSY CFT's  as well as in all the $CFT_{d-1}$ duals of the GB Gravity models \cite{My_thol,MPS}; see Sect.6 below for further details.} determines the coefficients $\cal A$, $\cal B$ and $\cal C$ in the 3-point stress-tensor function (\ref{TTT}). For $d=5$ these relations have the form \cite{MPS,My_thol}:
\begin{eqnarray}
   a=\frac{13{\cal A}-2{\cal B} -40{\cal C}}{72},\quad\quad\quad t_4=-\frac{15(17{\cal A}+32{\cal B}-80{\cal C})}{4(9{\cal A}-{\cal B}-10{\cal C})}\label{tac}
\end{eqnarray}
Equivalently it takes part of  the definitions of 
 the ``structure constants'' of particular channels in the $T_{ij}(x)T_{pq}(0)|0>$ OPE's (see for example Sect.6 of ref. \cite{anselmiope}). There exist many CFT's examples for which the perturbative UV/IR values of the $a$-central charge have been calculated \cite{osborn1,osborn3,anselmib,beyon}. Differently from the $c(d>3)$-charges they are always decreasing $a_{UV}>a_{IR}$ and their free field limits (when appropriately normalized) reproduce the effective number of the \emph{massless degrees of freedom}.

We find however that similarly to the well known two dimensional case \cite{x}, the simplest and the most useful definition of the  $a$-central function (for all the values of $d$) is as the \emph{pre-potential}
\begin{eqnarray}
 \beta(\s)=\frac{1}{g(\s)} \frac{da(\s)}{d\s}\label{zam}
\end{eqnarray}
for the $\beta$-function of the corresponding dual $QFT_{d-1}$ model. Starting from the explicit form (\ref{rg}) of the $\beta(d,\s)$ in terms of the superpotential W, and following the original two dimensional proof \cite{x}, we find that the $a$-function is given by:
\begin{eqnarray}
 a(d,\s)=\frac{(d-2)^{d-2}}{(-l_{pl}\kappa W)^{d-2}}\left(1-2\la\left(\frac{d-2}{d-4}\right)\frac{\kappa^2L^2W^2(\s)}{(d-2)^2}-3\m \left(\frac{d-2}{d-6}\right)\frac{\kappa^4L^4W^4(\s)}{(d-2)^4}\right)\label{afun}
\end{eqnarray}
up to an additive constant $a_0$. Observe that it indeed coincides  with the standard holographic definition \cite{rg,My_thol,MPS} obtained from the corresponding  Quasi-topological Gravity DW's equations, that leads to monotonically decreasing $a$-function, when the ``null energy'' condition for the scalar matter is fulfilled. The above relation (\ref{zam}) between the $\beta$ and the $a$-functions has an important consequence:
\begin{eqnarray}
    \frac{da}{dl}=-\beta^2(\s)g(\s) <0,\label{unit}
\end{eqnarray} 
that  provides the simplest $QFT$'s proof of the monotonic decreasing of the $a$-central function (\ref{afun}) with the RG scale $l$ for \emph{positive definite ``metric''} $g(\s)>0$.

Again as in eq. (\ref{cch}) above, the ``1-d Zamolodchikov's coupling space metric'' $g(\s)$\footnote{although it is not quite appropriate to introduce metrics in one-dimensional spaces, this terminology is frequently used (see for example refs.\cite{anselmi,freedtachy}) following the analogy with  the sigma-model like metrics $G_{ab}(\s_a)d\s^ad\s^b$ in the case of many couplings $\s_b$, $a,b=1,2,...,N$ } is involved in  the relations between $\beta$ and the $a$-function derivatives. Its presence reflects the properties of the particular holographic  ``renormalizaton scheme'' (called Wilsonian in Sect.2. of ref. \cite{freedtachy}), which turns out to be \emph{different} from the Zamodchikov's one (used in his proof of $2d$ $c$-theorem), that leads to \emph{constant metric} and ``gradient flows'' \cite{x}. As is well known the RG transformation $d\tilde{\s}=\sqrt{g(\s)}d\s$ interpolates between the holographic and the Zamolodchikov's (constant metric) RG scheme \cite{anselmi,freedtachy,anselmit}, thus  changing the form of the $\beta$, $a$,  and $c-$functions, but leaving the values of the UV critical exponents and also the form of $\frac{da}{dl}$ (\ref{unit}) \emph{unchanged}. Let us also mention the importance of the ``geometric'' analogue $W(\s)<0$ of the $g>0$ condition, we are currently requiring \cite{lovedw} in order to ensure that the  DW's $(a)AdS_d$ geometries describing the holographic RG flows  belong to the family of standard asymptotically $AdS_d$ space-times having one (UV-type) boundary at $y\rightarrow \infty$ and one horizon (of IR type) at $y\rightarrow -\infty$ \footnote{thus  avoiding $(a)AdS_d$  spaces of two boundaries or two horizons that occur when $W$ has zeroes as in $d=3$ examples  studied in refs.\cite{nmg,holo,vicosa,3} and also for the Janus-type DWs of ref. \cite{rg}.}.

It is worthwhile to mention that the $a$-Theorem statement is in fact equivalent of the $QFT_{d-1}$ unitarity condition of  positive definiteness of the norm of states created by the trace $\Theta(\s,x_i)=\beta(\s) \Phi_{\s}(x_i)$ of the stress-tensor, $i.e.$
$$<\Theta(\vec{x})\Theta(0)>=\beta^2(\s)<\Phi_{\Delta}(\vec{x})\Phi_{\Delta}(0)>=\beta^2g(\s)|\vec{x}|^{-2\Delta_{\Phi}}>0.$$ It requires that $\beta^2g>0 $ or equivalently  $\frac{da}{dl}<0$, $i.e.$ $a$-central function must be decreasing during the massless RG flow.

Taking into account eqs. (\ref{zam}) and (\ref{cbeta}), we next derive the following relation between the $a$, $g$ and $c$-functions: 
\begin{eqnarray}
     c(\s)=g\frac{da}{dg},\label{ca}
\end{eqnarray}
demonstrating that the $a(W)$-central function, when considered as a function of $g(\s)$,  appears to be a \emph{pre-potential} for the $c$-function as well. This fact provides an equivalent form of the $a$-Theorem\footnote{advocated by D.Anselmi in ref. \cite{anselmit,anselmirev}}: 
\begin{eqnarray} 
   \Delta a\equiv a_{UV}-a_{IR}=-\int_{\s_{UV}}^{\s_{IR}} \beta(\s)g(\s)d\s=-\int_{g_{UV}}^{g_{IR}} \frac{c(g)}{g}dg >0  \label{ans}
\end{eqnarray}
relating the \emph{RG flux} of the $a$-function $\Delta a$  to the invariant area of the $\beta(\s)$-function, that now can be replaced by $c/g$. Notice that for vanishing Lovelock couplings $\la=0=\m$, i.e for the $QFT_{d-1}$'s duals to $EH$-gravity-matter models, the $a$ and $c$-functions are proportional of the ``couplings metric'': 
\begin{eqnarray}
a_{EH}(W)=c_{EH}(W)=2g(W))/\kappa^2,\nonumber
\end{eqnarray}
which considerably simplifies all the relations we have derived above.

Two comments concerning the QFT's and gravitational meaning of the $a$-central function are now in order:

$\bullet$ The important physical identification \cite{My_thol,entjapa} of the $a$-central charges (valid for both even and odd dimensions) as representing  the universal coefficient in the leading contribution to the \emph{entanglement entropy} in the dual $CFT$'s(in specific boundary geometries)  have been recently confirmed by numerous $CFT$'s and holographic calculations \cite{entrop}. The explicit DW's solutions and related to them Holographic RG flows in the dual QFT's provide an efficient tool for studying the off-critical properties  of the entanglement entropy.

$\bullet$ The relation  between the UV and IR values of the $a$-central charge and the \emph{tensions} of the cibic QT Gravity DW's is based on the following suggestive total derivative form of QTG action (\ref{qtop}):
\begin{eqnarray}
 S^{{\mathrm{eff}}}_{GBL} = -\frac{2}{l_{pl}(d-2)^{d-1}} \int \! d^{d-1}x \, dy \; \frac{d}{dy}  \left[e^{(d-1)A(y)} \, (-l_{pl}\kappa W)^{d-1}\, a(W) \right],\nonumber
\end{eqnarray}
derived  in ref. \cite{lovedw}. Taking into account the appropriate boundary terms \cite{bterms1,bterms2} needed for the consistent definition of the QTG action and dividing by the covariant volume, one can easily calculate by the Brown-York method the corresponding DW's \emph{tensions}\footnote{see for example ref.\cite{nmg,vicosa} for the details concerning the simplest $d=3$ case.} in the terms of the  $a$-central charges $a_{UV}$ and $a_{IR}$ and of the values of the coupling space metrics $g(UV/IR)\sim f^{\frac{2-d}{2}}(UV/IR)$.

\section{Holographic $a/c$-Theorems}
\setcounter{equation}{0}

The critical behaviour of each $QFT_{d-1}$, dual to the Quasi-Topological Gravity (QTG) coupled to scalar matter (\ref{qtop}), is described by a set of distinct $CFT$'s, characterized by their central charges $c_k$ and $a_k$, the sign and the values of the critical exponents $s_k$, the corresponding structure constants $C_{\Phi\Phi\Phi}$ and  and by $\cal A$, $\cal B$ and $\cal C$. By construction they represent the physical and topological $QTG$'s vacua ($L_k:s_k,c_k,a_k,{\cal A}_k,{\cal B}_k,{\cal C}_k$). The particular features of the ``holographic RG evolution'' of all this $CFT_{d-1}$-data between two consecutive critical points $\s_k$ and $\s_{k+1}$ is geometrically described by the corresponding $DW_{k,k+1}=DW(\s_k,\s_{k+1})$ QTG domain wall, separating these vacua.

An important question to be answered concerns the specific restrictions we have to impose on the Lovelock couplings $\la$ and $\m$ and on the shape of the superpotential $W(\s)$ in order to define a physically consistent \footnote{since the ${\cal N}=1$ and ${\cal N}=0$ (super-symmetric) $SU(N_c)$ gauge theories  have \emph{not available} strong-coupling $QFT$ description, the consistency requirements on the corresponding holographic $\beta$, $a$ and  $c$-functions are crucial for the independent holographic definition of such $QFT$'s.} ``holographic'' $QFT_{d-1}$. A part of these consistency conditions, related to the \emph{critical} CFT-data as for example $c>0$, $a>0$, $\Delta_{\Phi}>0$ and $0<s_{UV}^k<(d-1)/2$, have been already studied in Sect.2. above (see also refs. \cite{My_qtop,My_thol,lovedw}). 

\subsection{On the content of $a/c$- Theorems}
This section is devoted to the problem of the additional \emph{off-critical requirements} on the corresponding  $a(l)$, $c(l)$ and $g(l)$-functions, that allow us to extend the unitarity consistency of the $CFT_{d-1}(UV/IR)$'s to the massless phases of the conjectured dual \emph{non-conformal} $QFT_{d-1}$. This set of conditions, known as ``$a/c$-\emph{Theorems}'' of Zamolodchikov's and Cardy's \cite{x,cardyth}, have been intensively studied by the standard Wilson Renormalization Group and by perturbative field-theoretical methods as well \cite{cappelli,osborn1,schwimi,polchi}. Their holographic versions \cite{rg,girardelo}, originally formulated for certain special $c=a$ ${\cal N}=4$ super-symmetric $QFT_4$ duals to $d=5$  Einstein (super)Gravity coupled to matter, have been recently extended to the case $c\neq a$ of most general (non-supersymmetric) $QFT_{d-1}$'s \cite{My_thol,cteor} involving appropriate ``higher derivatives'' generalizations of the EH-gravity. As we have mentioned in Sect.2.3 above, the proof of the holographic EH-, GB- and extended Lovelock's \emph{$a$-theorems} \footnote{ $i.e.$ the $d>3$ analogues of the Zamolodchikov's c-theorem} is almost straightforward and valid for all the matter stress-tensors that satisfy the null energy condition. The only important restriction turns out to be  the ``positive definiteness'' of the Zamolodchikov's couplings space metric $g(\s)>0$, $i.e.$ $W(\s)<0$. Another natural \emph{causality/stability/unitarity} requirement is given by the positivity  of the physical UV and IR $c$- and $a$- central charges  and of the corresponding $c$- and $a$- central functions: $c(\s)\geq 0$ and $a(\s)\geq 0$, as reviewed  in Sect.2. above\footnote{the list of the papers devoted to the investigations of these two conditions in different GB- and Lovelock- holographic models and especially of these ones  on the \emph{physically allowed} $a$-values is extensive \cite{espanha,2,bala,MPS}.}. The \emph{complete description} of the Holographic RG flows and of the specific features of the physically consistent dual $QFT_{d-1}$'s requires more detailed study of the analytic properties of these central functions.

One of the advantages of the superpotential method  is that it allows to explicitly realize  the $a$, $c$ and $\beta$-functions in terms of $W(\s)$. Together with the running coupling $\s(l)$, obtained from the RG eq. (\ref{rg}), we have all the ingredients necessary to establish the dependence of the desired analytic properties of $a$- and $c$-functions on the Lovelock couplings and on the $W$'s parameters. We begin with few definitions (valid for $d>4$ and $d\neq6$): 
\begin{eqnarray}
c(\s)&=&\left(\frac{d-2}{-l_{pl}\kappa W}\right)^{d-2}\left(1-\frac{1}{f_+^{top}}\frac{L^2\kappa^2W^2}{(d-2)^2}\right)\left(1-\frac{1}{f_-^{top}}\frac{L^2\kappa^2W^2}{(d-2)^2}\right),\nonumber\\
a(\s)&=&\left(\frac{d-2}{-l_{pl}\kappa W}\right)^{d-2}\left(1-\frac{1}{f_+^{a}}\frac{L^2\kappa^2W^2}{(d-2)^2}\right)\left(1-\frac{1}{f_-^{a}}\frac{L^2\kappa^2W^2}{(d-2)^2}\right),\nonumber\\
\frac{dc}{dl}&=&-\frac{2(d-2)^d}{\kappa^2l_{pl}^{d-2}}\left(\frac{W'(\s)}{W(\s)}\right)^2c(l)\left(1-\frac{1}{f_+^{c'}}\frac{L^2\kappa^2W^2}{(d-2)^2}\right)\left(1-\frac{1}{f_-^{c'}}\frac{L^2\kappa^2W^2}{(d-2)^2}\right),\label{cac}
\end{eqnarray}
We have introduced together with the topological scales $(L_{o,top}^2)_{\pm}=(L_{top}^2)_{\pm}f_{\pm}^{top}$ (see sect.2.1.) the following two intermediate scales $(L_{a}^2)_\pm$ and $(L_{c'}^2)_\pm$:
\begin{eqnarray}
f_{\pm}^{top}&\equiv &\frac{L^2}{(L_{top}^2)_\pm}=-\frac{1}{3\mu}\left(\la\mp\sqrt{\la^2+3\m}\right),\nonumber\\
f_{\pm}^{a}&\equiv& \frac{L^2}{(L_{a}^2)_\pm}=-\frac{1}{3\mu}\left(\frac{d-6}{d-4}\right)\left(\la\mp\sqrt{\la^2+3\m\frac{(d-4)^2}{(d-6)(d-2)}}\right),\nonumber\\
f_{\pm}^{c'}&\equiv& \frac{L^2}{(L_{c'}^2)_\pm}=-\frac{1}{3\mu}\left(\frac{d-4}{d-6}\right)\left(\la\mp\sqrt{\la^2+3\m\frac{(d-6)(d-2)}{(d-4)^2}}\right).\label{fac}
\end{eqnarray}
Notice that $f^{\eta}_{\pm}$ (where $\eta=top,a,c'$) are not always real (and positive) numbers and  depending on the signs and values of the $\la$ and $\m$ (and on the value of $d$) they might also represent the two complex conjugate solutions $f^{\eta}_{+}=(f^{\eta}_-)^*$  of the corresponding quadratic equations.

With the explicit form (\ref{fac}) of the $f^{\eta}_{\pm}$ at hand, and the conditions $c_{UV/IR}>0$ and $0<\Delta_{UV}^+< d-1$ already established in ref. \cite{lovedw}, we are going to derive the remaining restrictions on $\la, \m$ and $f(\s)=L^2\kappa^2W^2(\s)/(d-2)^2$ such that one of the following \emph{three versions} of the $a/c$ -Theorem are satisfied:

(1) \emph{Standard $a/c$-Theorem}:
\begin{eqnarray}
              c>0\quad,\quad a>0 \quad,\quad \frac{da}{dl}<0 \quad,\quad \frac{dc}{dl}<0 \label{stronger}
\end{eqnarray}
$i.e.$ both $a$- and $c$-central functions are positive and monotonically decreasing in a given \emph{massless} phase of the dual $QFT_{d-1}$ between two critical points $\s_{UV}$ and $\s_{IR}$\footnote{defined as two consecutive zeros of the $\beta-$function (\ref{rg})}. It represents a natural $c\neq a$ extension of the standard holographic $a=c$-theorem \cite{rg,My_thol,girardelo}.

(2) \emph{Modified c-Theorems}:
\begin{eqnarray}
       (2a)\quad\quad\quad\quad\quad\quad  c>0\quad\quad\quad a>0 \quad\quad\quad\quad \frac{da}{dl}<0 \quad\quad\quad\frac{dc}{dl}>0 \label{modifa}
\end{eqnarray}
$i.e.$ $a(l)$ is decreasing as always, but now $c(l)$ is monotonically \emph{increasing}; or else we can have
 \begin{eqnarray}
       (2b)\quad\quad\quad  c>0\quad\quad a>0 \quad\quad \frac{da}{dl}<0 \quad\quad \textrm{and}\quad \frac{dc}{dl}\quad \textrm{changing sign} \label{modifb}
\end{eqnarray}
$i.e.$ the $c(l)$-central function is \emph{non-monotonic}, due to  the existence of a critical scale $l_{cr}^{c}=(L_{c'}^2)_\pm$ such that $\frac{dc}{dl}(l_{cr}^c)=0$  (and critical coupling $\s_{UV}<\s(l_{cr})<\s_{IR}$), where the $c$-derivative changes its sign.

(3) \emph{$a$-Weaken Theorems}: The conditions of this weak form of the  theorem are the same as in (1),(2a) and (2b) above, \emph{except} that now the a-function can change its sign within the $(\s_{UV},\s_{IR})$ interval. This leads to the existence of another critical scale $l_{cr}^a=(L_{a}^2)_{\pm}$ (and mass $M_{cr}^2\sim 1/L_{a}^2$) defined by the a-function zeros $a(l_{cr})=0$. In fact such RG flows are not massless anymore, due to the critical mass scale up to where we have a consistent (causal and unitary) $QFT_{d-1}$.

\subsection{On the methods of proof}

Similarly to the derivation of the restrictions on the $f_k$ (and $h_k$) in Sect.2.1.(see ref. \cite{lovedw}) that lead to stable physical vacua in the different domains of admissible values of $\la$ and $\m$, it is convenient to keep our choose  of the fundamental  scale  as  $L^2=L_{0+}^2(\mu_+)=L_{0-}^2(\m_-)$. Let us remind that this fact, together with the existence of topological vacua $c(f_{\pm}^{top})=0$, determines  two distinct gravitational models (of equal fundamental scales): ($\la$,$\m_{\pm}$) characterized  by the particular  form of the Lovelock coupling $\mu$ as a function of $\la$ and of the corresponding $f^{\eta}_{\pm}(\la,\m_{\pm})\equiv f^{\eta}_{\pm}(\la)$ as well, see for example eqs. (\ref{h}) and (\ref{h tilde}). 

The crucial step in the proof of the theorem is to establish which of the $f^{\eta}_{\pm}(\la)$'s are real and positives, how they are ordered and finally, the dependence of this ordering on the values of $-\infty<\la <1/3$ . In fact in most of the cases it is enough to know the biggest and the smallest positives $f^{\eta}_{\pm}(\la)$'s. Given the explicit form (\ref{fac}) of $f^{\eta}$, it is not difficult to find the conditions that ensure their reality, $i.e.$ to derive the corresponding  restrictions on $\la$ and $\mu$ such that the expressions under the square roots are positive, namely:
\begin{eqnarray}
\la^2-3\mu X_{\eta}(d) \geq 0 ,\quad X_{top}=-1,\quad X_{a}=-\frac{(d-4)^2}{(d-2)(d-6)},\quad  X_{c'}=-\frac{(d-2)(d-6)}{(d-4)^2}\label{X}
\end{eqnarray}
In order to find the reality restrictions for the $\mu_{\pm}$ models, we substitute the $\mu_{\pm}$ values (\ref{mu pm}) in the eqs. $\la^2-3\mu X_{\eta}(d)=0$ that lead to the following quadratic equations for the ``critical'' values $\la_{\eta}^{(1,2)}$, where $f^{\eta}_{+}(\la)=f^{\eta}_{-}(\la)$:   
\begin{eqnarray}
\frac{81}{4}X_{\eta}^2\la^2+\left(\frac{81}{2}X_{\eta}+27\right)\la-9\left(\frac{3}{4}+X_{\eta}\right)=0,
\end{eqnarray}
Their  solutions have the following simple form:
\begin{eqnarray}
\la_{\eta}^{(1,2)}=-\frac{2}{3X_{\eta}^2}\left(\frac{3}{2}X_{\eta}+1\mp\left(1+X_{\eta}\right)^{3/2}\right),\nonumber
\end{eqnarray}
or more explicitly
\begin{eqnarray}
\la_{top}^{\pm}=\frac{1}{3},\quad\quad
\la_{\mu_{+}}^a=\frac{d(d-4)}{3(d-2)^2}<\frac{1}{3},\quad\quad
\la_{\mu_-}^a=\frac{1}{3}-\frac{4}{3(d-6)^2}<\la_{\mu_+}^a\nonumber\\
\la_{\mu_{\pm}}^{c'}=\frac{(d-6)(d-2)}{3(d-4)^4}\left(24+d(d-8)\mp\frac{16}{\sqrt{-(d-6)(d-2)}}\right)\in\mathbb{C}, \ d\ge7, \label{lambs}
\end{eqnarray}
Together with the well known  restriction $ \la \leq \la_{top}^{\pm}=1/3$ providing the $f_{top}^{\pm}$ reality, we realize that the $f_{a}^{\pm}(\la;\mu_{+})$ are real for $\la<\la_{\mu_+}^a$ and $f_{a}^{\pm}(\la;\mu_{-})$ - when $\la<\la^a_{\mu_-}$. For $d>6$ both $f_{c'}^{\pm}(\mu_+)$ and  $f_{c'}^{\pm}(\mu_-)$ are always real, while for $d=5$ this condition is satisfied for $\la \in (-9+16/\sqrt{3}, 1/3)$ only.

Although not always all of the $f^{\eta}_{\pm}$'s are real and positive, the factorization (\ref{cac}) of the three quadratic forms in simple multipliers (involving their roots) is the simplest and very well known method of studying the conditions under which they can have definite signs, say $c>0,  a\geq 0$ and $\frac{dc}{dl}\leq 0$. In the case of the standard $a/c$- Theorem, we have therefore to find the conditions such that : 
\begin{eqnarray}
\left(1-\frac{f(\s)}{f_{+}^{\eta}}\right)\left(1-\frac{f(\s)}{f_-^{\eta}}\right),\quad \textrm{with}\quad f(\s)\equiv \frac{L^2\kappa^2W^2(\s)}{(d-2)^2}\label{f_eta}
\end{eqnarray}
is positive for all the $\eta=top,a,c'$'s. There are four distinct cases for each one of the $\eta$'s: ($i$) $f^{\eta}_+>0$, $f^{\eta}_-<0$; ($ii$) $0<f^{\eta}_+<f^{\eta}_-$; ($iii$) $f^{\eta}_{\pm}<0$; ($iv$) $f^{\eta}_{\pm}\in\mathbb{C}$, $(f^{\eta}_+)^*=f^{\eta}_-$ that have to be studied separately.

In the case ($i$) the quadratic form (\ref{f_eta}) is positive definite only when the first multiplier is positive: $f(\s)<f_{+}^{\eta}$ for all the $\s\in(\s_{IR},\s_{UV})$. Due to the fact that  $|W(\s)|$ has in this interval one maximum $\s=\s_{IR}$, the above inequality is satisfied  along the entire RG flow if:
\begin{eqnarray}
f(\s_{IR})<f_{+}^{\eta} \quad \textrm{or equivalently } \quad L^{\eta}_{+} < L_{IR},\nonumber
\end{eqnarray}
thus introducing a certain minimal scale $L^{\eta}_{+}$ for this model\footnote{We recall that when $\s_{UV/IR}$ represent two consecutive extrema of $W(\s)$ such that $|W(\s_{UV})|<|W(\s)|<|W(\s_{IR})|$, the restrictions on $W(\s)$ within the interval $(\s_{IR},\s_{UV})$  can be written as an equivalent requirement on the corresponding (boundary/horizon) values $|W(\s_{IR/UV})|$.}.
We next consider the case ($ii$), where the condition that (\ref{f_eta}) is positive, requires that the two multipliers to have the same sign. When both are \emph{positives} we find again the above minimal scale requirement, while in the case  when both are \emph{negatives} the restriction $f(\s)>f_{-}^{\eta}$  defines the new region of \emph{maximal scale} $L^{\eta}_{-}$, $i.e.$ 
\begin{eqnarray}
f(\s_{UV})>f_{-}^{\eta} \quad \textrm{or equivalently } \quad L^{\eta}_{-} > L_{UV},\nonumber
\end{eqnarray}
Finally, in the remaining two cases ($iii$) e ($iv$) the eq. (\ref{f_eta}) is automatically positive and no other restrictions on the values of $W(\s)$ are needed .

The next step of the proof consists in the ordering  of all the $f_{\eta}^{\pm}$'s within the intervals of admissible $\la$ values, where they are \emph{real and positive}. The knowledge of their explicit forms (\ref{fac}) as certain functions of $\la$ and $d$ is indeed enough to solve analytically this problem. We find however its \emph{equivalent} graphical solution  to be  much more efficient. The plots of all the  $f_{\eta}^{\pm}(\la)$s for both $\mu_{\pm}$ models, for different values of $d$ and for specific intervals of values of $\la$ are given on Figs. (\ref{fig:1a})-(\ref{fig:1c}) and (\ref{fig:2a})-(\ref{fig:2b}). They provide the particular orderings of the $f_{\eta}$'s for all the cases of interest, to be used in the proof of the $a/c$-Theorems in the Sects.3.3 and 3.4 below. One representative example is given by :
\begin{eqnarray}
f^{c'}_-<f^{a}_-<0<f^a_{+}<f^{top}_+<f^{c'}_+<f^{top}_-,\nonumber
\end{eqnarray}
valid for $\mu_+$ model in five dimensions and within the interval $\la \in(1/4,1/3)$ only, as one can see from fig.(\ref{fig:1c}). Once the ordering of  all the  positive $f_{\eta}^{\pm}$'s  is established, the last step in the proof consists in deriving  the proper intersections  between  the  restrictions  we have found for the three distinct cases of $\eta=top,a,c'$'s.

The fact that two of the ``scale ratios'' $f^{\eta}_{\pm}(\la,\m)$ and also the corresponding  ``critical values'' $\la_{\eta}^{(1,2)}$ explicitly depend on the space-time dimension $d$, makes also necessary to consider \emph{separately} the proofs of the $a/c$ -Theorems in the following three cases: (a) $d=5$; (b) The GB-limit  $\mu=0$ and (c) $d\ge7$. As we are going to show in the next subsections they give rise to rather \emph{different} restrictions on the Lovelock couplings and on the values of the $W$ as well.

\subsection{$d=5$ Proof: Standard $a/c$-Theorem}

\subsubsection{$\mu_+$ model} 

The statement of the standard $a/c$ -Theorem for the $\m_+$ model now reads: 

\emph{ The $a(l)$- and $c(l)$-functions are both positive and monotonically decreasing during the RG flow between $\s_{UV}$ and $\s_{IR}$ only when the following restrictions}:
\begin{eqnarray}
 &&\bullet \ -\infty<\la<\frac{5}{27}: \ L_{IR}^2>\frac{L^2}{f^{top}_+(\la;\mu_+)},\quad f_{+}^{top}(\la;\mu_+)=\frac{1}{3\la}\left(1-\sqrt{1-3\la}\right),\nonumber\\
 &&\bullet\ \frac{5}{27}<\la<\frac{1}{3}: \ L_{IR}^2>\frac{L^2}{f^a_+(\la;\mu_+)},\nonumber\\
 &&f_{+}^a(\la;\mu_+)=\frac{1}{3\la+\sqrt{3\la\left(1+3\la+\frac{2}{3}\sqrt{1-3\la}\right)+\frac{2}{3}\left(1-\sqrt{1-3\la}\right)}}\label{ac-teor}
\end{eqnarray}  
\emph{are satisfied}.

\textit{Proof.} Following the $\mu_+$ model description given in Sect.2.1., we first consider the (a1) case corresponding to $\la\in (-\infty,\frac{1}{4})$ and $\mu_+(\la)\geq 0$. 

(a)\textit{Reality conditions.} It turns out that in this case  $f_{\pm}^{top}$ are always real: $f_+^{top}(\la)>0$, while $f_-^{top}(\la)<0$. 
We next realize that the condition $\la^2\geq\mu_+$, ensuring the reality of $f_{\pm}^a(\la,\mu_+)$'s, can be rewritten as 
$$(\la-\frac{5}{27})(\la+1)\geq 0.$$ 
Together with their explicit form (\ref{fac}), it leads to the conclusion that both $f_{\pm}^a(\la)$ are real and \emph{negative} for $\la<-1$ and real and \emph{positive} for $\la \geq 5/27$.\footnote{ Notice that at the ``degenerate points" $\la_{\pm cr}^a=-1,5/27$ the topological and a-scales do coincide, say $f_+^a(5/27)=f_-^a(5/27)=f_+^{top}(5/27)=9/5$, etc. and  the corresponding dual $CFT$'s have vanishing central charges $a(\la_{cr})=0=c(\la_{cr})$.} Finally, we find that the requirement $\la^2 \geq 9\mu$ for the reality of $f_{\pm}^{c'}(\la,\mu_+)$ implies : 
$$(\la+9+\frac{16}{\sqrt{3}})(\la+9-\frac{16}{\sqrt{3}})\geq 0.$$ 
Then it becomes evident that both $f^{c'}_{\pm}<0$  are real and negative within the interval $\la\in (-\infty,-9-\frac{16}{\sqrt{3}})$, while both are positive $0<f^{c'}_+<f^{c'}_-$ when  $\la\in(-9+\frac{16}{\sqrt{3}},1/4)$.

\begin{figure}[ht]
\centering
\subfigure[]{
\includegraphics[scale=0.7]{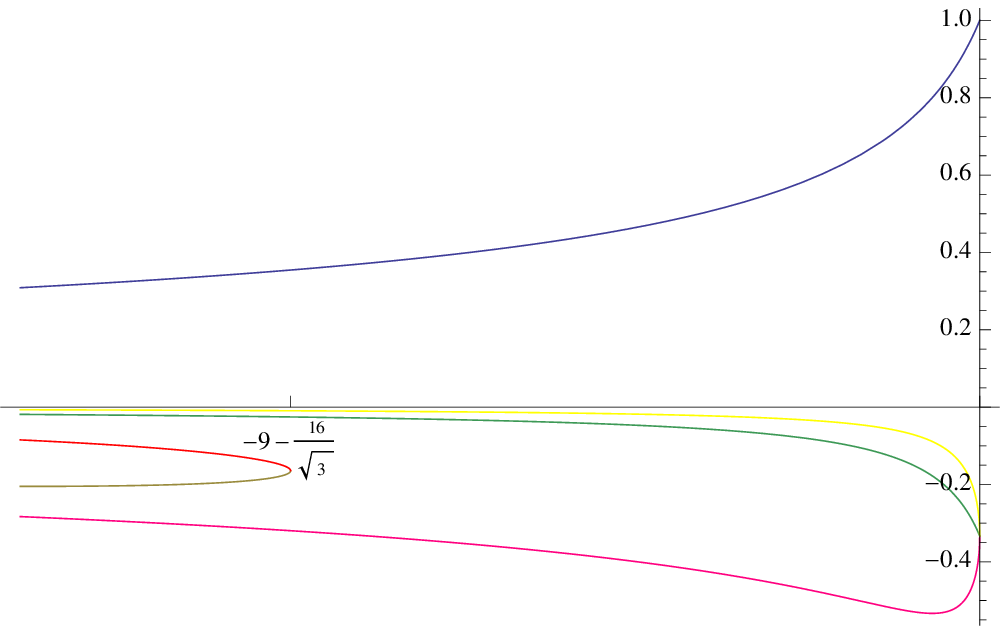}
\label{fig:1a}
}
\subfigure[]{
\includegraphics[scale=0.7]{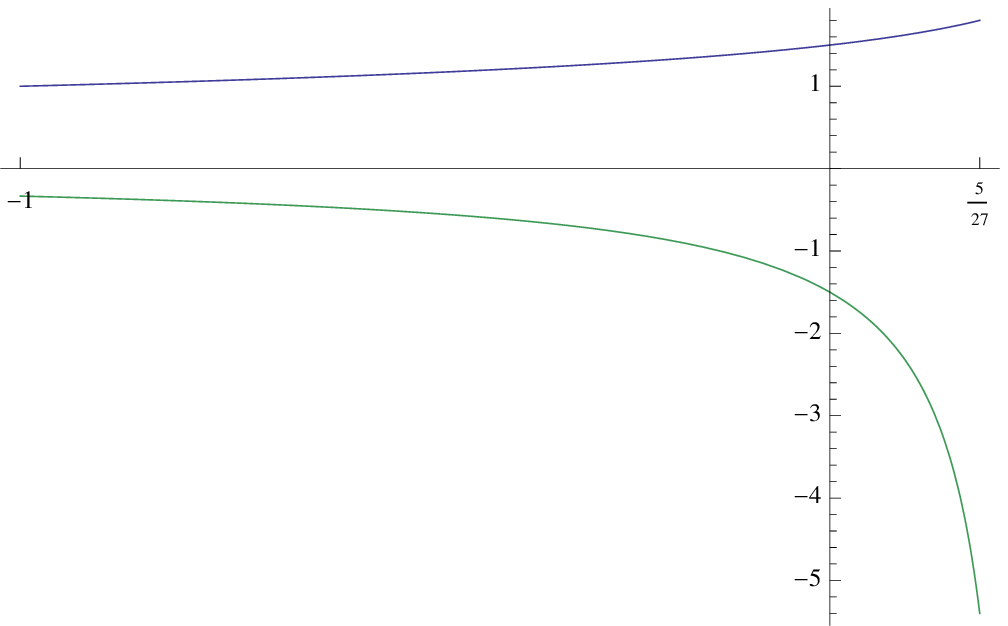}
\label{fig:1b}
}
\subfigure[]{
\includegraphics[scale=0.7]{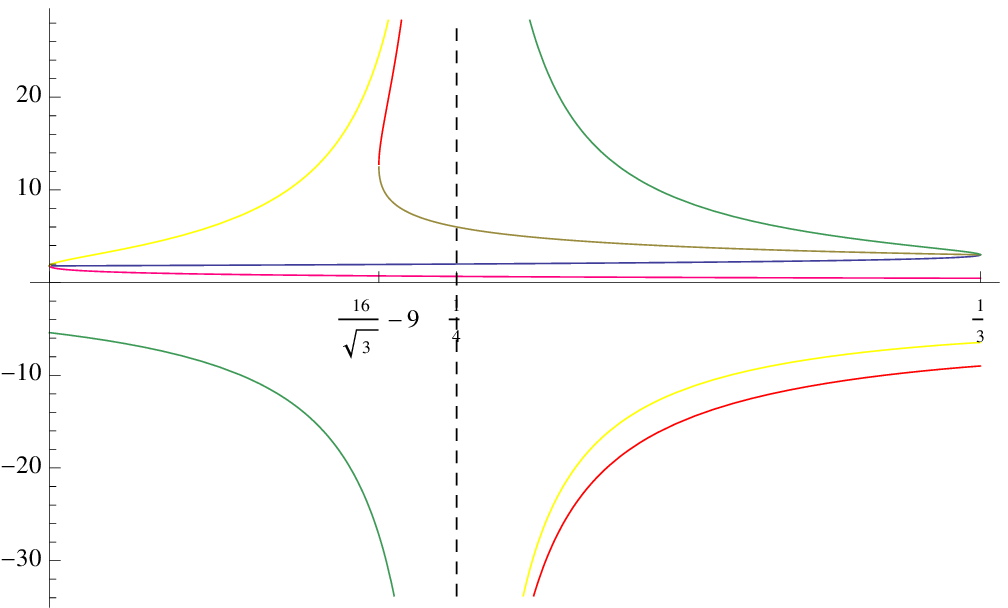}
\label{fig:1c}
}
\label{fig:1}
\caption{The $f^{\eta}_{\pm}(\la;\mu_+)$ curves for $d=5$: $f^{top}_+$ is plotted in blue; $f^a_+$ in pink; $f^{c'}_+$ is beige; $f^{top}_-$ in green; $f^{a}_-$ in yellow and  $f^{c'}_-$ in red. The $\la$ intervals are as follows: on Fig.(a) $\la<-1$ with ($-1,0$) as reference point(r.p.); on Fig.(b) $-1<\la<5/27$ with ($0,0$) as r.p.; and  on Fig.(c) $5/27<\la<1/3$ with  ($5/27,0$) as r.p.}
\end{figure}

(b)\textit{Ordering}. According to our discussion in sect.3.2., the simplest method for establishing of the ordering of all the real positive $f_{\pm}^{\eta}$'s given by eqs. (\ref{fac}) and (\ref{ac-teor}) is the graphical one. As one can see from their plots on the Figs. (\ref{fig:1a})-(\ref{fig:1b}), for the values of $\la$ within the interval $-\infty<\la<\frac{5}{27}$, only $f^{top}_+$ is positive, while for  $\la\in(\frac{5}{27},-9+\frac{16}{\sqrt{3}})$ we find that:
\begin{eqnarray}
f^{top}_-<0<f^a_{+}<f^{top}_+<f^a_{-},\label{laf2}
\end{eqnarray}
Finally as shown on Fig.(\ref{fig:1c}) for $\la \in(-9+\frac{16}{\sqrt{3}}, \frac{1}{4})$  the following ordering :
\begin{eqnarray}
f^{top}_-<0<f^a_{+}<f^{top}_+<f^{c'}_+<f^{c'}_-<f^a_{-},\label{laf3}
\end{eqnarray}
takes place. 

(c)\textit{$f_{UV/IR}$ restrictions.} Remember that the restrictions on the values of $f_{UV}$ and $f_{IR}$ (or $L_{UV/IR}$), for which the standard $a/c$-Theorem is satisfied  are given by  the common solutions of the following inequalities:
\begin{eqnarray}
    (f-f_-^{top})(f-f_+^{top})<0,\quad\quad (f-f_-^{a})(f-f_+^{a})>0,\quad\quad (f-f_-^{c})(f-f_+^{c})>0,\label{munegat}
\end{eqnarray}
Taking into account eqs. (\ref{laf2}) and (\ref{laf3}), we conclude that:
\begin{eqnarray}
 \bullet\ -\infty<\la<\frac{5}{27}:\quad 0<f_{UV}<f_{IR}\leq f_+^{top},\quad\quad 
 \bullet\ \frac{5}{27}<\la<\frac{1}{4}:\quad 0<f_{UV}<f_{IR}\leq f_+^{a}. \nonumber
\end{eqnarray}  
represent all the solutions of the requirements (\ref{munegat}).

We next consider the $\mu_+$ model within the interval  $\frac{1}{4}<\la<\frac{1}{3}$, where $\mu_+<0$ and both $f_{\pm}^{top}$ are now \emph{positive}, $i.e.$ the cases (a2) and (a3) of sect.2.1. Similar arguments as the ones given above, together with the plots of the curves $f_{\pm}^{\eta}(\la,\mu_+)$ presented on Fig.(\ref{fig:1c}) lead us to the conclusion that:  
\begin{eqnarray}
f^{c'}_-<f^{a}_-<0<f^a_{+}<f^{top}_+<f^{c'}_+<f^{top}_-\label{a23}
\end{eqnarray}
Due to the fact that now $\mu+$ is negative, the conditions for validity of the $a/c$-Theorem have the form (\ref{munegat}), but now with inverted signs, $i.e.$:
\begin{eqnarray}
    (f-f_-^{top})(f-f_+^{top})>0,\quad\quad (f-f_-^{a})(f-f_+^{a})<0,\quad\quad (f-f_-^{c})(f-f_+^{c})<0,\label{muposit}
\end{eqnarray} 
Nevertheless, their solution is again given by 
\begin{eqnarray}
\bullet\ \frac{1}{4}<\la<\frac{1}{3}:\quad\quad 0<f_{UV}<f_{IR}\leq f_+^{a}, \nonumber
\end{eqnarray}  
reflecting the corresponding new $f_{\pm}^{\eta}(\la,\mu_+)$'s ordering (\ref{a23}). 

Combining all these results we arrive at the statement of the standard $a/c$ -Theorem announced  at the beginning of this section.

\subsubsection{$\mu_-$ model}
 The $\mu_-$ model corresponds to particular values (\ref{mu pm}) of  $\mu_-(\la)$ for $\la \in (-\infty,1/3)$. 
 
\textit{Reality conditions.} As shown in Sect.2.1., the  $f^{top}_{\pm}(\la;\mu_-)$ given by (\ref{h tilde}) are both negative for all the \emph{negative} values of $\la $. They become real and positive $0<f_+^{top}<f_-^{top}$ for $\la \in (0, 1/3)$. Taking into account the explicit form (\ref{fac}) of $f_{\pm}^a(\la,\mu_-)$ and $f_{\pm}^{c'}(\la,\mu_-)$, we  next realize that for $\la<1/3$ they are \emph{always real}. 

\begin{figure}[ht]
\centering
\subfigure[]{
\includegraphics[scale=0.7]{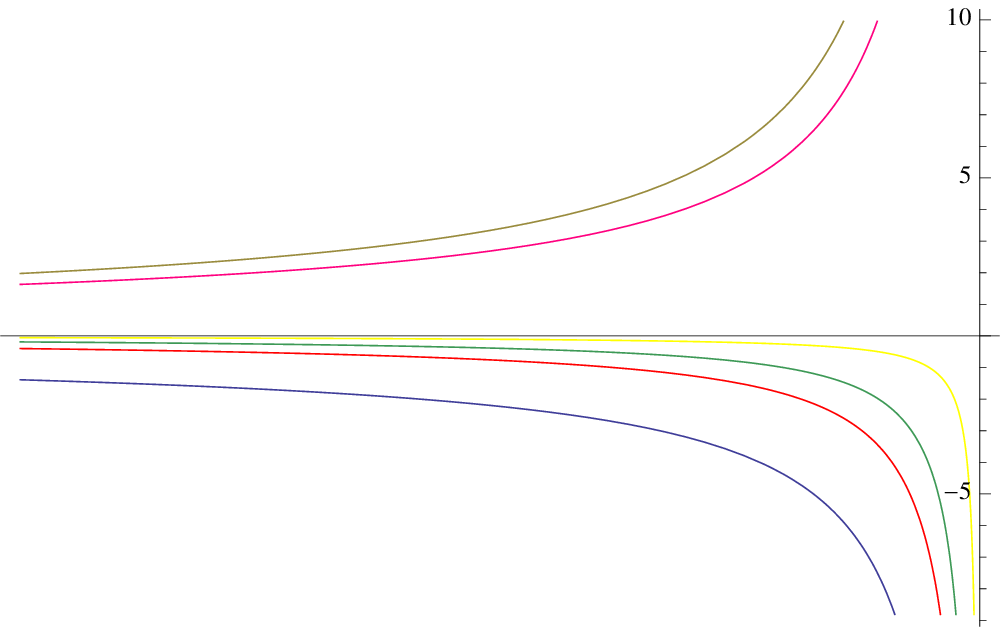}
\label{fig:2a}
}
\subfigure[]{
\includegraphics[scale=0.7]{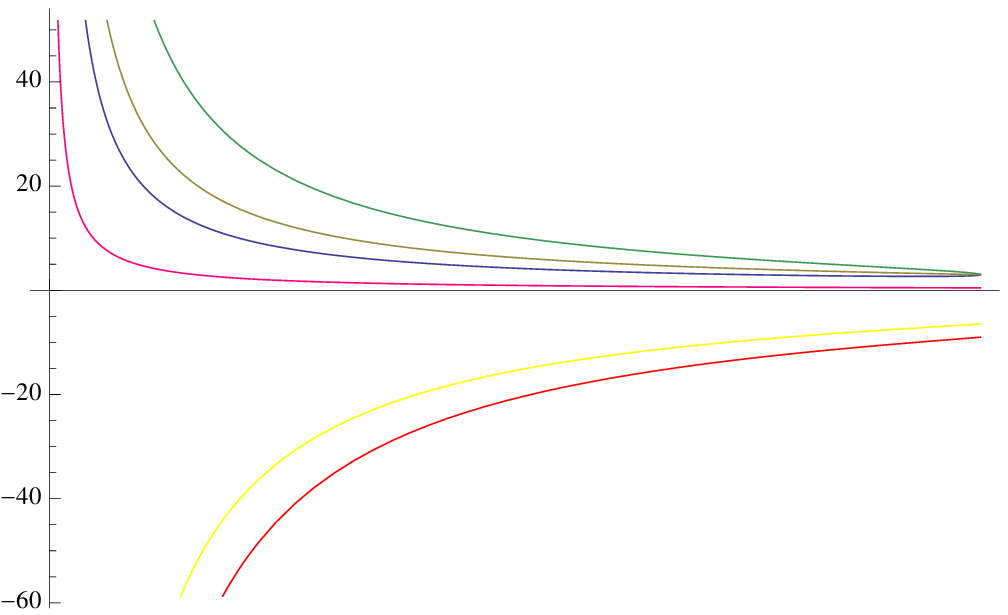}
\label{fig:2b}
}
\label{fig:2}
\caption{The $f^{\eta}_{\pm}(\la;\mu_-)$ curves for  $d=5$: $f^{top}_+$ is plotted in blue; $f^a_+$ in pink; $f^{c'}_+$ in beige; $f^{top}_-$ in green; $f^{a}_-$ in yellow and $f^{c'}_-$ in red. Fig.(a) corresponds to $\la<0$ and the Fig.(b) to $0<\la<1/3$.}
\end{figure}

\textit{Ordering.} As one can see from  the Figs. (\ref{fig:2a}) and (\ref{fig:2b}),
the following orderings of $f_{\pm}^{\eta}(\la,\mu_-)$'s take place: for $\la<0$ (see (\ref{fig:2a})) we have
\begin{eqnarray}
  f^{top}_+<f^{c'}_{-}<f^{top}_-<f^a_-<0<f^a_+<f^{c'}_+,
\end{eqnarray}
while for  $0<\la<\frac{1}{3}$ (see \ref{fig:2b}) we find that $f^{c}_-$ and $f^{a}_-$ are  now negative and the remaining $f^{\eta}_{\pm}$'s are ordered as follows:
\begin{eqnarray}
f^{c'}_-<f^a_-<0<f^a_+<f^{top}_+<f^{c'}_+<f^{top}_-.
\end{eqnarray}

\textit{$f_{UV/IR}$ restrictions.} Similarly to the $\mu_+$ model in the region where $\mu_+<0$, the restrictions imposed on $f_{UV/IR}$ by the standard $a/c$ -Theorem for the $\mu_-$ model  are again given by eqs. (\ref{muposit}). They are satisfied for  all the values of $f_{UV/IR}$ such that $0<f_{UV}<f_{IR}\leq f_+^{a}$. Hence the  $a/c$ -Theorem statement for the five dimensional $\mu_-$ model takes the following simple form: 

\textit{The condition}
\begin{eqnarray}
&&\bullet \ -\infty<\la<\frac{1}{3}: \ L_{IR}^2>\frac{L^2}{f^a_+(\la;\mu_-)}\label{mumin}.
\end{eqnarray}
\textit{guarantees that both $a-$ and $c-$central charges are positive and monotonically decreasing during the RG flow from $\s_{UV}$ to $\s_{IR}$}.

\subsubsection{Topological versus $a$-scales}

The description of the models with arbitrary \emph{negative} values of the Lovelock couplings $\la$ and $\mu$ (not necessarily related as in the case of $\mu_-$ model) requires separate discussion. Due to the fact that now $f^{top}_{\pm}$ are either both negative or complex, we realize that $c(f)>0$ for all $f$'s without any restrictions. Hence in the \emph{absence of topological vacua} no natural smallest or/and largest  scales $L^{top}_{\pm}$ exist. Indeed one can introduce as such a fundamental scale the smallest of the physical vacua scales by imposing $f_{IR}=1$. We have to remind however that the requirement $a(f)>0$ (in the case when one of the $f_{\pm}^a$ is positive) introduces  certain restrictions on the values of physical scales, namely $f_{UV/IR} \leq f_-^a$ . As one can see from the explicit forms (\ref{fac}) of $f_{\pm}^a$ and $f_{\pm}^{c'}$  for negative $\la$ and $\mu$ we have that: $0<f_-^a <f_-^{c'}$. Then the Standard $a/c$ -Theorem conditions lead to the following requirement:
\begin{eqnarray}
 0<f_{UV}<f_{IR} \le f_-^a \quad\quad \textrm{or}\quad\quad  L_{UV}>L_{IR} \geq L_-^a, \label{atheo}
\end{eqnarray} 
with $f_-^a=L^2/(L_-^a)^2$ and $ f_{IR}=L^2/L_{IR}^2$, thus introducing a  \emph{minimal a-scale} $L_-^a$. Therefore the most appropriate normalization in this case is to impose the condition $f_-^a=1$, which fixes the fundamental QT-Gravity scale as $L=L_-^a$. Similarly to the topological scale normalization, described in Sect.2.1., that leads to the specific relations (\ref{mu pm}) between the Lovelock couplings $\la$ and $\mu$, the choice $f_-^a=1$ give rise  to the following \emph{linear} relation:
\begin{eqnarray}
     |\mu|= \frac{1}{3}+2|\la|, \quad\quad\la<0,\quad\mu<0 \label{a-model}
\end{eqnarray}
Then the $d=5$ $a/c$-Theorem requirements read: $ 0<f_{UV}<f_{IR}<1$. Notice that again we get the same UV/IR-physical scales restriction (\ref{atheo}), which confirms the fact that the form of the restrictions on the physical scales \emph{does not depend on the choice of the fundamental QTG scale $L$}.

An important comment concerning  the nature of the $a$-scale is  now in order. Notice that $L_-^a$ has  rather \emph{different} gravitational and $QFT_{d-1}$ meaning compared to the one of the topological and physical scales. The $L_{\pm}^{top}$ and $L_{UV/IR}$ scales are related to the cosmological constants $\Lambda_{UV/IR}= -(d-1)(d-2)/2L^2_{UV/IR}$ of the $AdS_d$ vacua solutions of QT Gravity-matter model (\ref{qtop}). According to the $AdS/CFT$ correspondence, these ``vacua scales'' define the central charges of the dual $CFT_{d-1}^{UV/IR}$ (see Sect.2.2. above) and geometrically they do correspond  to the boundary/horizon's of the $(a)AdS_d$ DWs space-times.  Instead, the $a$-scale $L_-^a$, which is characterized by the \emph{vanishing} $a(f_-^a)=0$ central charge, in general is not related neither to the Lovelock gravity vacua $\s_k$ nor to the zeros of the $\beta-$ function of the dual $QFT_{d-1}$\footnote{excepts in the very special degenerate cases, when the $a$-scale coincides with one of the physical scales $L_-^a=L_{IR}$ or/and with one of the topological scales $L_-^a=L^{top}_+$. Such ``$a$-critical points'' $\s_{cr}^a$, $i.e.$ when $a(\s_{cr}^a)=0$, do correspond to a QT Gravity vacuum dual to $CFT_{d-1}(\la=5/27)$ of vanishing central charges.}.

Let us remind that the interpretation of $a(l)$ as an entanglement entropy in the dual $CFT_{d-1}$ \cite{entjapa,entrop,My_thol} indicates that $a(\s_{cr}^a)=0$ is the end point of the consistent causal description in  \emph{both} the gravitational and its dual $QFT$ models. In this context we should  mention that our (smooth)  DW's solutions for a large class of superpotentials $W(\s)$ in the cubic Quasi-Topological Gravity (\ref{qtop}), \emph{do not provide}  arguments for an eventual interpretation of the zeros of  $a(\s)$-central function as a new saddle point of the effective gravitational action, describing \emph{first order} phase transitions in the dual $QFT_{d-1}$, as suggested in ref. \cite{Myers-new} in the case of thin walls of GB gravity.

\subsection{Comments on $d=5$ $c$-Modified and $a$-Weaken Theorems}

The conditions of validity of the Standard $d=5$ $a/c$-Theorem established in Sect.3.3. above, provide certain $\la$- and $\mu$-dependent restrictions on the values of the Superpotential and equivalently on the ``physical scales'' $L_{UV/IR}$, during the RG flow, such that  both $a(l)$- and $c(l)$-central functions are positive and monotonically decreasing. The most important consequence of our analysis of the requirements (\ref{munegat}) and (\ref{muposit}) is that: 

\emph{The QTG-induced $d=5$ $c(l)$-central function is always decreasing, $i.e.$ $\frac{dc}{dl}<0$  for all the allowed values of the Lovelock couplings and therefore no Modified Holographic $c$-Theorem can exists in five dimensions.} 

This statement establishes a very strong \emph{non-perturbative} restriction on the  properties of the holographic c-function of the $a\neq c$  $QFT_{4}$'s duals to the $d=5$ QT Gravity-matter model (\ref{qtop}), thus \emph{excluding} the possibility of existence of consistent  dual $QFT_4$ models of increasing or non-monotonic $c(l)$-functions\footnote{notice that according to refs.\cite{anselmi,anselmit,anselmib}, based on perturbative 1-loop calculations in a class of ${\cal N}=1,2$ supersymmetric $QFT_4$, such behaviour of $c(l)$ is not in principle forbidden.}. 

We next consider the $a$-Weaken version of the $d=5$ Theorem, when the restriction (\ref{ac-teor}) for the $\mu_+$ model, valid within the interval $\frac{5}{27}<\la<\frac{1}{3}$, is replaced by:
\begin{eqnarray}
0<f_{UV}\leq f_+^{a} \leq f_{IR} \leq f^{top}_+,\label{mupweak}
\end{eqnarray}
thus permitting to the $a(l)$-central function to change its sign at a specific ``$a$-scale'' $l_{cr}^a=L_+^a$. The corresponding ``critical coupling''  $\s_{cr}^a\in(\s_{IR},\s_{UV})$, determined by the real solutions of the equation $a(\s_{cr}^a)=0$, is dividing the former massless phase $(\s_{IR},\s_{UV})$ in two parts: $(\s_{IR},\s_{cr}^a)$ of \emph{negative} $a(l)\leq 0$ and the ``physical'' one $(\s_{cr}^a,\s_{UV})$ with $a(l)>0$, where all the conditions of the Standard $a/c$-Theorem are satisfied. Indeed the consistent RG evolution of the $CFT_4$ data terminates at $\s_{cr}^a$. Due to the fact that  the former IR critical point $\s_{IR}$ is now placed in the ``non-physical'' region and also due to  the presence of the new finite critical scale (and mass) $l_{cr}^a=L_+^a =1/M_{cr}^a$, the $QFT_4$-phase $(\s_{IR},\s_{UV})$ is \emph{not massless anymore} (see Sect.7.4 below  for more details and examples). The conclusion is that: when the parameters of the superpotential $W(\s)$ are satisfying the following weaker restrictions:
\begin{eqnarray}
(L_{top}^2)_+<L_{IR}^2\le(L_{a}^2)_+<L_{UV}^2,\label{aweak}
\end{eqnarray}  
then the $a$-Weaken form of the Theorem holds for the $\mu_+$-model within the interval $\la\in(5/27,1/3)$. Note that we are excluding the case  $L_{top}^2<L_{IR}^2<L_{UV}^2\le L_{a}^2$ at all, $i.e.$ when the $a(l)$ is negative within the entire interval $(\s_{IR},\s_{UV})$, since no consistent RG flows (and dual $QFT_4$'s) can be defined for those values of the coupling. 

Let us also mention the remarkable fact that the $QFT_4$'s duals to the $\mu_+$-model considered within the interval $\la\in(-\infty,5/27)$ \emph{have to satisfy the Standard Holographic $a/c$-Theorem}. Due to the absence of critical $a$-scales for these values of the Lovelock couplings, the $a(l)-$central function is always positive and monotonically decreasing and the only permitted Holographic RG flows  $UV\rightarrow IR$ are the massless ones. Therefore in the case of the $\mu_+$-model  there exists one special point $\la_{cr,+}^a=5/27$,\footnote{ where the topological and $a$-scales coincide : $f_+^a(5/27)=f_-^a(5/27)=f_+^{top}(5/27)=9/5$. This model describes a particular $CFT_4$ with vanishing central charges $a(\la_{cr}^a)=0=c(\la_{cr}^a)$.} which splits the interval of the allowed values of $\la$ in two subintervals $(-\infty, 5/27)$ and $(5/27,1/3)$. For the values of $\la$ belonging to each one of these intervals one can realize  qualitatively different Holographic RG flows  and  the corresponding dual $QFT_4$'s  have rather different phase structure as it is shown in Sect.7.4. below. Similar statement is valid for the $\mu_-$ model with $\la_{cr}^a=0$.

\section{Gauss-Bonnet $a/c$-Theorems}   
\setcounter{equation}{0}

The Gauss-Bonnet Gravity ($i.e.$ the $\mu=0$ QTG model) coupled to scalar matter for $d \geq 5$  is known to represent the simplest consistent gravity model, which in the context of the $AdS/CFT$ correspondence give rise to specific $a \neq c$ $CFT_{d-1}$'s. The requirements of the corresponding GB  ``$a$-Theorem'', together with the  $c(l)>0$ positivity restrictions  have been  established in refs.\cite{hidro,2,My_thol}. The  recent construction of the GB flat domain walls \cite{lovedw} provides an efficient tool  for the description of the Holographic RG flows in the corresponding dual $QFT_{d-1}$ (see sect.7.3. below). In this section we shall derive the conditions we have to impose on the matter superpotential $W$, necessary  for the validity of the $d\geq 5$ GB Holographic $a/c$-Theorems.  

As is well known, the monotonic decreasing of the $a(l)-$ central function  is automatically satisfied once the condition $W<0$ is imposed. The starting point of our analysis of the $a(l)$ and $c(l)$ positivity conditions, as well as  of the monotonicity of $c(l)$, are  the following explicit forms of these quantities:
\begin{eqnarray}
&&c(\s)=\left(\frac{L}{(d-2)l_{pl}\sqrt{f}}\right)^{d-2}\left(1-2\la f\right),\quad a(\s)=\left(\frac{L}{(d-2)l_{pl}\sqrt{f}}\right)^{d-2}\left(1-2\la \big(\frac{d-2}{d-4}\big)f\right),\nonumber\\
&&\frac{dc}{dl}=-\frac{2(d-2)^d}{\kappa^2l_{pl}^{d-2}}\left(\frac{W'(\s)}{W(\s)}\right)^2 c(l)\left(1-2\la \big(\frac{d-4}{d-2}\big)f\right)\label{gbac}
\end{eqnarray} 
We next  introduce the GB counterparts $f_{GB}^{\eta}=L^2/(L_{GB}^{\eta})^2$ of the ``critical" values of $f(\s)$ that are determined by the corresponding zeros of $c(f_{GB}^{top})=0=a(f_{GB}^a)$ and $\frac{dc}{dl}(f_{GB}^{c'})=0$:
\begin{eqnarray}
f_{GB}^{top}=\frac{1}{2\la},\quad \quad f_{GB}^{a}=\frac{1}{2\la}\left(\frac{d-4}{d-2}\right),\quad\quad f_{GB}^{c'}=\frac{1}{2\la}\left(\frac{d-2}{d-4}\right)\label{fGB}
\end{eqnarray} 
Notice that for  $\la>0$  and $d \geq 5$ the $f_{GB}^{\eta}$'s are ordered as follows:
\begin{eqnarray}
       0<f_{GB}^a\leq f_{GB}^{top}\leq f_{GB}^{c'}\label{gborder}    
\end{eqnarray}
Then the conditions of validity of the \emph{standard} $d\geq 5$ GB $a/c$-Theorem:
\begin{eqnarray}
       0<f\leq f_{GB}^a,\quad \quad 0<f\leq f_{GB}^{top},\quad\quad 0<f\leq f_{GB}^{c'}\label{gbteo}    
\end{eqnarray}
take the following simple form:
\begin{eqnarray}
       0<f_{UV}<f_{IR}\leq f_{GB}^a=\frac{1}{2\la}\left(\frac{d-4}{d-2}\right),\quad\quad L_{UV}>L_{IR}\geq L_{GB}^a\geq L_{GB}^{top}\label{gbcond}    
\end{eqnarray}

For negative values of $\la<0$, all the conditions (\ref{gbteo}) for validity of the standard $a/c$-Theorem: $a>0$, $c>0$ and $da/dl<0$, $dc/dl<0$ are automatically satisfied without any restrictions on the physical scales $L_{UV/IR}$. In the absence of topological or $a$-scales, the natural choice of the normalization is the smallest physical vacua scale $L=L_{IR}$.

$\bullet$ \textit{Comments on the normalization.} It is worthwhile to remind  the  restrictions on the GB coupling $\la$ related to the reality and positivity  conditions imposed on the solutions $f_{\pm}$ of the \emph{GB vacua equation}: $h = f ( 1 - \la f)$ with $h>0$. In the case we are considering  $\la > 0$, the additional requirement is given by: $\la\leq 1/4h$. We next observe that the $h(f^{top}_{GB}) = 1/4\la=L^2/L^2_{0top}$  defines the maximal value of $h(f)$. Therefore we can relate the ``fundamental scale'' $L$ (\ref{qtop}) to the minimal scale $L_{IR}$ introduced by the matter interaction, $i.e.$ we can take  $h_{IR}=h(f_{IR}) = 1$ as ``normalization'':  
$$h(f^{top}_{GB}) > h_{IR}=1 > h_{UV} \quad\quad \textrm{or equivalently}\quad\quad \la < 1/4$$ 
Notice that the above ``physical vacua normalization'' together with the $a/c$-Theorem condition (\ref{gbcond}) leads to the following \emph{stronger} restriction on the values of $\la$:
\begin{eqnarray}
0< \la \leq \frac{d(d-4)}{4(d-2)^2}<\frac{1}{4}, \label{larest}
\end{eqnarray}
where we have used the explicit form of the solution for the IR vacua equation: 
\begin{eqnarray}
f_{IR}=\frac{1-\sqrt{1-4\la h_{IR}}}{2\la}\label{firsol}
\end{eqnarray}
 One can use an alternative normalization $h(f_{top}) = 1$ (see \cite{lovedw}), based on the topological scale $L=L^{0top}_{GB}$, that \textit{fixes}  $\la = 1/4$. In this case we get as a consequence of the requirement (\ref{gbcond}) instead of eq. (\ref{larest}) a stronger restriction on the allowed values of $h_{IR}=L_{0top}^2/L_{0IR}^2$:
\begin{eqnarray}
0< h_{UV}<h_{IR}\leq \frac{d(d-4)}{(d-2)^2}< h_{top}=1, \label{hrest}
\end{eqnarray}
In fact for $\la>0$ both normalizations: $h_{top}=1$ and $h(f_{IR})=1$ are  equivalent and they reproduces the same restrictions on the physical scales $L_{UV/IR}$. 

For negative $\la$ we have not topological scales at all (since $c>0$ for all the $\la<0$) and the only available normalization is the one related to the IR vacua scale,$i.e.$ $h_{IR}=1$.

$\bullet$ \textit{$a$-Weaken GB Theorem.} Similarly to the case of  QT Gravity Holographic $a/c$-Theorems, one can consider the $a$-Weaken form of the Theorem for positive values of $\la<1/4$:
\begin{eqnarray}
(L_{top}^2)_{GB}<L_{IR}^2\le\left(\frac{d-2}{d-4}\right)(L_{top}^2)_{GB}<L_{UV}^2\nonumber 
\end{eqnarray}
thus allowing the $a(l)$-central function to change its sign at some $\s=\s_{cr}^a\in (\s_{IR},\s_{UV})$. For these $a$-Weaken conditions, the corresponding dual $QFT_d$ has a consistent description only for the couplings within the interval $\s\in(\s_{cr}^a,\s_{UV}$), which is not any more describing a massless phase, due to the existence of the $a$-mass scale $M_{GB}^a=1/L^a_{GB}$.

We conclude the discussion of the GB Holographic $a/c$-Theorems by mentioning the important fact that for all the allowed values of GB coupling $\la\in(-\infty,1/4)$, for arbitrary dimensions $d\geq 5$ and for rather general forms of the matter superpotential $W(\s)<0$ \emph{no modified $c$ -Theorem can take place, $i.e.$ the only possibility we can consistently realize is the following: $a>0$, $c>0$ and the $a$- and $c$- central functions are both monotonically decreasing}.

\section{Holographic $d\ge7$ $a/c$- Theorems}       
\setcounter{equation}{0}

\subsection{$\mu_{\pm}$ models Standard Theorem}

The proof of the $d \ge7$ $a/c$ -Theorem follows the same logics and we use the same methods of the $d=5$ case. Let us begin with the statement of the Standard $a/c$-Theorem valid for $d \ge 7$. According to our analysis of the conditions for validity of the Theorem, presented in App. \ref{app_A.1}, for the $\mu_+$ model we get the  following requirements:

\emph{The conditions that $a(l)$- and $c(l)$-central functions of  the dual $QFT_{d-1}$'s are both positive and monotonically decreasing during the RG flow between two consecutive critical points $\s_{UV}$ and $\s_{IR}$ are given by}:
\begin{eqnarray}
&&(1)\bullet \quad -\infty<\la<\frac{1}{4}: \quad  f_{IR}<f^a_{+}(\la;\mu_+),\\
&&(2)\bullet \quad \frac{1}{4}<\la<\frac{d(d-4)}{3(d-2)^2}:\quad f_{IR}<f^a_{+}(\la;\mu_+) \textrm{ or } f_{UV}>f^{c'}_{-}(\la;\mu_+),\\
&&(3)\bullet \quad \frac{d(d-4)}{3(d-2)^2}<\la<\frac{1}{3}:\quad f_{IR}<f^{top}_{+}(\la;\mu_+) \textrm{ or } f_{UV}>f^{c'}_{-}(\la;\mu_+).
\end{eqnarray}
The main difference in the Proof of this Theorem, when compared to the $d=5$ case, comes from the new intervals of $\la$ where $f_{\pm}^{\eta}$ are real positive or complex. They are a consequence of the different sign of the factor  $\left(\frac{d-4}{d-6}\right)$ multiplying $\mu$ in both $\frac{dc(\s)}{dl}$ and $a(\s)$-functions. This fact also causes modifications in the form (\ref{munegat}) and (\ref{muposit}) of $f^{\eta}$'s restrictions (\ref{munegat}) and (\ref{muposit}) for $\eta=a,c'$, while keeping unchanged the $c>0$ one. Another important change concerns the new orderings of the  $f^{\eta}_{\pm}$'s. Let us also mention  the presence of the new  \emph{intermediate} scale\footnote{ defined by $f^{h}= - \la / 2 \m_+ + \sqrt{ \left( \la^2 + 4 \m_+ \right)/4 \m_+^2}$ which represents the greatest root of corresponding $h(f) = 0$ equation} $L^h_-$, introduced  by $f^h=L^2/(L_-^h)^2$ in the regions of negative $h<0$ that leads to certain differences between the cases $d=7, d=8$ and $d\ge 9$ as described by eqs. (\ref{seven}) and (\ref{nine}) of App. \ref{apexA}. It turns out that the requirements of the corresponding \emph{standard} $a/c$-Theorem remain in fact the same for all $d \ge 7$. However when the \emph{modified $c$-Theorem} is considered - thus permitting, say of increasing $c(l)$, which imposes $ f_{+}^{c'}<f<f_{-}^{c'}$ - one can find different restrictions for $f_{UV/IR}$ in the \emph{maximal} scale region depending on the space-time dimensions and on the different intervals of values of $\la$, as for example in the $d=7$ case: 
\begin{eqnarray}
\bullet \la \in (7/25,8/27): \quad f^h<f<f^{c'}_-,\quad\quad\quad \bullet \la \in (8/27,1/3): \quad f^{top}_-<f<f^{c'}_-\nonumber
\end{eqnarray}
where we have now both \emph{the minimal and maximal scale restrictions together}. The discussion of the conditions of validity of the holographic modified $c$-Theorem is presented in Sect.5.2. below.

Similar considerations lead to the following set of restrictions on the allowed values of $f_{UV/IR}$ in the case of \emph{the $\mu_-$ model}:

(1)In the region of \emph{negative} $\la<0$ and $\mu<0$, according to the definitions (\ref{fac}) all the $f^{\eta}$'s for $d \ge 7$ are either negative or complex. Therefore we have \emph{no any restrictions for the validity of the Standard Theorem, and no natural fundamental scale exists}, differently from the $d=5$ case, where we have $f^a_->0$ that introduces $L_-^a$ as a minimal fundamental scale.

(2)In the interval $\la\in (0,1/3)$, the conditions under which the Standard $a/c$-Theorem  is satisfied  are given by:  
\begin{eqnarray}
\bullet \ &&d=7,8: \quad 0<\la<\frac{1}{3}: \quad\quad\quad  f<f^{top}_+(\la;\mu_-) \quad\textrm{or}\quad f>f^{c'}_-(\la;\mu_-),\label{mumenosoito}\\
\bullet \ &&d \ge 9 :  \quad 0<\la<\frac{(d-8)(d-4)}{3(d-6)^2}: \quad f<f^{a}_+(\la;\mu_-) \quad \textrm{or} \quad f>f^{c'}_-(\la,\mu_-)\label{mumenosteo}\\
\bullet \ &&d \ge 9 :  \quad \frac{(d-8)(d-4)}{3(d-6)^2}<\la<1/3: \quad f<f^{top}_+(\la;\mu_-) \quad \textrm{or}\quad f >f^{c'}_-(\la;\mu_-)\label{mumenosteoc}
\end{eqnarray}
where $f$ is denoting $f_{IR}$ or $f_{UV}$ in the case of minimal/maximal scale correspondingly.
The detailed derivation of the above requirements is presented in App. \ref{app_A.2}.

\subsection{Maximal scale and Modified $c$-Theorem}

The  analysis of the conditions under which the Standard $d \geq 7$ Holographic $a/c$-Theorem holds, reveals one interesting new feature, namely the existence of distinct domains of values of the Lovelock couplings $\la$ and $\mu_{\pm}(\la)$, where the allowed physical scales $f_{UV/IR}$ are determined by certain \emph{maximal scale} restrictions. Thus the corresponding dual  $QFT_{d}$'s with $a\neq c$ can have arbitrarily large values of both central functions. These new properties reflect the possibility that for $d \geq 7$ we can  have both  $f_{\pm}^{top}(\la)$'s and  $f_{\pm}^{c'}(\la)$'s positive. Therefore we can appropriately choose the parameters of the matter superpotential $W(\s)$ such that the $a/c$-Theorems can be satisfied in two well separated regions of \emph{minimal or maximal scales} restrictions  as it shown in Sects.5.1.and App. \ref{apexA}. The common feature of all these $d \geq 7$ models is  that the \emph{minimal scale} restrictions are always present and they are  almost identical to those of the $d=5$ models and of the $d\geq 7$ GB models.

Another important fact to be mentioned is that the proofs of the Standard $d \geq 7$ Theorems provide all the ingredients needed in the descriptions of the specific restrictions that give rise to the corresponding \emph{$a$-Weaken and Modified $c$-Theorems}.

\textit{$a$-Weaken Theorem.} According to the analysis of Sect.3 and  App. \ref{apexA}, the conditions of the $a$-Weaken Theorem can be realized always when we have at least one $a$-scale  ($i.e.$ one positive $f_{\pm}^a$ is present) and in the region of the \emph{minimal scale only}. Notice that when both $f_{\pm}^a >0$ are positive, the following ordering : $0<f_+^a <f_+^{top}<f_-^a< f_-^{top}$ takes place. This leads to the important conclusion that: \emph{it is impossible to realize the $a$-Weaken theorem in the region of the maximal scale}, $i.e.$ for $f>f_-^{top}$. We next separate the cases of one or two a-scales, where the conditions of the $a$-Weaken Theorem :
\begin{eqnarray}
    0<f_{UV}\leq f_+^a\leq f_{IR}\leq f_+^{top}
\end{eqnarray} 
can be  satisfied. Taking into account the results presented in App. \ref{apexA}, we deduce the following list of  models   
\begin{eqnarray}
\bullet \mu_-\textrm{model} (d \geq 9): 0<\la \leq \frac{(d-8)(d-4)}{3(d-6)^2},\quad\quad
\bullet \mu_+\textrm{model} (d \geq 7): -\infty<\la \leq \frac{d(d-4)}{3(d-2)^2}\nonumber
\end{eqnarray}
that admit the $a$-Weaken version of the Theorem. Notice that for the $\mu_-$-model and for $d=7,8$  \emph{no $a$-Weaken Theorem holds}. 

\textit{Modified $c$-Theorem}. As we have already mentioned, the modifications permitting monotonically increasing  and non-monotonic $c(l)$-central functions are allowed only in the case when the maximal scale $L_-^{c'}$  exists, $i.e.$ for  $f_-^{top/h} < f_{UV/IR}<f_-^{c'}$. Then the standard $a/c$-Theorem requirements $f_-^{c'}\leq f_{UV}<f_{IR}$ can be replaced by one of the following new conditions:
\begin{eqnarray}
&\bullet& \quad \textrm{increasing}\quad c(l):  f_-^{top/h} \leq f_{UV}<f_{IR}\leq f_-^{c'}\nonumber\\
&\bullet& \quad \textrm{non-monotonic}\quad c(l):  f_-^{top/h}\leq f_{UV}\leq f_-^{c'}<f_{IR}\nonumber 
\end{eqnarray}
that give rise to the $c$-Modified version of the Theorem. According to the results of App. \ref{apexA}., one can realize these modifications in the case of  $\mu_-$-model for all the values of $\la \in(0,1/3)$ and for all $d \geq 7$. Instead for the $\mu_+$ model one can have $QFT_{d-1}$ models satisfying c-Modified Holographic Theorem for all the $d \geq 7$  only  within the intervals:
\begin{eqnarray}
&&\bullet \frac{1}{4}<\la<\frac{8}{27} :\quad\quad  f^h\leq f_{UV}<f_{IR}\leq f_-^{c'} \quad \textrm{or}\quad f^h\leq f_{UV}< f_-^{c'}\leq f_{IR}\nonumber\\ 
&&\bullet \frac{8}{27}<\la<\frac{1}{3} :\quad\quad  f_-^{top}\leq f_{UV}<f_{IR}\leq f_-^{c'}\quad \textrm{or}\quad f_-^{top}\leq f_{UV}< f_-^{c'}\leq f_{IR}\nonumber 
\end{eqnarray}
Observe that for  $\la$ within the interval $(-\infty,1/4)$, we have \emph{no maximal scale at all} and therefore no Modified c-Theorems can be realised for the $\mu_+$-model. Note also that in all the cases of monotonically \emph{increasing} $c(l)$-central function, the $f_{UV/IR}$'s  have to respect \emph{the two scales restriction}, $i.e.$ their values are restricted between the minimal  $L_-^{c'}$ and the maximal scales, say $L_-^{top}$ or $L_-^h$,\footnote{remember that the definition of the new maximal ``h-scale'' in the case of negative $h<0$ is given by $f^{h}=L^2/(L_-^h)^2$}. Instead in the case of \emph{non-monotonic} $c(l)$ we have only one relevant scale - the maximal $L_-^{top/h}$ one. 

Our final comment is about the differences between the three distinct conditions on the $c(l)$ monotonicity  properties, in the case when the maximal scale does exist. Due to the lack of clear physical interpretations of the $L_-^{c'}$ as a fundamental maximal scale, it is evident that the most consistent physical description of the dual $QFT_{d-1}$ is represented by the case of \emph{non-monotonic} $c(l)$ , where the $a/c$-Theorem restrictions can be formulated  in  terms of the topological scale $L_-^{top}$ only. The description of the dual $QFT$'s that in the region of maximal scale are satisfying the conditions of the Standard ($i.e.$ $dc/dl<0$)  or of the Modified $dc/dl>0$ Holographic Theorems is involving additional scales that have not well established Quasi-Topological Gravity meaning. Instead in the case when $W$ is satisfying the minimal scale restrictions, the Standard $a/c$-Theorem is the only one that can be formulated  in the terms of the well defined physical scales.  

It is worthwhile to also mention the fact that the case $d=5$  is the only space-time dimension, for which \emph{no maximal scale exists} and we can  realize the restrictions imposed by the Standard  $a/c$-Theorem only.


\section{Energy Fluxes Positivity and $a/c$-Theorems}
\setcounter{equation}{0}

The RG evolution of the central charges $a$ and $c$  and of the conformal dimensions $\Delta_{\Phi}$, established by the Holographic $a/c$-Theorems in Sects.3-5, represents an important ingredient in the description of the  massless phases in the $QFT_{d-1}$'s duals to Quasi-Topological Gravity (\ref{qtop}). Based on the explicit form of the holographic $\beta_W$-function (\ref{rg}), they provide the conditions on the \emph{physical scales} $L_{UV/IR}$ and on the gravitational couplings $\la$ and $\mu$ ($i.e.$ on the $L_{\pm}^{top}$) in order to have certain desired relations between $CFT_{UV}$ and $CFT_{IR}$ data and preserving the unitarity consistency of the \emph{both} $CFT$'s. This section is devoted to the analysis of the \emph{additional restrictions} on the values of $L_{UV/IR}$, $\la$ and $\mu$ imposed by the requirements of unitarity and causality of the $CFT_{UV/IR}$'s, coming from the \emph{remaining part} of the $CFT$'s data and to the RG features of this new conformal data. Namely, the problem concerns the investigation (1) of the properties of the structure constants of the $T_{ij}(x_n)T_{kl}(0)$ OPE's, $i.e.$ the parameters $\cal A$,$\cal B$ and $\cal C$  of the stress-energy 3-point functions  and (2) of the new restrictions on the $f_{UV/IR}$  related to  the energy fluxes positivity (p.e.f.) conditions that $\cal A_{UV/IR}$, $\cal B_{UV/IR}$ and $\cal C_{UV/IR}$  have to satisfy \cite{maldahof,2,parna,hofman,BM,MPS}.

\subsection{More on $CFT_{d-1}$ data: positive energy fluxes conditions}

The energy flux (per unit solid angle $\Omega_{d-3}$) measured in the future null infinity in the direction $n^i$ is given by the following ``energy one-point functions'':
\begin{eqnarray}
    <{\cal E}(\vec{n})>_{\cal O}= \frac{<0|{\cal O}^{\dagger}{\cal E}(\vec{n}){\cal O}|0>}{<0|{\cal O}^{\dagger}{\cal O}|0>} \label{efun}
\end{eqnarray}
As explained in refs.\cite{maldahof,2,BM}\footnote{we are closely following the results and the definitions  of \cite{2}, especially those of Sect.3 and App. \ref{apexA}, but with slightly different normalization of $c$ and $a$ and with $d$ replaced by $d-1$.},
in each unitary $CFT_{d-1}$ and for the states $|p>\sim {\cal O}_{p}|0>$ created by the spacial components of $T_{ij}$ of helicities p=2,1,0      ($i.e.$ tensor, vector and scalar w.r.t. $SO(d-3)$ subgroup), the corresponding energy fluxes can be calculated by taking appropriate limits of the integrated 3- and 2-point functions of the stress tensor. The final result (see eqs. (3.6) and (3.32-34) of ref. \cite{2}) can be written in the following simple form: 
\begin{eqnarray}
E_s=<{\cal E}(\vec{n})>_{scalar}&=&\frac{E}{\Omega_{d-3}}\left(1+\frac{d-4}{d-2}t_2+\frac{d(d-3)-2}{d(d-2)}t_4\right),\nonumber\\
E_v=<{\cal E}(\vec{n})>_{vector}&=&\frac{E}{\Omega_{d-3}}\left(1-\frac{d-4}{2(d-2)}t_2-\frac{2}{d(d-2)}t_4\right),\nonumber\\
E_t=<{\cal E}(\vec{n})>_{tensor}&=&\frac{E}{\Omega_{d-3}}\left(1-\frac{1}{d-2}t_2-\frac{2}{d(d-2)}t_4\right),\label{pefcon}
\end{eqnarray}
where $E$ denotes the total energy and the new coefficients $t_2$ and $t_4$ are parametrized by $\cal A$, $\cal B$ and $\cal C$\footnote{the corresponding  explicit formulas $t_K(\cal A,\cal B,\cal C)$, $K=2,4$ are given by  eq. (3.9) of ref. \cite{2} with $d\rightarrow d-1$ and they lead to the equivalent form (A.17-18) of $<{\cal E}(\vec{n})>_{p}$ in terms of $\cal A$, $\cal B$ and $\cal C$ only.}. When $c\neq 0$, one can choose equally well $a$, $c$ and $t_4$ as independent parameters. 

It is then evident that the natural requirement of \emph{positivity of the energy fluxes} related to the  above considered ``tensor, vector and  scalar states'' 
\begin{eqnarray}
E_{s}\geq 0,\quad\quad\quad E_{v}\geq 0,\quad\quad \quad E_{t}\geq 0 \label{pefcondi}
\end{eqnarray} 
lead to certain important restrictions on the values of $t_2$ and $t_4$, which together with the $c\geq 0$ condition provide strong requirements on  the physically allowed values of $\cal A$, $\cal B$ and $\cal C$. For example, for  all the GB induced CFT's and, more generally, for all the ${\cal N}=1$ super-symmetric $CFT_{d-1}$ the parameter $t_4^{GB}=0$ is always vanishing \cite{maldahof,hofman} and the corresponding p.e.f. conditions take the form: 
\begin{eqnarray}
-\frac{d-2}{d-4}\leq t_2\leq \frac{d-1}{2}\label{tpef}
\end{eqnarray}

The Holographic gravitational d-dimensional counterparts of the above  field-theoretical $CFT_{d-1}$ p.e.f. requirements (\ref{pefcon}) are known to coincide with the causality conditions needed to avoid the propagation of superluminal signals (out of the light-cone of the $AdS_d$ boundary) for both the black-holes and shock wave backgrounds in the dual GB or QT Gravity models \cite{BM,hofman,2,MPS}. The problem of the causal consistency of the considered DW's backgrounds as well as of their stability require further investigations of the properties of the linear fluctuations of the metrics and of the matter field $\s$  that are out of the scope of the present paper. 

Let us also mention that the  ``new $CFT$ data'' (as well as the ``old'' one): $\cal A$, $\cal B$, $\cal C$, $a$ and $c$  turns out to have  an equivalent ``holographic'' realization in terms of the fundamental scale $L$  and of the $f_{UV/IR}=L^2/L_{UV/IR}^2$ only. Their explicit form indeed depends on the details of the given GB and QTG models \cite{My_thol,espanha}. Therefore the p.e.f. conditions lead to certain restrictions on the critical values of $f$ or $W(\s)$ and on the $\la$ and $\mu$ values as well. They are introducing  new upper bounds $f_{max}=f_{\pm}^{ep}$,
\begin{eqnarray}
0<f_{UV}<f_{IR}<f_{max}\nonumber
\end{eqnarray}
to be compared with the minimal scale $f_{\pm}^a$ and $f_{\pm}^{top}$-bounds (\ref{fac}), derived in Sect.3. Taking into account the well known $AdS/CFT$ identification $(\frac{L_k}{l_{pl}})^{d-2}\approx N^2_c$ of the vacua scales $L_{UV/IR}$ with  the color number $N_c^{UV/IR}$  of the corresponding $SU(N_c)$  CFT's we can find a lower bound for $N_c^{IR}$ at the IR-end of the RG flow.

\subsection{Holographic GB $t_2$-Theorem}

6.2.1.\textit{New minimal scales for the $a/c$-Theorem.} The $CFT_{d-1}$'s duals to GB Gravity are characterized by $t_4=0$ and as a consequence all the remaining data: $t_2$ and $\cal A$, $\cal B$, $\cal C$, can be parametrized by the central charges $a$ and $c$ only:
\begin{eqnarray}
t_2=\frac{(d-1)(d-2)}{(d-3)}\left(1-\frac{a}{c}\right) \label{tdve}
\end{eqnarray}
Then the p.e.f. requirements (\ref{tpef}) take the following form
\begin{eqnarray}
\frac{(d-1)(d-4)}{d(d-4)+1}\leq \frac{c}{a}\leq \frac{d-1}{2} \label{ca}
\end{eqnarray}
thus restricting the ratio of the central charges $c/a$ to belong to the well known ${\cal N}=1$ SUSY unitary $CFT_{d-1}$'s window \cite {maldahof,My_thol,2,espanha}. Taking into account the explicit form (\ref{gbac}) of the GB central charges in terms of the $f_{UV/IR}$, we next derive the new upper bounds $f_{\pm}^{ep}$ imposed by the p.e.f. conditions:
\begin{eqnarray}
&\bullet&  \la>0 :\quad 0<f_{UV}<f_{IR}<\frac{(d-3)(d-4)}{2\la(d^2-5d+10)}=f_+^{ep}<f_{GB}^a,\quad L_{UV}^2>L_{IR}^2>L_{ep}^2=\frac{L^2}{f_+^{ep}},\nonumber\\   
&\bullet&  \la<0 :\quad 0<f_{UV}<f_{IR}<\frac{(d-3)}{2|\la|(d+1)}=f_-^{ep},\quad \quad\quad L_{UV}^2>L_{IR}^2>L_{ep}^2=\frac{L^2}{f_-^{ep}},\label{fepr}
\end{eqnarray}
They introduce specific new constraints on the allowed physical scales $L_{UV/IR}$, characterizing the UV- and IR- $CFT_{d-1}$'s that represents the limiting critical points $\s_{UV/IR}$ of the massless RG flow in consideration. Similarly to the restrictions (\ref{larest}) on the values of $\la_{GB}$, introduced by the $a$-scale $f_-^{a}$ (see sect.4), the above  p.e.f. conditions (\ref{fepr}) and (\ref{ca}) can be rewritten in an equivalent form :
\begin{eqnarray}
   -\frac{(d-3)(3d-1)}{4(d+1)^2}\leq \la h_{IR}\leq \frac{(d-3)(d-4)(d^2-3d+8)}{4(d^2-5d+10)^2}, \label{lapef}
\end{eqnarray}
which in the normalization $h_{IR}=1$ is reproducing  eq. (3.48) of ref. \cite{2}. Let us remind that the remaining constraints on the $f_{UV}$ and $\la h_{UV}$ are a simple consequence of the requirements $0<s_{UV}<(d-1)/2$ and $s_{IR}<0$, that guarantee the existence of RG flow and its direction. They require that  $L_{IR}$ must be the smallest physical scale:
$$ L_{UV}>L_{IR},\quad\quad 0<f_{UV}<f_{IR},\quad\quad 0<h_{UV}<h_{IR},\quad\quad W<0$$
which reflects the properties of the critical points $W'(\s_{cr})=0$, namely that $\s_{UV}$ must be a maximum and $\s_{IR}$ the nearest local minimum of $W(\s)$.

The above discussion that leads us to the requirements (\ref{tdve}), (\ref{ca}), (\ref{fepr}) and (\ref{lapef}) imposed on the $CFT_{UV/IR}$ data by the p.e.f. conditions (\ref{pefcon}), together with the arguments presented in Sect.4., can be summarized as follows: 

\emph{The Holographic GB $a/c$ Theorem statement - the central charges $a(l)$ and $c(l)$ are both positive and monotonically decreasing during the massless RG flows - is valid  for the $CFT_{UV/IR}$ satisfying the p.e.f. constraints as well, but only when the stronger p.e.f. restrictions on the physical scales $L_{UV/IR}$ and $\la$ take place.}

\vspace{0.5cm}
6.2.2.\textit{ The $\frac{a}{c}$ and  $t_2$- Theorems}. We next consider few consequences concerning the RG flows of the $CFT$'s 3-point functions data. Instead of the ${\cal A}/{\cal B}/{\cal C}$-analogues of the Holographic  $a/c$-Theorems we chose to study the $t_2(l)$  RG evolution\footnote{We recall that in the GB case all the stress-tensor ``3-points data''  ${\cal A}$,${\cal B}$ and ${\cal C}$ are well known functions of the two central charges $a\neq c\neq 0$.}. The eqs. (\ref{gbac}) and (\ref{firsol}) allows us to rewrite the ratio of the central charges (for $c\neq 0$) in the following form:
\begin{eqnarray}
\frac{a}{c}=\frac{d-2}{d-4}-\frac{2}{(d-4)\sqrt{1-4\la h}}\label{acratio}
\end{eqnarray}
Then the corresponding $\frac{a}{c}$ -Theorems are a straightforward consequence of the RG flow conditions $0<h_{UV}<h_{IR}<h_{ep}$:
\begin{eqnarray}
\bullet\quad \la>0 :\quad\quad\quad \frac{a_{UV}}{c_{UV}}\ge \frac{a_{IR}}{c_{IR}}\quad\quad\quad\quad\bullet\quad \la<0 :\quad\quad\quad \frac{a_{UV}}{c_{UV}}\le \frac{a_{IR}}{c_{IR}},\label{ratioth}
\end{eqnarray}
which is in fact a statement about the RG properties of $t_2$: it is \emph{increasing for positive $\la$} and \emph{decreasing for negative $\la$}.
An alternative proof of the $t_2$ -Theorem is based on its explicit form in terms of the $f$-variable:
\begin{eqnarray}
  t_2(f,\la)=\frac{4\la f(d-1)(d-2)}{(1-2\la f)(d-3)(d-4)},\quad\quad\quad\quad \frac{dt_2}{df}=\frac{4\la (d-1)(d-2)}{(1-2\la f)^2(d-3)(d-4)}\label{t2th}
\end{eqnarray}
and it again leads  to the same conclusions.

Few comments concerning the eventual physical ($d=5$) applications of the above results are now in order:

$\bullet$ \textit{On the RG evolution of $E_p$.}  The holographic $t_2$ -Theorem, together with the explicit forms (\ref{pefcon}) of the corresponding energy fluxes (with $t_4=0$), allows us to make conclusions about the RG evolution of these energy fluxes as well: 
$$E_{s}(UV)< E_{s}(IR),\quad\quad\quad E_v(UV)> E_v(IR)\quad\quad\quad E_t(UV)> E_t(IR).$$ 
Notice however the important difference between the $a$ and $c$ central charges and the $t_2$ and the $E_p$'s (p=s,v,t) properties. Namely, the eqs. (\ref{gbac}) represent the off-critical $QFT_{d-1}$ definitions of the central functions $a(l)=a(f(\s))$ and $c(l)=c(f(\s))$ based on the holographic $\beta_W(\s)$-function, while the corresponding formulas determining $t_2$ and $E_p$ in terms of $a$ and $c$ (or as functions of $f$) have only well established \emph{critical} meaning, $i.e.$ they are valid for the $CFT_{UV/IR}$ only. The corresponding off-critical expressions for $t_2(\s)$ and $E_p(\s)$ can be derived by the conformal perturbations theory. It is expected that the critical values $t_2(UV/IR)$ and $E_p(UV/IR)$ receive relevant ``bulk'' contributions from the non-vanishing trace $\Theta(\s,x_i)=\beta(\s) \Phi_{\s}(x_i)$ of the stress-tensor, thus involving the structure constants of the following  3-point functions of $\Theta(\vec{x})$ and $T_{ij}(\vec{x})$:
 $$<T_{ij}(\vec{x})\Theta(\vec{y})T_{kn}(0)>,\quad\quad<\Theta(\vec{x})\Theta(\vec{y})\Theta(0)>,\quad\quad<\Theta(\vec{x})T_{ij}(\vec{y})\Theta(0)>.$$   
They can be easily calculated within the frameworks of the $CFT_{UV}$ by applying the $OPE$'s, discussed in Sect.2.:
$$ C_{\Theta\Theta\Theta}=\beta^3(\s)g(\s) C_{\Phi\Phi\Phi},\quad\quad\quad\quad C_{\Theta T\Theta}=\Delta_{\Phi}(\s)\beta^2(\s)g(\s),$$ 
where $\Delta_{\Phi}(\s)$ is the anomalous dimension of the relevant operator $\Phi_{\s}(x)$, the  $C_{\Phi\Phi\Phi}$ is the  $\Phi_{\s}(x)\Phi_{\s}(0)$ OPE's structure constant, given by eq. (\ref{str}), and  $g(\s)=C^{(2)}_{\Phi}$ is the  2-point function normalization constant (\ref{cbeta}). Notice that the above integrated 3-point functions involving $\Theta$ give relevant contributions  to the off-critical bulk viscosity $\zeta(\s)\neq 0$ as well \cite{bviscos}. 

Although the derivation of the off-critical positive energy fluxes conditions is an open problem, the above established $t_2$ and $E_p$ -Theorems  contain an important information relating the UV- to IR- \emph{critical values} of the corresponding quantities. However they have a quite different status and \emph{can not} be considered as statements determining the monotonic properties of the \emph{yet unknown} off-critical expression for $t_2(\s)$ and $E_p(\s)$ during the massless RG flows.

$\bullet$  \textit{$\eta/s$ -Theorem}? Let us remind that one of the main motivations for studying the $CFT's$ and, related to them, \emph{non-conformal} $QFT$'s, dual to certain $d=5$ GB Gravity solutions - black holes, shock waves, DW's etc. - is their use in the description of the hydrodynamics of the strong coupled quark-gluon plasma \cite{hydro,hidro,shydro,BM,MPS}. As we have mentioned in the Introduction, the practical question to be answered concerns the possible violations of the ``universal lower bound'' $\eta/s = 1/4\pi$ of the ratio of shear viscosity to entropy density. The calculations based on the GB black hole solutions (duals to finite temperature CFT's) lead to the following simple form of this ratio in terms of the corresponding central charges \cite{2,My_thol,bala}:
\begin{eqnarray}
\left(\frac{\eta}{s}\right)_{UV} \approx \frac{1}{4\pi}\left(\frac{a}{c}\right)_{UV}=\frac{1}{4\pi}\left(3 -\frac{2}{\sqrt{1-4\la h_{UV}}}\right),\nonumber
\end{eqnarray}
which suggests that a similar formula is valid for the IR values of  $\left(\frac{\eta}{s}\right)_{IR}$ as well. Although the considered RG flows interpolate between \emph{zero temperature} $CFT_{UV/IR}$'s, one might expect to have \emph{decreasing} $\eta/s$ for positive $\la$, just as the above established $\frac{a}{c}$ -Theorem is claiming. One argument in favour of such conclusion is  that both the zero and finite temperature  $CFT$'s share the same central charges and they do satisfy the same p.e.f. conditions \cite{hofman}. Another argument comes from the fact that the finite temperature $T=1/\beta_{thg}$  analogues  of the considered DW's called ``thermal gas'' solutions\footnote{they have the same form as the zero temperature DW's \cite{lovedw}, but with the 4-d Minkowski part replaced by a circle $S_1$ of radius $\beta_{thg}=1/T$ for the time and by 3-d torus for the space $x_i$-part.}, are successfully used  in the description of certain finite $T$ $QCD_4$-like and $sQGP$ models up to certain not very high temperatures $T\in(0,T_{min})$, see for example refs.\cite{ven-kir} and the references therein. The complete description of the off-critical behaviour of $\eta/s$ as a function of the RG scale (or of the coupling constant $\s$) by using or not the Holographic RG methods is however an open problem, out of the scope of the present paper.

\subsection{New minimal scales in $d=5$ QT Gravity $a/c$-Theorems}

The critical UV holographic features of $d=5$ Quasi-Topological Gravity (\ref{qtop}) without matter interactions, together with the off-critical properties of the $a(l)$ central function, have been established in refs. \cite{MPS}. Due to the fact that both $t_2$ and $t_4$ are now different from zero, the p.e.f. conditions (\ref{pefcon}) are restricting their values within the triangle formed by the three curves $E_s(t_2,t_4)=0$, $E_v(t_2,t_4)=0$ and $E_t(t_2,t_4)=0$. We are interested in  the new minimal scales arising from the p.e.f. requirements and imposing further stronger restrictions on the values of $f_{UV/IR}$. Our starting point are the following $d=5$ holographic expressions for $t_2$ and $t_4$:
\begin{eqnarray}
  t_2(f_{cr})=\frac{24f_{cr}(\la-87\mu f_{cr})}{1-2\la f_{cr}-3\mu f_{cr}^2},\quad\quad\quad t_4(f_{cr})=\frac{3780\mu f_{cr}^2}{1-2\la f_{cr}-3\mu f_{cr}^2},\quad f_{cr}=f_{UV/IR},\label{tquatro}
\end{eqnarray}
derived in sect.4 of ref. \cite{MPS}. We next substitute eqs. (\ref{tquatro}) into eqs. (\ref{pefcon}) and as a results we get the following consequences of the p.e.f. conditions for $f_{cr}$:
\begin{eqnarray}
1-10\lambda f+189\mu f^2\ge0,\quad\quad 1+2\lambda f-855\mu f^2\ge0,\quad\quad 1+6\lambda f+1317\mu f^2\ge0,\label{fcond}
\end{eqnarray}
where the positivity of the $c-$ central charge has been already taken into account. Let us denotes the roots of the above quadratic forms as
\begin{eqnarray}
f_{\pm}^{ep,1}&=&\frac{5}{189\mu}\left(\lambda\mp\sqrt{\lambda^2-\frac{189}{25}\mu}\right),\\
f_{\pm}^{ep,2}&=&\frac{1}{855\mu}\left(\lambda\mp\sqrt{\lambda^2+855\mu}\right),\\
f_{\pm}^{ep,3}&=&-\frac{3}{439\mu}\left(\lambda\mp\sqrt{\lambda^2-\frac{439}{3}\mu}\right),\label{roots}
\end{eqnarray}
which in the case when they are real and positive introduce  new \emph{p.e.f. scales} $L_{\pm}^{ep,k}=L/\sqrt{f_{\pm}^{ep,k}}$ with $k=1,2,3$.

In order to implement the p.e.f. requirements in the context of the $d=5$ $a/c$-Theorems (as stated in sect.3.) we need to know  few properties of these $f_{\pm}^{ep,k}(\la,\mu_{\pm})$, namely: (i) the intervals of values of $\la$, where the corresponding roots $f_{\pm}^{ep,k}$ are \emph{real and positive} for each one of the models $\mu_+$ and $\mu_-$ separately;(ii) which is the smallest one  $f_{min}^{ep}$ in between all these  $f_{\pm}^{ep,k}>0$'s within each of the $\la$-regions established above; (iii) whether these $f_{min}^{ep}$'s are smaller or bigger then the minimal of the other $a$-, $top$- and $c'$- scales $f^{\eta}_{pm}>0$ already used in the proof of the Theorem. The methods to perform this analysis are the same as those introduced in sect.3.2 and as always the graphical one is the most efficient. In the case when the smallest of these  new p.e.f. scales turn out to be  smaller then the ``old ones'' $f_{\pm}^{\eta}$, the validity of the $a/c$-Theorem compatible with the $UV/IR$- p.e.f. conditions imposes \emph{stronger} restrictions on the allowed $f_{UV/IR}$ values: $ 0<f_{UV}<f_{IR}<f_{min}^{ep}$. 

\vspace{0.5 cm}

\begin{figure}[ht]
\centering
\subfigure[]{
\includegraphics[scale=0.7]{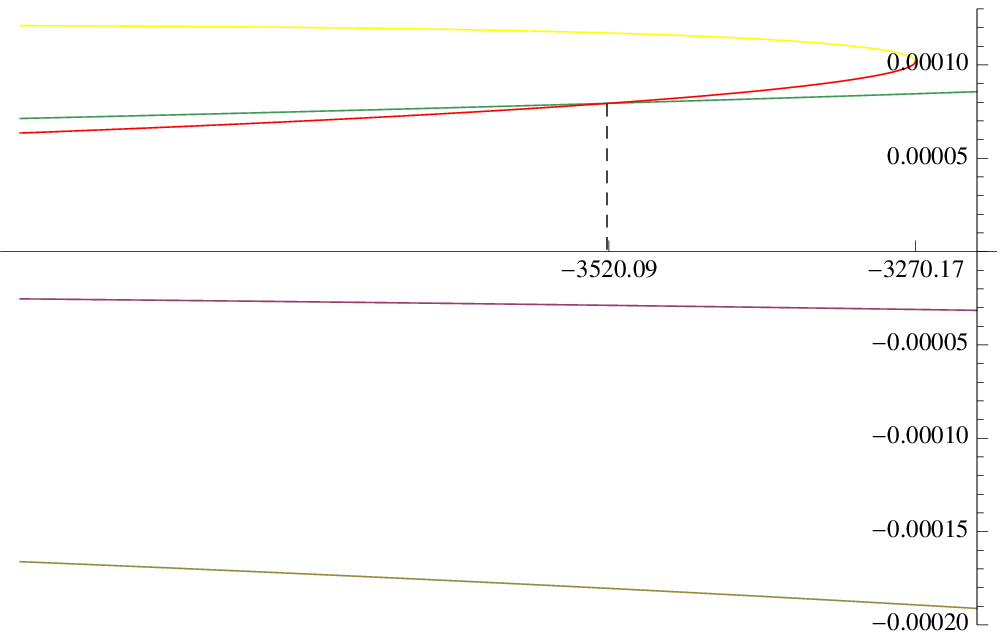}
\label{fig:pres1a}
}
\subfigure[]{
\includegraphics[scale=0.7]{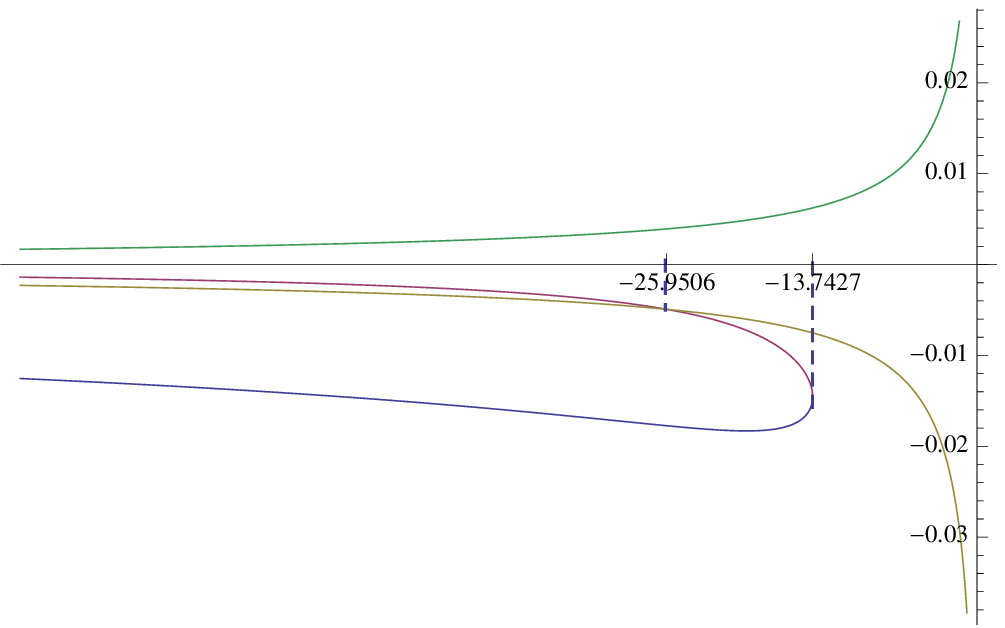}
\label{fig:pres1b}
}
\subfigure[]{
\includegraphics[scale=0.7]{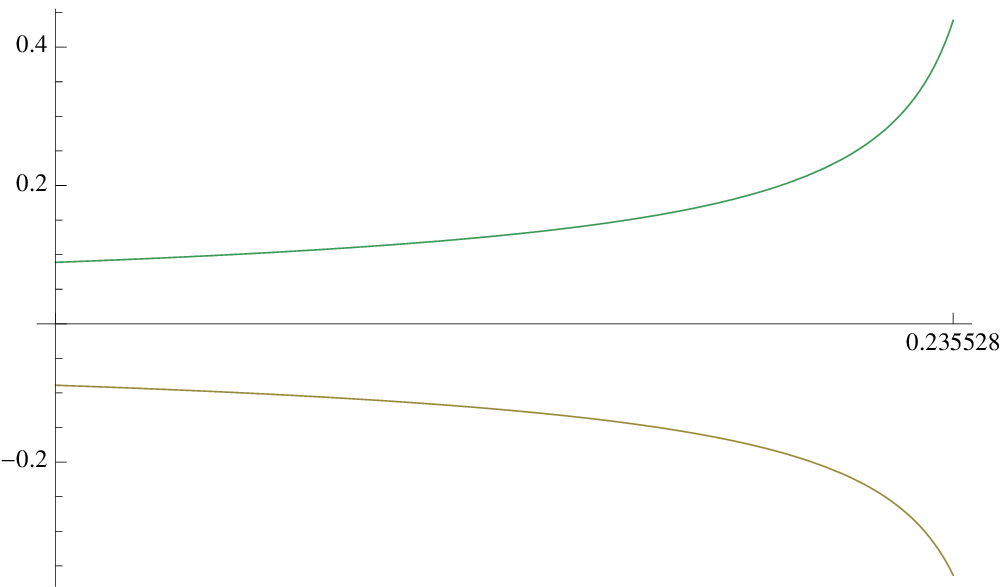}
\label{fig:pres1c}
}
\subfigure[]{
\includegraphics[scale=0.7]{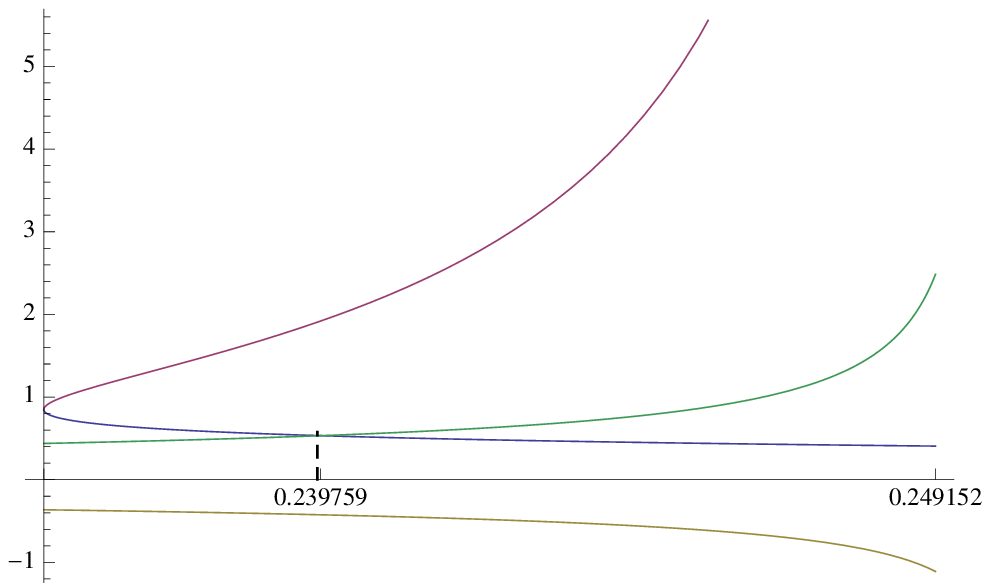}
\label{fig:pres1d}
}
\label{fig:pres1}
\caption{The $f^{ep,i}_{\pm}(\la;\mu_+)$ curves for $\la \in (-\infty,0.249152)$: $f_+^{ep,1}$ is depicted in blue; $f_-^{ep,1}$ in purple; $f_+^{ep,2}$ in beige; $f_-^{ep,2}$ in green; $f_+^{ep,3}$ in yellow and  $f_-^{ep,3}$ in red. The $\lambda$ values are within different intervals: on Fig.$(a)$ $-\infty<\la<-3270.17$ (the $f_+^{ep,1}$ is missed being too negative, namely $f_+^{ep,1}<f_+^{ep,2}<0$); on Fig.$(b)$, $-3270.17<\la<0$; on Fig.$(c)$, $0<\la<0.235538$ and on Fig.$(d)$, $0.235538<\la<0.249152$.}
\end{figure}

\begin{figure}[ht]
\centering
\subfigure[]{
\includegraphics[scale=0.7]{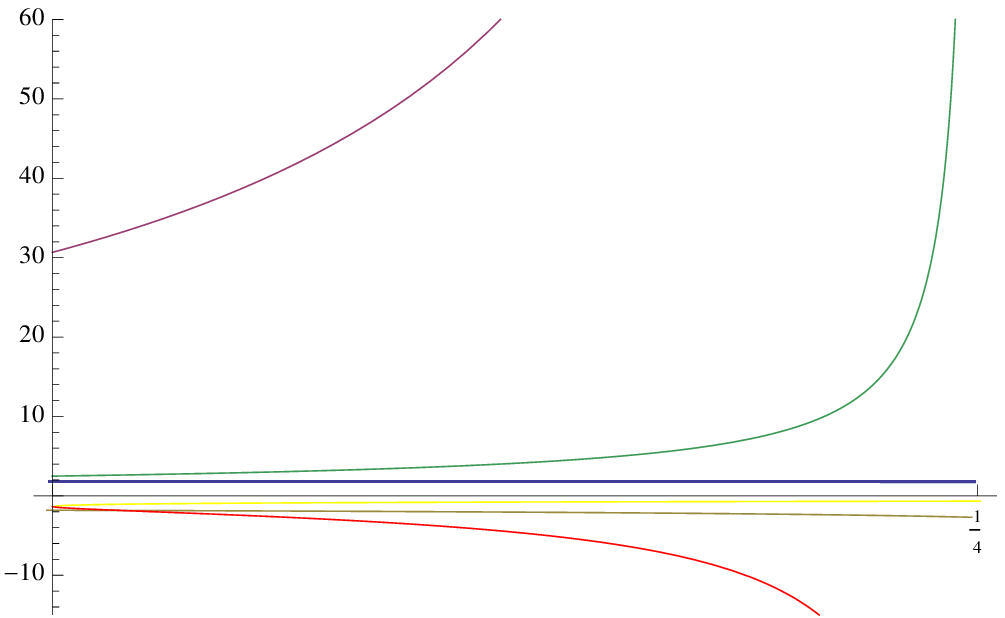}
\label{fig:pres2e}
}
\subfigure[]{
\includegraphics[scale=0.7]{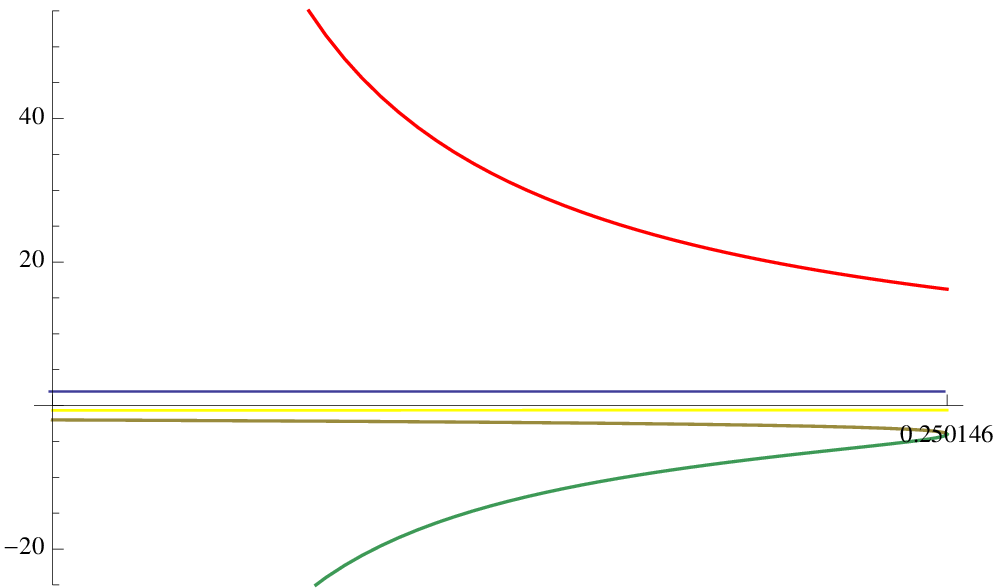}
\label{fig:pres2f}
}
\subfigure[]{
\includegraphics[scale=0.7]{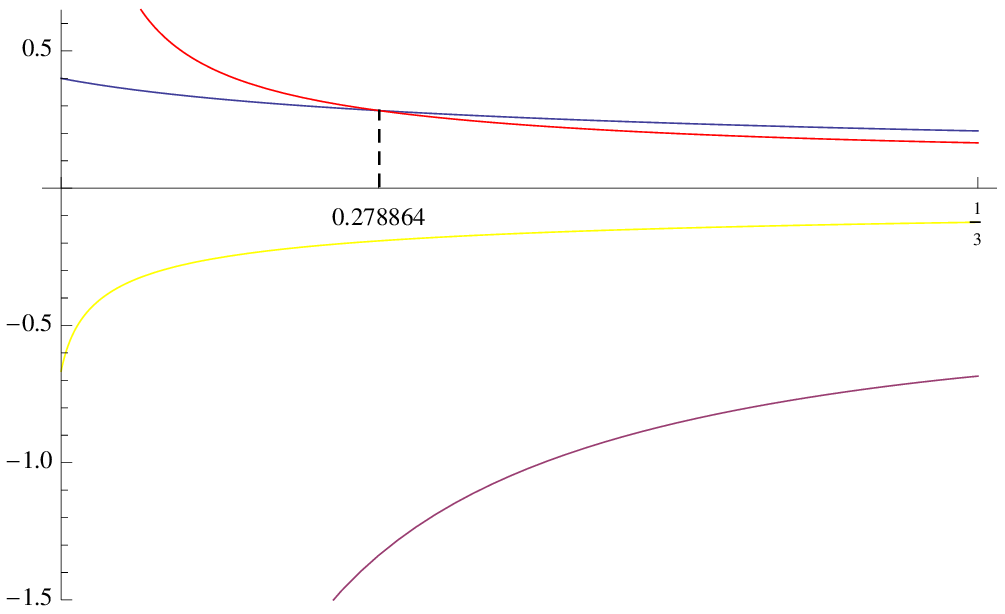}
\label{fig:pres2g}
}
\label{fig:pres2}
\caption{The $f^{ep,i}_{\pm}(\la;\mu_+)$ curves for $\la \in (0.249152,1/3)$: $f_+^{ep,1}$ is depicted in blue; $f_-^{ep,1}$ in purple; $f_+^{ep,2}$ in beige; $f_-^{ep,2}$ in green; $f_+^{ep,3}$ in yellow and $f_-^{ep,3}$ in red. The $\lambda$ values on different figures belong to different intervals: on Fig.$(a)$, $0.249152<\la<1/4$; on Fig.$(b)$, $1/4<\la<0.250146$ (here the ``most negative'' curve $f^{ep,1}_-$ is missing again) and on Fig.$(c)$, $0.250146<\la<1/3$.}
\end{figure}

6.3.1.\textit{The $\mu_+$ model.} As one can see from the graphics of the functions $f_{\pm}^{ep,k}(\la,+)$ plotted on Figs.(\ref{fig:pres1a})-(\ref{fig:pres1d}) and (\ref{fig:pres2e})-(\ref{fig:pres2g}), the $f_{\pm}^{ep,1}\in R$ only within the intervals
\begin{eqnarray}
 -\infty <\la\leq 13.7427 \quad\quad \textrm{and}\quad  -2.235528<\la<\frac{1}{3},\nonumber
\end{eqnarray}
while the $f_{\pm}^{ep,2}$ are real only for $\la\in (-\infty, 0.250146)$, and the $f_{\pm}^{ep,3}\in R$ are real only when
\begin{eqnarray}
-\infty<\la<-3270.17 \quad\quad\textrm{and}\quad\quad 0.249152<\la<\frac{1}{3}.\nonumber
\end{eqnarray}
Their ordering in the regions where they are positive is clearly demonstrated by the corresponding graphics. For example, 
within the interval $\la \in(-\infty,-3520.09)$  we conclude by comparing the six curves presented on fig.(\ref{fig:pres1a}) that
\begin{eqnarray}
 f_+^{ep,1}<f_+^{ep,2}<f_-^{ep,1}<0<f_-^{ep,3}<f_-^{ep,2}<f_+^{ep,3}\nonumber
\end{eqnarray}
and therefore the the minimal scale in this interval is given by $f_-^{ep,3}$. 

Analysing the existence of ``critical'' $\la$-points, where  two of the curves $f_{\pm}^{ep,k}$ are crossing each other, as shown on the Figs.(\ref{fig:pres1a})-(\ref{fig:pres1d}, we realize that the $\la$-interval $(-3520.09, 0.239759)$ can be  divided  in five different subintervals, corresponding to different $f_{\pm}^{ep,k}$'s orderings: 
\begin{eqnarray}
\bullet &&-3520.09<\lambda<-3270.17:\quad\quad f_+^{ep,1}<f_+^{ep,2}<f_-^{ep,1}<0<f_-^{ep,2}<f_-^{ep,3}<f_+^{ep,3},\nonumber\\
\bullet && -3270.17<\lambda<-25.9505:\quad\quad f_+^{ep,1}<f_+^{ep,2}<f_-^{ep,1}<0<f_-^{ep,2},\nonumber\\
\bullet&& -25.9505<\lambda<-13.7427:\quad\quad f_+^{ep,1}<f_-^{ep,1}<f_+^{ep,2}<0<f_-^{ep,2},\nonumber\\
\bullet&& -13.7427<\lambda<0.235528:\quad\quad f_+^{ep,2}<0<f_-^{ep,2},\nonumber\\
\bullet&& 0.235528<\lambda<0.239759:\quad\quad f_+^{ep,2}<0<f_-^{ep,2}<f_+^{ep,1}<f_-^{ep,1},\nonumber
\end{eqnarray}
Notice also that some of the $f_{\pm}^{ep,k}(\la,+)$'s that are real within one of these subintervals  become complex in  others ones\footnote{the curves corresponding to complex values of $f_{\pm}^{ep,k}$, in fact, do not appear on the figures.}. The common feature of all these cases is that the smallest scale is always given by $f_-^{ep,2}$. 

As one can see from the plots on Figs.(\ref{fig:pres1d})-(\ref{fig:pres2f}), similar phenomena take place in the next interval $\la \in(0.239759,0.278864)$ : 
\begin{eqnarray}
\bullet&& 0.239759<\lambda<0.249152:\quad\quad f_+^{ep,2}<0<f_+^{ep,1}<f_-^{ep,2}<f_-^{ep,1},\nonumber\\
\bullet&& 0.249152<\lambda<1/4:\quad\quad f_-^{ep,3}<f_+^{ep,2}<f_+^{ep,3}<0<f_+^{ep,1}<f_-^{ep,2}<f_-^{ep,1},\nonumber\\
\bullet&& 1/4<\lambda<0.250146:\quad\quad f_-^{ep,1}<f_-^{ep,2}<f_+^{ep,2}<f_+^{ep,3}<0<f_+^{ep,1}<f_-^{ep,3},\nonumber\\
\bullet&& 0.250146<\lambda<0.278864:\quad\quad f_-^{ep,1}<f_+^{ep,3}<0<f_+^{ep,1}<f_-^{ep,3},\nonumber
\end{eqnarray}
where the minimal scale turns out to be $f_+^{ep,1}$. Finally, as one can see from Fig.(\ref{fig:pres2g}) in the last interval we have 
\begin{eqnarray}
\bullet&& 0.278864<\lambda<1/3:\quad\quad f_-^{ep,1}<f_+^{ep,3}<0<f_-^{ep,3}<f_+^{ep,1},\nonumber
\end{eqnarray}
Therefore now we can recognize  $f_-^{ep,3}$ as representing the minimal scale. 

Notice the importance of the special points $\la_{cr}^{i,j}$ of the intersection of two of the curves $f{\pm}^{ep,i}$ and $f_{\pm}^{ep,j}$, say at $\la_{cr}^{2,3}=-3520.09$ we have that $f_-^{ep,2}=f_-^{ep,3}$, etc. They are determining the limiting points of the $\la$ intervals, where at least one of the $f^{ep,k}$ is positive. The comparison (both the analytical and the graphical ones) of the minimal of the $f_{\pm}^{ep,k}(\la+)>0$'s  with the minimal of the other  $f_{\pm}^{\eta}$'s scales, within the above established ``physically allowed'' regions of $\la$, demonstrates that \emph{the p.e.f. scales are always smaller then all of the $f_{\pm}^{\eta}$ scales}. However in the different intervals defined by the corresponding $\la_{cr}^{i,j}$ we can have \emph{different} minimal p.e.f. scales. Finally, the stronger form of the $d=5$ $a/c$-Theorems for the $\mu_+$ model, with the p.e.f. requirements implemented, takes the following form:   
\begin{eqnarray}
\bullet &&  -\infty<\lambda<-3520.09 \quad (f_-^{ep,2}=f_-^{ep,3}), \quad\quad\quad f_{IR} <f_-^{ep,3},\nonumber\\
\bullet &&-3520.09<\lambda<0.239759 \quad (f_-^{ep,2}=f_+^{ep,1}), \quad\quad f_{IR} <f _-^{ep,2},\nonumber\\
\bullet && 0.239759<\lambda<0.278864 \quad (f_+^{ep,1}=f_-^{ep,3}),\quad\quad f_{IR} < f_+^{ep,1},\nonumber\\
\bullet&& 0.278864<\lambda<1/3,\quad\quad\quad\quad\quad\quad\quad\quad\quad\quad f_{IR} < f_-^{ep,3}.\label{peftheo}
\end{eqnarray}
As we have shown in Sects.3 and 6.2., the Theorem can be also formulated as  a set of restrictions on the allowed values of IR-scale $L_{IR}$, as for example 
\begin{eqnarray}
 L_{IR}>L_-^{ep,3},\quad\quad \textrm{or}\quad\quad L_{IR}>L_-^{ep,2}\quad\quad \textrm{or}\quad \quad L_{IR}>L_+^{ep,1}.\nonumber
\end{eqnarray}

\begin{figure}[ht]
    \centering
    \includegraphics[scale=0.8]{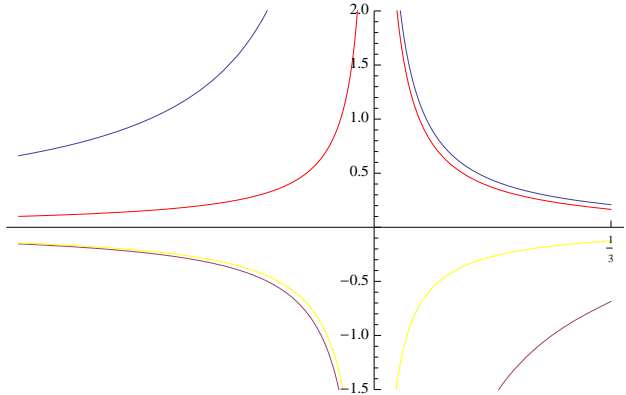}
    \begin{quotation}
    \caption[ed]{\small The curves $f^{ep,i}_{\pm}(\la;\mu_-)$: $f_+^{ep,1}$ is depicted in blue, $f_-^{ep,1}$ in purple, $f_+^{ep,3}$ in yellow and $f_-^{ep,3}$ in red.}
     \label{fig:pres3}
    \end{quotation}
\end{figure}

6.3.2. \textit{The $\mu_-$ model.} Similar analysis based on the plots of the $f_{\pm}^{ep,k}(\la,\mu_-)$ curves presented on Fig.(\ref{fig:pres3}), demonstrate that in this case the $f_{\pm}^{ep,2}$ are complex, both the $f_+^{ep,1}$ and $f_+^{ep,3}$ are negative and the following ordering takes place:
\begin{eqnarray}
     f_+^{ep,1}<f_+^{ep,3}<0<f_-^{ep,3}<f_-^{ep,1}<f_+^a\label{order}
\end{eqnarray}
Then the stronger form of the $a/c$ -Theorems reads:
\begin{eqnarray}
f_{IR}<f^{ep,3}_-, \ \forall \ \lambda<1/3 \label{pefminus}.
\end{eqnarray}

6.3.3. \textit{On the $\la$ restrictions.} An important remark concerns the additional p.e.f. restrictions on the $\la$ values similar to the ones of the GB models (\ref{lapef}). We have to remind that in the considered case of QT Gravity we are working with \emph{the fundamental scale $L$ always fixed to be equal to the bare topological scale $L=L_-^{0top}$, i.e $h_-^{top}=1$}. The proper definitions of the $\mu_{\pm}$ models, $i.e.$ the explicit form of the $\mu_{\pm}(\la)$-functions, and correspondingly the allowed ranges of values for $\la$ (see sect.2.1.) are with respect to this particular choice of $L^2$. It is different from the normalization $h_{UV}=1$ used in ref. \cite{MPS}, where the p.e.f. conditions (\ref{fcond}) lead to specific additional restrictions on $\la$, displayed on fig.1. of ref. \cite{MPS}. In our \emph{topological normalization} the same arguments as those of \cite{MPS} - namely, solving the vacua equation $h_{IR}=f_{IR}(1-\la f_{IR}-\mu f_{IR}^2)$ and next replacing the corresponding solutions $f_{IR}^{(k)}(\la,\mu_{\pm})$ into eqs. (\ref{fcond}) - result in restrictions on $\la h_{IR}$ and \emph{not on the $\la$ alone}. Hence once the choice $h_-^{top}=1$ is made, the p.e.f. requirements are imposing in fact further restrictions on the allowed values of $h_{IR}<h_{max}$ and on the $h_{UV}<h_{IR}$ as well. Let us emphasise once more that the different normalizations $h^{top}=1$, $h_{UV}=1$ and $h_{IR}=1$ although imposing different restrictions on $\la$ and $\mu$, are in fact equivalent.

Let us also mention that the implementation of the p.e.f. requirements into $d\geq 7$ $a/c$-Theorems for the $QFT_{d-1}$ duals of QT Gravity is an open problem. It requires further  holographic calculations of the explicit forms of the $t_2(f,\la,\mu)$ and $t_4(f,\la,\mu)$, which are yet unknown. 

\vspace{0.5cm}
6.3.4. \textit{Towards the holographic $t_4$-Theorems}. The fact that all the available ``2- and 3-points'' $CFT_{d-1}$ data: $ a$, $c$, $t_2$ and $t_4$,\footnote{and related to them energy fluxes $E_p(t_2,t_4)$ and the structure constants $\cal A$, $\cal B$, $\cal C$ and $C_{\Theta\Theta\Theta}$ as well.} are holographically parametrized by the $f_{UV/IR}$ values, together with the off-critical identification $f(\s)\propto W^2(\s)$ of $f(\s)$ with the matter superpotential $W(\s)$, allows to establish the RG evolution of all the $CFT$'s characteristics. Similarly to the GB case studied in Sect.6.2., we can derive the analogues of the $a/c$-Theorems for this new data. The simplest consequence concerns the RG evolution of the ratio $\frac{a}{c}$ which, together with the $t_4(f,\la,\mu)$, determines the properties  of $t_2$ and of  all the energy fluxes $E_p(f)$ as well. Considering its derivative :
\begin{eqnarray}
\frac{d}{df}\left(\frac{a}{c}\right)=\frac{4(-\la+6\mu f -9\la \mu f^2)}{(1-2\la f-3\mu f^2)^2}\nonumber
\end{eqnarray}
we conclude that its sign is governed by the corresponding numerator. Instead of the exhaustive list of all the possibilities for the different  signs of $\la$ and $\mu$ and of the roots $\tilde{f^{\pm}}$ of the eq. $(-\la+6\mu f -9\la \mu f^2=0)$, we give here few representative examples:
\begin{eqnarray}
&\bullet& \quad \mu>0,\quad\quad \la<0 : \left(\frac{a}{c}\right)_{UV}<\left(\frac{a}{c}\right)_{IR}\nonumber\\
&\bullet& \quad \mu<0,\quad\quad \la>0 : \left(\frac{a}{c}\right)_{UV}>\left(\frac{a}{c}\right)_{IR}\quad\textrm{for}\quad f_{IR}<\frac{1+\sqrt{1+\frac{\la^2}{|\mu|}}}{3\la}\label{acctheo}
\end{eqnarray}

The complete description of the RG evolution of $t_4(f(\s))$ requires certain new  holographic calculation  to be performed by the conformal perturbation theory methods or/and by using the linear fluctuations around the QT Gravity DW's solutions. As we have already mentioned in sect.6.2 in the case concerning the ``$t_2(\s)$-Theorem'', it is expected that the $t_4(\s)$ in the dual non-conformal $QFT_4$ will get new contributions from the off-critical terms related to the non-vanishing trace $\Theta(\s,x_i)=\beta(\s) \Phi(x_i)$ of the stress-tensor. Nevertheless the knowledge of $t_4(UV/IR)$ given by eqs. (\ref{tquatro}), together with the UV-to-IR flow condition $f_{UV}<f_{IR}$, allows us to derive the following relations between its critical values:
\begin{eqnarray}
&\bullet&  \mu>0,\quad \la<0 : t_4(UV)<t_4(IR),\quad \bullet\quad  \mu>0,\quad \la>0 : t_4(UV)<t_4(IR)\quad \textrm{for}\quad f_{IR}<\frac{1}{\la},\nonumber\\ 
&\bullet&  \mu<0,\quad \la>0 : t_4(UV)>t_4(IR)\quad\textrm{for}\quad f_{IR}<\frac{1}{\la},\quad \bullet\quad \mu<0,\quad \la<0 : t_4(UV)>t_4(IR)\nonumber
\end{eqnarray}
The eventual monotonic properties of $t_4(\s)$ during the massless RG flow remain indeed an open problem.

\section{Holographic RG Flows and Phase Transitions}
\setcounter{equation}{0}

The off-critical $(a)AdS_d/QFT_{d-1}$ correspondence, based on the d-dimensional Lovelock-like generalizations of the Einstein Gravity  coupled to scalar matter with an appropriate superpotential $W<0$, is expected to provide  non-perturbative solutions for certain $a\neq c$ \emph{ unitary} $QFT_{d-1}$'s  in case  the conditions of the $a/c$-Theorem  and the positive energy fluxes requirements are fulfilled. The problem addressed in the present section concerns the description of the phase structure of such dual $QFT_{d-1}$'s, by applying  the Holographic RG methods \cite{VVB,rg} to the case of the cubic Quasi-Topological Gravity \cite{My_qtop,My_thol,c_th,lovedw}.

Let us define it more precisely:  Given a superpotential $W(\s)<0$ with few extrema, to deduce from the \emph{geometric data} -vacua and flat DW's solutions \cite{lovedw} - the corresponding \emph{ $QFT$-data}: the characteristics of the $CFT$'s representing UV and IR critical points, the RG evolution of this data, the specific properties of the massless and massive phases and the nature of the phase transitions between them. 
Although the rules of the off-critical version of the $AdS_d/QFT_{d-1}$ correspondence, reviewed in Sect.7.1. below, are indeed universal 
and provide a consistent description of the $QFT$ dual  to the QT Gravity models independently on the particular form of the superpotential $W<0$, we choose to study the simplest representative example of \emph{quartic Higgs-like superpotential}. As it is demonstrated in sects.7.2, 7.3 and 7.4. below, the corresponding  dual $QFT_d$'s turns out to have  interesting and rather realistic off-critical behaviour.

\subsection{On the Off-Critical Holography Dictionary}

7.1.1. \textit{RG equations.} Let us  briefly recall the standard Wilson RG methods\footnote{see for example the excellent textbooks \cite{cardy}, \cite{muss}} we are going to use in the description of the  changes in the behaviour of the effective action and of the reduced free energy $F(\s)\sim e^{(d-1)l}$, of the correlation length $\xi(\s)\sim e^{-l}$ and  of certain correlation functions at the neighbourhood of each critical point $\sigma_k^*$. The scaling properties of their singular parts are determined by the well known RG equations: 
 \begin{eqnarray}
 &&\beta(\sigma)\frac{dF_s(\sigma)}{d\sigma} + (d-1)F_s(\sigma)=0,\quad\quad \beta(\sigma)\frac{d\xi(\sigma)}{d\sigma} =\xi(\sigma),\nonumber\\
 &&|x_{12}|\frac{\partial G_{\Phi}(x_{12},\sigma)}{\partial|x_{12}|}+\beta(\sigma)\frac{\partial G_{\Phi}(x_{12},\sigma)}{\partial\sigma} + 2(d-1+\frac{d\beta(\sigma)}{d\sigma})G_{\Phi}(x_{12},\sigma)=0\label{fs}
\end{eqnarray}
whose solutions can be easily found for each given $\beta$-function. In the case of simple zeros $n_k=1$ of  $\beta(\s)$, we can use its linear ``near-critical'' approximation  $\beta(\sigma) \approx -s_{k}(\sigma- \sigma_k^*)$, which together with eq.(\ref{rg}) provide the explicit expressions (\ref{criti}) for the exact values of the critical exponents  $s_{k}=-\frac{d\beta}{d\sigma}(\sigma^*_{k})\neq 0$ of the dual $QFT_d$. It is straightforward to check that at this approximation the RG equations (\ref{fs}) do reproduce the well known  \emph{second order phase transitions} scaling laws:
\begin{eqnarray}
 G_{\Phi}^{(k)}(x_{12},\sigma)=<\Phi_{\sigma}(x_1)\Phi_{\sigma}(x_2)>_k\approx \frac{e^{-\frac{|x_{12}|}{\xi_k}}}{|x_{12}|^{2(d-1-s_k)}}\nonumber\\
F^{(k)}(\sigma)\approx \left(\sigma- \sigma_k^*\right)^{\frac{d-1}{s_{k}}}, \quad\quad\quad \xi_k\approx(\sigma - \sigma_k^*)^{-\frac{1}{s_k}}\label{sl} ,
\end{eqnarray} 
at the neighbourhood  of each critical point  $\sigma_k^*$. The ``higher order'' zeros $n_k>1$ correspond to qualitatively different critical ``scaling'' behaviours, characterized by the specific essential singularities (involving the new parameters $\rho_k(n_k)$) of the correlation length and of  the free energy:
\begin{eqnarray}
F^{(k)}(\s, n_k)\approx \exp [(d-1)\rho_{k}(n_k)(\sigma- \sigma_k^*)^{1-n_k}] \label{marg}
\end{eqnarray}
known to describe infinite order phase transitions.

\vspace{.5cm}
7.1.2. \textit{Basic Rules of the Off-Critical $(a)AdS_d/QFT_{d-1}$ correspondence.}  We next remind the basic ingredients and some of the particular features of the $a\neq c$ \emph{non-conformal Holography} with a special  emphasis on  the restrictions that the a/c-Theorems requirements impose on the phase structure of the duals $QFT_{d-1}$: 

(1) \emph{CFT data}. Those of the stable extrema $\s^{cr}_k$ of the superpotential $W(\s)$ that represent simple zeros of the $\beta_W-$function (\ref{rg}), correspond to second order phase transitions. At such critical values $\s^{cr}_k$ of the coupling, the considered  dual $QFT_{d-1}$ becomes conformal invariant and its critical behaviour is described by a set of unitary $CFT^{(k)}_{d-1}$=($c_k, a_k, \Delta_k, C^{(k)}_{\Delta}, t_2, t_4$), duals to the \emph{stable physical vacua} of the Quasi-Topological Gravity. Both the $CFT_{d-1}(UV)$ and $CFT_{d-1}(IR)$ related by the massless RG flows are constrained to satisfy all the unitarity and p.e.f. conditions (\ref{news}) and (\ref{pefcon}).

(2) \emph{Massive and Massless RG flows}. The  non-constant solutions $\s_{k,k+1}(l)=\s(l;\s_k,\s_{k+1})$ of the RG eqs.(\ref{rg}), representing the way the coupling constant $\s(l)$ of the dual $QFT$ is running  between two consecutive critical points, describe the RG flow  and the phase transition that occurs at the UV-critical point $\s_k=\s_{UV}$. The same solution when inverted, $i.e.$ $l(\s)=-\int\frac{d\s}{\beta(\s)}$ reproduces  the DW's scale factor $e^{-2l}=e^{2A(\s)}$. In the case of polynomial superpotentials of $N$ extrema $\s_k$, the explicit solutions\footnote{for marginally degenerated critical points $s_{UV}=0$, $i.e.$ for second or higher order zeros of $\beta$, new terms containing essential singularities  appear, as shown in sect.4.3. of ref. \cite{lovedw}.} \cite{lovedw}, representing $\xi(\s)$ and $F(\s)$ can be easily found from eqs. (\ref{fs}):
\begin{eqnarray}
\xi(\s) = G_W(\s)\prod_{k=1}^{N}\big(\frac{\s-\s_k}{\s_0-\s_k}\big)^{-\frac{1}{s_k}},\quad \sum_{k=1}^{n}\frac{1}{s_k}=0,\quad\quad \s \in R\label{corlen}
\end{eqnarray}
where $G_W(\s)<\infty$ is a certain known non-singular function, the critical exponents $s_k$ are given by eqs. (\ref{criti}) and $\s_0\in(\s_k,\s_{k+1})$ denotes the ``RG -initial'' value of the coupling $\s$. 

The analytic form of the $\xi(\s)$ and $F(\s)$\footnote{as well as  of the corresponding 1- and 2-point correlation functions $<\Phi_{\s}(x_i)>$ and $<\Phi_{\s}(x_i)\Phi_{\s}(y_j)>$}  for all the values of the coupling $\s \in R$, $i.e.$ within all the physically allowed intervals $p_k=(\s_k,\s_{k+1})\in R$ called \emph{phases}, determine few  distinct RG evolutions, identified as  massive $p_k^{mass}$ or massless $p_k^{ml}$ phases. In the \emph{massless} phase the RG scale $L_{rg}(l)$ increases from $L_{rg}^{UV}=0$  to  $L_{rg}^{IR}=\infty$, $i.e.$ $\s(l=-\infty)=\s_{UV}$ and $\s(l=\infty)=\s_{IR}$, while in the \emph{massive} phase it is reaching some specific finite scale $L_{rg}^{max}=1/M_{ms}$ for infinite values of the coupling:
\begin{eqnarray}
&&massless ( UV \rightarrow IR ): 0<L_{rg}\le\infty \quad \xi(\sigma_{UV}^*)\approx \infty,\quad  \xi(\sigma_{IR}^*)\approx 0;\quad \s(\pm\infty)=\sigma_{IR/UV}, \nonumber\\    
&&massive (UV\rightarrow\infty): 0<L_{rg}\le L_{rg}^{ms}\quad \xi(\sigma_{UV}^*)\approx \infty,\quad \xi(\sigma\approx\infty)\approx L_{rg}^{max}\nonumber
\end{eqnarray}
The inverse $M_{ms}$ of the maximal scale $L_{rg}^{max}$  defines the smallest  mass gap in the energy spectrum and as a consequence of the eqs. (\ref{fs}) the corresponding 2-point correlation function manifest \emph{exponential decay} $e^{-M_{ms}|x_{12}|}$, typical for the IR limit of the free massive particle propagator. This behaviour has to be compared to the one of the \emph{ massless RG flows}, where in order to reach the maximal coupling's distance $|\sigma_{IR}-\sigma_{UV}|$ the scale $l$ must run over the entire  interval $L_{rg}\in (0,\infty)$.

The \emph{massless phases}  are geometrically described by the $AdS_d(L_{UV})/AdS_d(L_{IR})$ DW's, with the  UV-critical point representing  its $AdS_d(UV)$ boundary. Instead in the \emph{massive phases} $p^{mass}=(\s_{UV},\infty)$, the running coupling $\s(l)$ gets its maximal value for \emph{ a finite} RG distance $L_{rg}^{max}$ corresponding  to $\s(L_{rg}^{max})=\infty$. Such behaviour is known to have a geometrical description in terms of the \emph{singular} DW's of $AdS_d/n.s.$-type, whose boundary is again related to certain UV vacuum, but its IR region terminates at a naked singularity $y=y_0$, $i.e.$ $R(y_0)=-\infty$.  Observe that both \emph{the regular and the singular}  DW's are described by the same scale factor $e^{2A}(\s)\sim \xi^{2}$ as in eq. (\ref{corlen}) above, considered in different intervals of values of $\sigma$. The finite value of the  correlation length $\xi(\s=\infty)\sim 1/M_{ms}$ in the massive phase is a consequence of the special property of the critical exponents: $\sum_{k=1}^{n}\frac{1}{s_k}=0$ (see ref. \cite{lovedw}).

We should also mention that for \emph{polynomial} superpotentials $W_{N}(\s)$ with $N\geq 2$, the GB and the QTG Holographic RG flows and those   based on  the pure EH gravity-mater models are of \emph{quite a different nature}. Namely, the EH's induced $\beta_{EH}-$function has $\s_{ns}^{cr}=\infty$ as a degenerate critical point, $i.e.$ $\beta_{EH}(\infty)=0$  with $s_{ns}=0=c^{EH}_{ns}$, which corresponds to the EH's solutions of  naked singularity\footnote{As is well known it has not an appropriate $AdS/CFT$ description since the large cosmological constants, $i.e.$ CFT's of small and zero central charges, are out of the validity of the $AdS/CFT$ correspondence \cite{witt}}. Such a problem is \emph{absent} in the considered GB- and QTG-matter models, due to the fact that now we have always that $\beta_{GB/QTG}(\infty)=\infty$. Hence the naked singularity is not a critical point anymore, but an ending point of the massive phase of the dual $QFT$.

(3) \emph{Phase Transitions.} When the superpotential has few extrema, we can order the positions $\s_k$ of all the physical and topological vacua, that give rise to a set of consecutive intervals $(\s_k , \s_{k+1})$, $k =1,2,\ldots,n$, whose ends determine the b.c.'s of a specific DW. These  DWs solutions define a finite ``chain'' of consecutive DW's of common boundaries and/or horizons\cite{holo,lovedw}. Each one of the individual $DW_{k,k+1}$'s of this chain is characterized by the specific ``initial'' (in $y$) value $\s(0)=\s_0 \in (\s_k , \s_{k+1})$ of the matter field, that serves as a coupling constant initial condition $\s_0$ for the RG eqs. (\ref{rg}) and (\ref{corlen}). 
For different superpotentials and for different values of $\la$ and $\m$ we can have distinct sequences of DW's corresponding to qualitatively different $(a)AdS$ geometries: the \emph{stable} ones $AdS_d(IR)/AdS_d(UV)$, the \emph{singular} ones $AdS_d(UV)/n.s.$, etc. In the dual $QFT$ such DW chain represents  a set of different massive and massless phases defining its \emph{phase structure}. The change of the IR -type of b.c.'s that occurs at each one of the boundaries represents the \emph{transition} between two different type of $(a)AdS_d$ geometries: for example the simplest two DWs chain $AdS_d(IR)/AdS_d(UV)/n.s.$ describes the massless-to-massive second order phase transition, etc.

The above discussion makes evident that the  complete phase structure of the dual $QFT_{d-1}$ can be  described by few different perturbed $CFT_{d-1}(\s^k_{UV})$'s (\ref{pert}) - as many as different UV critical points we have. Therefore in order to establish the validity of the  off-critical $(a)AdS_{d}/QFT_{d-1}$ correspondence in a region of the coupling space including one critical point  $\s^k_{UV}$ it is sufficient to find certain perturbed $CFT_{d-1}$, whose $CFT$ data and near-critical behaviour of the $\Phi_{\s}$ correlation functions coincide with the ones derived from the Quasi-Topological Gravity-matter model with an  appropriate superpotential $W(\s)$. Notice that the \emph{non-perturbative} $\beta-$function (\ref{rg}) contains all the information about the complete phase structure and on the nature of the phase transitions for all the allowed values of the coupling $\s$, but indeed in the specific for $AdS/CFT$  large UV-scales $L_{UV}\gg 1$ ( $i.e.$ $N_c^2\gg 1$) approximation.

\subsection{Quartic Higgs-like Superpotential} 

There is a wide variety of  superpotentials $W(\s)$ that lead to physically consistent Holographic RG flows and interesting phase structures of the corresponding dual $QFT_d$'s. We next consider the simplest  representative example of the quartic superpotential 
$W(\s) = - B [ ( \s^2 - x_0 )^2 + D ]$ of inverted ``double-well'' type. For $B$, $D$ and $x_0$ all positive, $i.e.$ for $W(\s)<0$, it allows an explicit analytic construction of the  physical DWs\footnote{Due to the reflection symmetry $W(-\s) = W(\s)$,  we restrict our analysis  to the case  $\s > 0$ only.}(see sect.4 of ref. \cite{lovedw} for details).

The extrema $W'(\s) = 0$ of the superpotential for $\s\geq 0$  denoted by  $\s_{{\mathrm{IR}}} = 0$ and $\s_{{\mathrm{UV}}} = \sqrt{x_0}$ are  candidates for representing the physical vacua\footnote{in this section and in the App. B we are fixing for simplicity $\kappa=1$.}:
$$W_{{\mathrm{IR}}} = - B( x_0^2 + D ) \; , \;\; f_{{\mathrm{IR}}} = \frac{L^2 B^2( x_0^2 + D )^2}{(d-2)^2}  \; ; \;\; W_{{\mathrm{UV}}} = - B D \; , \;\; f_{{\mathrm{UV}}} = \frac{L^2 B^2 D^2 }{ (d-2)^2} \; , $$
where $W_{{\mathrm{IR}}} = W(\s_{{\mathrm{IR}}})$, etc. and $f_\ir = L^2/L_\ir^2 $. Here $L^2$ is given by the normalization established in Sect.3., $i.e.$ $L^2 = L^2_{0top}$. Therefore we can easily relate $B,D$ and $x_0$ to the corresponding vacua scales:
\begin{equation}
B x_0^2= \frac{(d-2)}{L} \left( \sqrt{f_{{\mathrm{IR}}}} - \sqrt{f_{{\mathrm{UV}}}} \right)  \; , \;\;\;\; B D = \frac{(d-2)}{L} \sqrt{f_{{\mathrm{UV}}}}  \; , \label{B and BD}
\end{equation}
As a result the $a/c$-Theorem and p.e.f. requirements on the $f_{UV/IR}$ (derived in Sects.3-6) are now easily transformed into certain conditions on the parameters of $W(\s)$. Further restrictions on these parameters are imposed by the BF unitarity condition and by the requirement that the operators $\Phi_{\s}^{UV}(x_i)$, driving the RG flows, are \emph{relevant}, namely that $0<s_{UV}<d-1$.

As we have demonstrated in Sect.2.1., the complete vacua structure of the considered models also includes the \emph{topological vacua} and depending  on the signs of the GB and Lovelock couplings $\la$ and $\mu$ we can have few of them\footnote{ in the case of the QT Gravity  for $\m>0$ we have only one such vacua, two - for $\m<0$ and $0<\la < 1/3$ and no one when both $\la$ and $\m$ are negative.}. Their  ``positions'' $\s_{top}^\pm$ are given by all the real solutions of the following equation:
\begin{equation}
\left(\frac{ (\s_{top}^\pm)^2 }{x_0} - 1 \right)^2 = \frac{ \sqrt{f_0^{top}} - \sqrt{f_{{\mathrm{UV}}}} }{   \sqrt{f_{{\mathrm{IR}}}} - \sqrt{f_{{\mathrm{UV}}}}},   \label{sigma top eq}
\end{equation}
where $f_0^{top}$  denotes $f_{GB}^{top}=\frac{1}{2\la}$ for the GB case and the corresponding $f_{\pm}^{top}$ for the cubic QT gravity models  (see sect.2.1 above). The sign of the denominator is the sign of $B$, hence in order that the r.h.s. of the last equation to be positive, $i.e.$   $B > 0$, the conditions  $f_{{\mathrm{UV}}} < f_{{\mathrm{IR}}}$ and $f_{{\mathrm{UV}}} < f_\pm$ have to be satisfied.

It is worthwhile to mention that the Holographic RG methods reviewed in this subsection are perfectly valid in the case of cubic Quasi-Topological Gravity-matter models of more complicated polynomial or non-polynomial   Superpotentials \cite{rg,vicosa}, as well as when the matter sector involves more then one scalar fields \cite{rg,lovemany}. The explicit form of the corresponding DW's scale factors for other choices of the few  extrema superpotential, as for example the super-gravity induced one $W(\s) = B \cosh (\kappa\s) [ 2\delta - \cosh(\kappa\s)]$ and of the periodic one $W=Bcos(\kappa \s)-D$, share many of the properties of  DW's of the ``double-well'' superpotential, studied here. The relevance of the \emph{quadratic} superpotential  must be pointed out. As is well known, the construction of the corresponding DW's \cite{lovedw} allows to describe the near-critical behaviour of the $QFT_{d-1}$ dual to GB or QT Gravity models, based on \emph{arbitrary matter superpotentials}. 

\subsection{Phase Transitions in  QFT's duals to GB Gravity}

According to the Holographic RG rules, the main features of the massless and massive RG flows in the $QFT$'s dual to GB Gravity with quartic Superpotential \cite{lovedw}, can be extracted from the analytic properties of the  DW's scale factor (see sect.4.4. of ref. \cite{lovedw}) or equivalently from the zeros and the singularities of the correlation length in the dual model:
\begin{eqnarray}
&&\xi(\s)\equiv e^{A(\s)} = e^{A_\infty} (\s^2)^{- 1/2s_{{\mathrm{IR}}}} \; \mid \! \s^2 - x_0 \! \mid^{- 1/s_{{\mathrm{UV}}}} G_{top}(\s), \quad \; G_{top}= \prod_{j = 1}^{4}  \mid \! \s^2 - \s_j^2 \! \mid^{- 1 / s_{top}^j} \; ,\nonumber \\
&& s_{{\mathrm{UV}}} = 16 B x_0 L_{{\mathrm{UV}}} ( 1 - L_{top}^2 / L_{{\mathrm{UV}}}^2 ) ; \;\; s_{{\mathrm{IR}}} = - 8 B x_0 L_{{\mathrm{IR}}} ( 1 - L_{top}^2 / L_{{\mathrm{IR}}}^2 ) ;\nonumber \\
&& s_{top}^j = - 64 (d-2)^{-1}  B^2 x_0^{-2} L_{top}^2 (\s^2_j - x_0)^2 \, \s^2_j \,; \; j = 1, \dots , 4;\label{gbsol} 
\end{eqnarray}
Following ref. \cite{lovedw}, we have introduced the parameters $\s_j^2 \equiv u_j + x_0  $  with    $  u_1 = - u_2 = u_+ \; ; \;\; u_3 = - u _4 = \tilde{u}_+  $  given by: 
\begin{eqnarray}
u_+ = \sqrt{D}\sqrt{ \frac{L_{UV}}{ L_{top}} - 1} \; ; \;\; \tilde{u}_+ =  i \sqrt{D} \sqrt{ \frac{L_{UV}}{L_{top}} + 1} , \label{lapostop}
\end{eqnarray}
that are related to the positions of the ``topological vacua'', $i.e.$ all the (real or/and complex numbers, depending on the values of $\la$) algebraic solutions of eqs. (\ref{sigma top eq}) above. We further parametrize the normalization constant $exp(A_\infty)\equiv \frac{1}{M}$ in terms of the RG ``initial value'' $\s_0$ of the coupling $\s$ by choosing the standard Wilson RG  normalization $\xi(\s_0)=1$.
   
The restrictions imposed by  the GB $a/c$-Theorem on the DWs properties and equivalently on the nature of the corresponding $QFT_{d-1}$'s phases, strongly depend on the sign of $\la$. Therefore the description of the phase transitions in these two cases requires a separate discussion.

\subsubsection{Negative $\la$ models: massless-to-massive phase transition}

For \emph{negative values} of $\la$ the requirement of positivity of both central functions $a>0$ and $c>0$ does not lead to any restrictions on the $f_{UV/IR}$ values. The p.e.f. requirement (\ref{fepr}) however introduces \emph{a minimal scale} $L_-^{ep}$ and it constrains $f_{UV/IR}$'s into to the interval: 
\begin{eqnarray}
0<f_{UV}<f_{IR}<\frac{(d-3)}{2|\la|(d+1)}.\label{gbpefc}
\end{eqnarray}
Since all the $f_{GB}^{\eta}$, given by eq. (\ref{fGB}), are now \emph{negative numbers}, we have no topological vacua  at all. Notice that the GB domain walls scale factors $exp(2A(\s))\sim\xi^2(\s)$ are given again by eqs. (\ref{gbsol}), but with $L_{top}=L\sqrt{2\la}$ replaced by $i|L_{top}|=iL\sqrt{2|\la|}$,\footnote{in our normalization $L=L_{0top}$.}. Although both  $s_{top}^j$ and $\s_j^2$ become complex numbers (see App. \ref{apexB}):
\begin{eqnarray}
\s_1^2&=&x_0+\sqrt{D}\sqrt{\frac{L_{UV}}{iL_{top}}-1}=(\s_3^2)^*=(\s_1^2)^R+i(\s_1^2)^I,\nonumber\\
\s_2^2&=&x_0-\sqrt{D}\sqrt{\frac{L_{UV}}{iL_{top}}-1}=(\s_4^2)^*=(\s_2^2)^R+i(\s_2^2)^I,\quad \s_j^2=(\s_{j+2}^2)^* \nonumber\\
s_{UV}&=&16Bx_0L_{UV}\left(1+\frac{L_{top}^2}{L_{UV}^2}\right),\quad \quad s_{IR}=-8Bx_0L_{IR}\left(1+\frac{L_{top}^2}{L_{IR}^2}\right),\nonumber\\
s_{top}^{j}&=&\frac{64B^2L_{top}^2}{(d-2)}\s_j^2(\s_j^2-x_0)^2=(s_{top}^{j+2})^*=(s_{top}^{j})^R+i(s_{top}^{j})^I, \ j=1,2 \label{slaneg}
\end{eqnarray}
the  function $G_{top}^{-}(\s)$  remains real and for  \emph{negative} $\la$ it has no singularities:
\begin{eqnarray}
G_{top}^-=\prod_{j=1}^2\left[(\s^2-(\s_j^2)^R)^2+((\s_j^2)^I)^2\right]^{-\frac{s_j^R}{(s_j^R)^2+(s_j^I)^2}}\times\nonumber\\
\times \exp\left({\frac{2s_j^I}{(s_j^R)^2+(s_j^I)^2}\arctan\left(\frac{(\s_j^2)^I}{\s^2-(\s_j^2)^R}\right)}\right)\label{Gnegtop}
\end{eqnarray}
The corresponding scale factor is an example of a specific chain $AdS_d(IR)/AdS_d(UV)/n.s.$ of two DWs, that give rise to the following \emph{phase structure} of the dual $QFT$: 

$\bullet$ \textit{Massless phase.} The standard  $AdS_d(IR)/AdS_d(UV)$- domain wall  connects the two physical GB vacua $\s_{UV/IR}$. Their dual are unitary $CFT_{d-1}(UV/IR)$'s, which are supposed to respect all the a/c-Theorem and p.e.f. requirements (see sect.6.2.). For positive values of $B$ and $x_0$, satisfying the condition:  
$$ 0<16 B x_0 L_{{\mathrm{UV}}} ( 1 +L_{top}^2 / L_{{\mathrm{UV}}}^2 )<\frac{(d-1)}{2},$$ 
the operator $\Phi_{\s}^{UV}(x_i)$ of dimension $\Delta_{UV}=d-1-s_{UV}<d-1$ is \emph{relevant}. As one can conclude from the explicit form (\ref{gbsol}) of the correlation length, the RG evolution from $CFT(UV)$ to $CFT(IR)$, i.e  when $\s\in(\s_{IR},\s_{UV})$, describes the \emph{massless phase} of the $QFT_d$ dual to the GB-matter model for negative $\la$ within the interval (\ref{lapef}).

\begin{figure}[ht]
    \centering
    \includegraphics[scale=0.7]{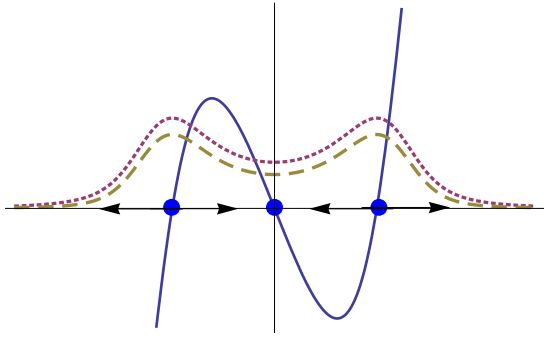}
    \begin{quotation}
    \caption[ed]{\small Massless-to-massive phase transition in $QFT_4$ dual to GB model with $\la<0$: $-\beta(\s)$-function is depicted in blue; $c(\s)$-function in beige and $a(\s)$-function in purple.}
     \label{fig:hell}
    \end{quotation}
\end{figure}

$\bullet$ \emph{Massive phase.} The singular DW of $AdS_d(UV)/n.s.$-type defined within the interval $(\s_{UV},\infty)$  shares the same boundary $AdS_d(UV)$ as the $AdS_d(IR)/AdS_d(UV)$ - domain wall. Hence the boundary conditions for the matter field $\s(y)$ of both DW's, determined by the UV-critical exponent $s_{UV}$, do coincide. They present however rather different behaviours at the deep IR-region: $\xi_{GB}(\s_{IR})=0$ in the massless phase has to be compared to the following finite value of the correlation length: 
\begin{eqnarray}
\xi_{GB}(\s \rightarrow \infty)=\frac{1}{M_{GB}},\quad M_{GB}=(\s_0^2)^{- 1/2s_{{\mathrm{IR}}}} \; \mid \! \s_0^2 - x_0 \! \mid^{- 1/s_{{\mathrm{UV}}}} G_{top}^-(\s_0),\label{gbmass}
\end{eqnarray}
with $\s_0\in(\s_{UV}, \infty)$, indicating the presence of the finite mass gap $M_{GB}(\la,L_{UV},d,\s_0)$ in the energy spectrum of the dual $QFT_{d-1}$. Therefore the interval $(\s_{UV},\infty)$ can be identified as the \emph{massive} phase of this $QFT$. Notice the crucial role of the
following important identity :
\begin{eqnarray}
\frac{1}{2s_{IR}}+ \frac{1}{s_{UV}}+2\sum_{j=1}^2\frac{s_j^R}{(s_j^R)^2+(s_j^I)^2}=0\nonumber
\end{eqnarray}
in the derivation of the mass formula (\ref{gbmass}). The Holographic RG flows describing the evolution of the  $CFT_{UV}$ data, $i.e.$ the  $a(\s)\neq c(\s)$ central functions and $\beta(\s)$, shown on fig.(\ref{fig:hell}), demonstrate all the features  typical of the second order \emph{massless-to-massive  phase transition} occurring at $\s=\s_{UV}$ in the dual $QFT_{d-1}$.

$\bullet$ \textit{The dual $QFT_{d-1}$ identification}. The verification of the validity of the conjectured duality between GB domain walls and  the RG flows of certain $QFT$ consists in the comparison of the available perturbative $QFT$ data with the Holographic RG data. As usually, the  GB vacua solutions\footnote{For large values of the corresponding $CFT$-central charges, $i.e.$ $L_{UV}/l_{pl}>L_{IR}/l_{pl}>|L_{top}|/l_{pl}\gg1$, where the $AdS/CFT$ correspondence is valid} are recognized as certain ( ${\cal N}=1$ supersymmetric) $CFT_{d-1}$'s defined by their central charges $c_{UV}$, $a_{UV}$ and by the relevant operator(s) $\Phi_{\s}^{UV}(x_i)$ of dimension $\Delta_{UV}$, whose OPE's has the form (\ref{ope}) with structure constant: 
\begin{eqnarray}
C_{\Phi\Phi\Phi}^{UV}=\frac{24(d-2)\Gamma((d-1)/2)}{D\pi^{(d-1)/2}}\sqrt{x_0}\left(1-\left(\frac{L_+^{top}}{L_{UV}}\right)^2\right).\nonumber
\end{eqnarray}
The problem is to find  an appropriate (super-symmetric) $CFT_{d-1}$, possessing  all these specific \emph{geometrically induced} properties. 
In the cases when such perturbative $OPE$'s information is available, on can  identify the non-conformal $QFT_{d-1}$, having the phase structure established above, as certain \emph{perturbed} $CFT$ (\ref{pert}). The argument for such conclusion  comes from the well known fact \cite{cardy} that starting with such p$CFT_{d-1}$ action, one can easily reproduce the perturbative $\beta(\tilde{\s})-$function (\ref{betapert}) near to the UV-critical point $\s_{UV}$  with $\tilde{\s}=\s-\s_{UV}$ and $s^{+}=s$, as explained in Sect.2.2.3. The particular form of the solutions of the corresponding RG equations (\ref{rg}) 
\begin{eqnarray}
\tilde{\s}(l)&=&\frac{s\s_0 e^{sl}}{C_{\Delta}[\s_0(e^{sl}-1)+s/C_{\Delta}]},\nonumber\\
\xi(\tilde{\s})&=&\left(\frac{\tilde{\s}-\frac{s}{C_{\Delta}}}{\tilde{\s}}\right)^{1/s}\left(\frac{\sigma_0}{\sigma_0-\frac{s}{C_{\Delta}}}\right)^{1/s},\quad \frac{1}{M}=\xi(\tilde{\s}\rightarrow\infty)\sim\left(\frac{\s_0-s/C_{\Delta}}{\s_0}\right)^{-1/s}\label{pertfase}  
\end{eqnarray}
confirms that  such p$CFT$ (for small $|\s-\s_{UV}|<1$ ) indeed describes a second order phase transition \emph{almost identical} to the one obtained from the holographic $\beta$ -function (\ref{rg}). Notice that for the inverted Higss superpotential we are considering, the OPE's of the irrelevant $\Phi_{\s}^{IR}(x_i)$ operator of dimension $\Delta_{IR}$ (from the $CFT^{IR}$ representing IR critical point) are quite different from the UV-ones: the quadratic term in eq. (\ref{ope}) is now absent due to the vanishing of the IR-structure constant $C^{IR}_{\Phi\Phi\Phi}=0$.

It is worthwhile to mention that the complete identification of the dual $QFT$'s as certain $pCFT$ indeed  requires further comparison between the corresponding $\Phi_{\s}(x_i)$'s correlation functions, obtained from the analysis of small fluctuations around the DW's background  in GB Gravity-matter model with the ones calculated by the conformal perturbations techniques.

\subsubsection{Positive $\la$ models: massless RG flows}

The GB Gravity-matter model in the case of positive GB-coupling $\la$ within the interval (\ref{lapef}) permits few topological vacua $\s_{top}^j$ defined by the zeros of the c-function $c(f_{GB}^{top})=0$.  The condition $c\geq 0$ for causality ($i.e.$ absence of ghosts) of the GB Gravity\footnote{and equivalently one of the requirements for the unitarity of the dual $CFT_{d-1}$} introduces a natural \emph{fundamental minimal scale} $L_{GB}^{0top}=L_{GB}^{top}/\sqrt{2\la}$, which however is \emph{different} form the smallest physical p.e.f. scale  $L_+^{ep}\geq L_{GB}^{top}$, as shown in sect.6.2. The remaining  a/c-Theorem  and p.e.f. conditions  are satisfied when the physical vacua scales $L_{UV/IR}$ are restricted as follows:
\begin{eqnarray}
       L_{UV}>L_{IR}\geq L_{+}^{ep}=\frac{L^{0top}_{GB}}{\sqrt{f_{+}^{ep}}}\label{gbposit}    
\end{eqnarray}
which has as a consequence that  $s_{UV}$ is positive and all the others $s_{IR}$ and $s_{top}^j$ are negative. Then the DW's scale factor (\ref{gbsol}) describes (for $\s>0$) a ``chain'' of two different GB domain walls of common boundary: the physical one  $AdS_d(IR)/AdS_d(UV)$ 
and  ``phys-top'' one  $AdS_d(UV)/AdS_d(top)$, which
 is relating the physical UV vacua to the topological one $\s_{top}$. The behaviour of the correlation length (\ref{gbsol}), namely $\xi(\s_{UV})\approx \infty$ and $\xi(\s_{IR})=0=\xi(\s_{top})$, indicates that both DW's might describe \emph{massless RG flows}. However only the ``physical vacua'' phase $\s \in (\s_{IR},\s_{UV})$ satisfies all the $CFT(UV/IR)$'s consistency conditions: positivity of the a-central function and the p.e.f.'s. Notice that the c-central function is  \emph{positive and monotonically decreasing} in  both phases $(\s_{IR},\s_{UV})$ and $(\s_{UV},\s_{top})$. 

\begin{figure}[ht]
\centering
\subfigure[]{
\includegraphics[scale=0.7]{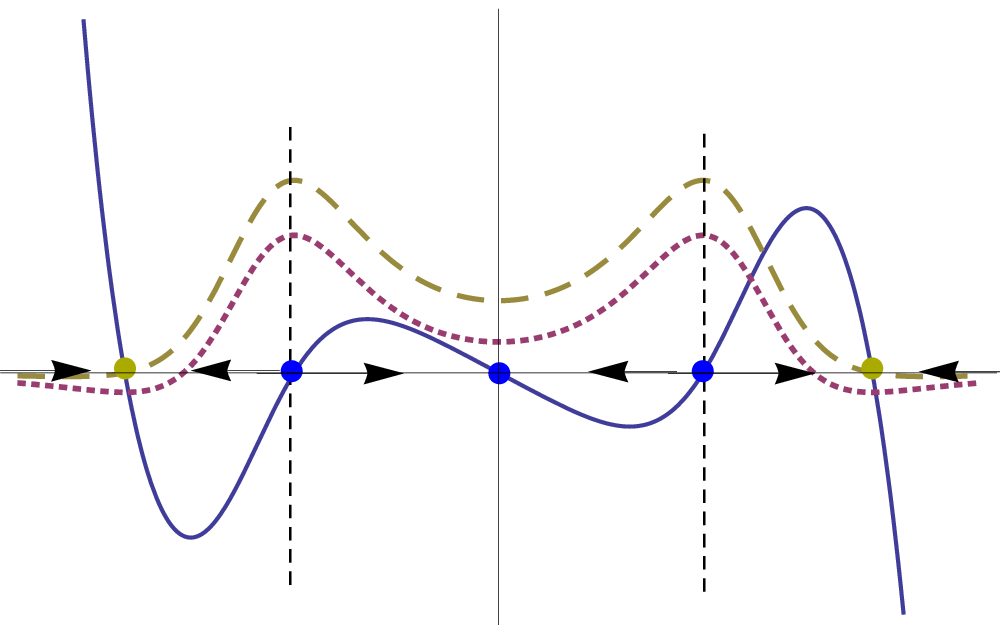}
\label{fig:7a}
}
\subfigure[]{
\includegraphics[scale=0.7]{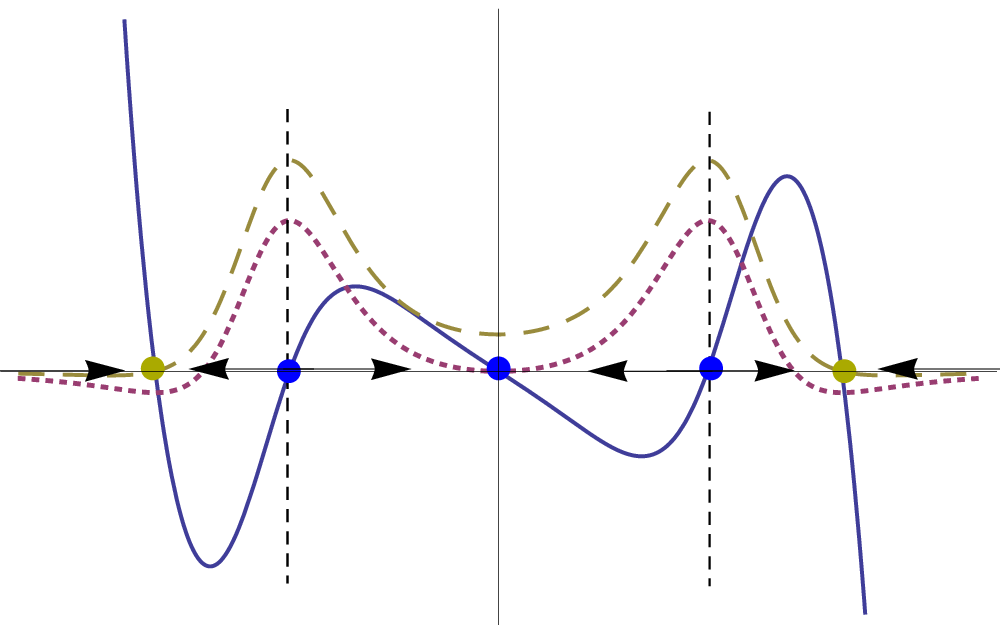}
\label{fig:7b}
}
\subfigure[]{
\includegraphics[scale=0.7]{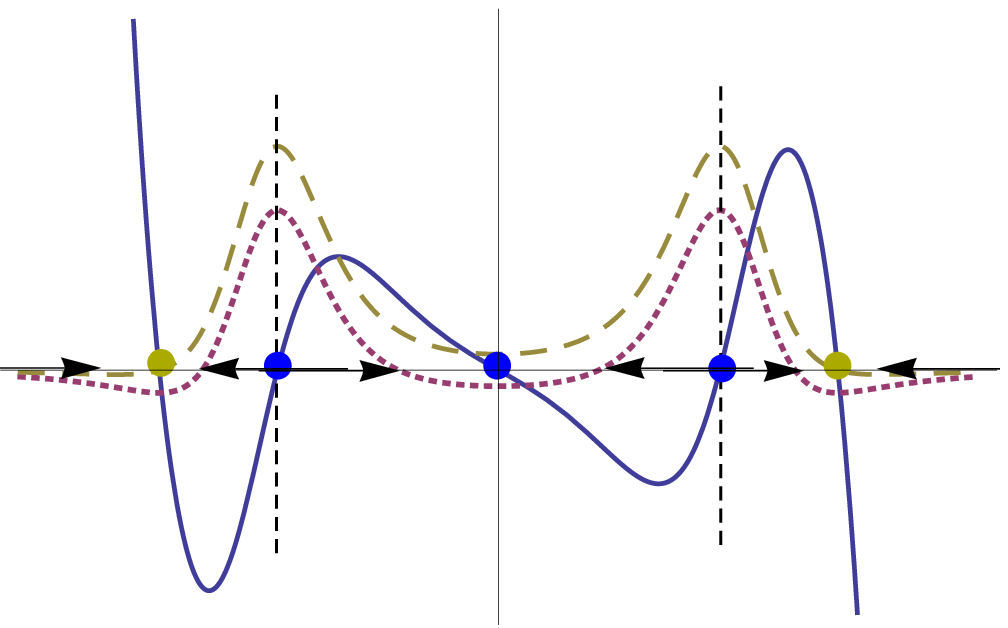}
\label{fig:7c}
}
\subfigure[]{
\includegraphics[scale=0.7]{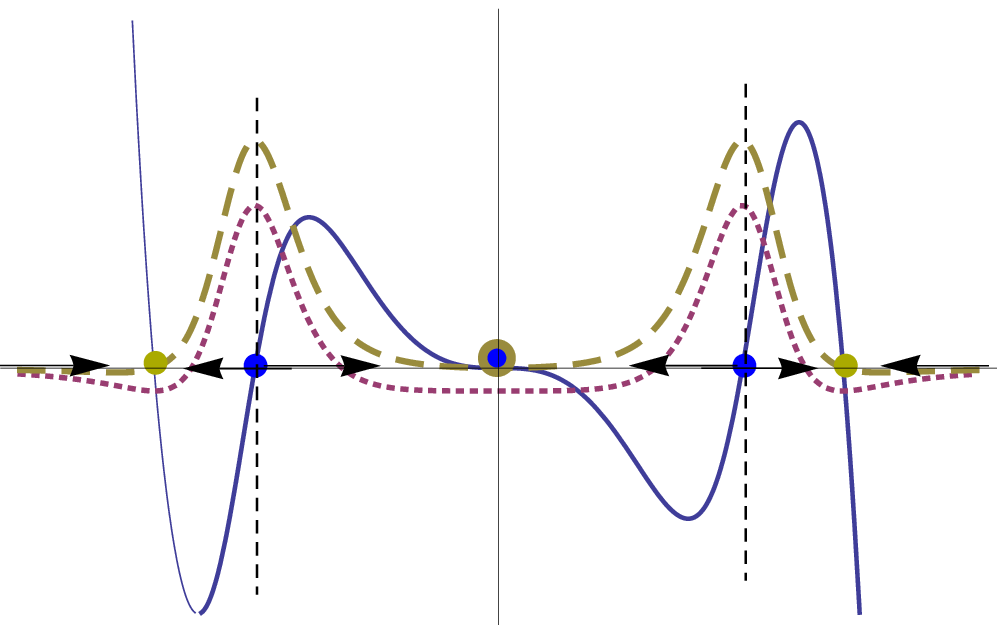}
\label{fig:7d}
}
\label{fig:5}
\caption{RG flows in $d=5$ $\mu_+$-model with $\la=\frac{1}{4}$: the $-\beta(\s)$-function is the continues blue curve; the $c(\s)$-function is the beige traced one and the  $a(\s)$-function is the purple doted curve; the critical points that are extrema of $W(\s)$ are depicted in blue, while the beige ones represent  topological $c=0$ $CFT_4$'s. The IR-scale $L_{IR}$ restrictions: on Fig.(a) $L^2_{IR}>L^2/f^{a}_+(\la;\mu_+)$; on Fig.(b) $L^2_{IR}=L^2/f^{a}_+(\la;\mu_+)$; on Fig.(c) $L^2/f^{top}_+(\la;\mu_+)<L^2_{IR}<L^2/f^{a}_+(\la;\mu_+)$ and on Fig.(d)  $L_{IR}^2=L^2/f^{top}_+(\la;\mu_+)=L^2_{top}$.}
\end{figure}

We next realize that the $CFT_{top}$ is characterized by $c_{top}=0$ and $a_{top}<0$. Therefore the ``phys-top" phase $\s\in (\s_{UV},\s_{top})$ is divided in two parts by the ``critical point'' $\s_{cr}^a$ with $a(\s_{cr}^a)=0$, which is reached when the RG scale gets its critical value  $l_{cr}=L_{GB}^a$, as shown on Fig.(\ref{fig:7a}). The physically consistent part $(\s_{UV},\s_{cr}^a)$ of this phase, where $a>0$ is still positive, thus represents a natural mass scale $M_{GB}^a=1/L_{GB}^a$. The complete description of such eventually physical ``$a$-massive'' phases is out of the scope of the present paper. The degenerate case $L_{IR}=L^a_+ $ of coinciding IR- and $a$-scales, demonstrated on Fig.(\ref{fig:7b}), is an example where the zero $\s_{cr}^a$ of $a(\s)$- central function is representing the \emph{IR critical point}. However such values of $L_{IR}$ \emph{do not satisfy} the GB p.e.f. conditions (\ref{fepr}) and (\ref{gbposit}). Therefore the corresponding dual $QFT$'s have to be discarded as physically inconsistent. Finally, in the example shown on Fig.(\ref{fig:7c}), $i.e.$ for $L^2/f^{top}_+(\la;\mu_+)<L^2_{IR}<L^2/f^{a}_+(\la;\mu_+)$  the $a(\s)$- central function changes its sign within the former massless phase $(\s_{UV},\s_{IR})$, thus violating the p.e.f. conditions and the $a/c$- Theorem. 

\vspace{0.5cm}

$\bullet$ \textit {Marginal degenerations}. Let us also mention the limiting case $L_{IR}=L_{GB}^{top}$, when the UV-vacuum $\s_{UV}^2=x_0$ coincides with two of the topological vacua $\s_j^2=x_0$ for $j=1,2$,\footnote{remember that $\s_j^2$ and $s_{top}^j$ for $j=3,4$ are complex conjugate of each other, as one can see from their definitions (\ref{lapostop}) for $\la>0$ and do not represent topological vacua at all}. The scale factor and $\xi(\s)$  now exhibit an \emph{essential singularity} at $\s^2=x_0$:
\begin{eqnarray}
\xi(\s)=e^{A(\sigma)}\propto (\sigma^2)^{-\frac{1}{2s_a^{(0)}}}\exp{\left(-\frac{D}{32x_0BL(\sigma^2-x_0)^2}\right)}\prod_{i=3}^4|\sigma^2-x_i|^{-\frac{1}{s_{gr}^i}}\label{margb} 
\end{eqnarray}
Although it  involves complex quantities, the last term can be rewritten in  a real form similar to the one of eq. (\ref{Gnegtop}). This qualitatively new behaviour is known to be specific for the infinite order phase transitions, which is expected to take place in the dual $QFT$ . It can be confirmed also by  perturbative RG calculations in the dual  $QFT$ with $s_{UV}=0$, similar to the ones presented at the end of the previous subsection. Notice however that in this particular example of marginally degenerated critical point, graphically represented on Fig.(\ref{fig:7d}), the p.e.f. conditions (\ref{gbposit}) \emph{are violated}. Hence it does not represent physically consistent unitary $QFT_{d-1}$. 

\begin{figure}[ht]
\centering
\subfigure[]{
\includegraphics[scale=0.7]{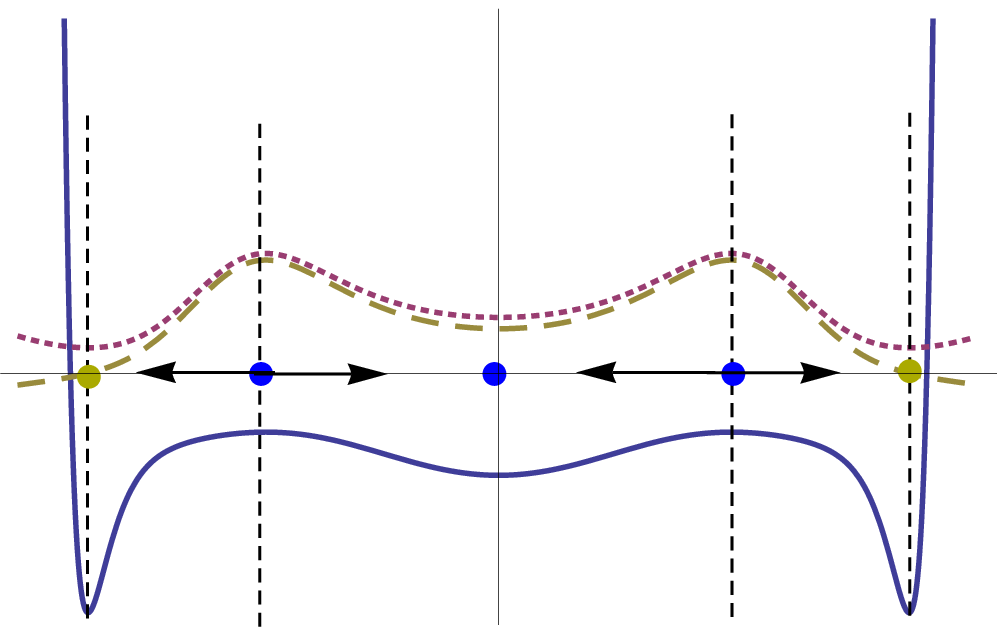}
\label{fig:6a}
}
\subfigure[]{
\includegraphics[scale=0.7]{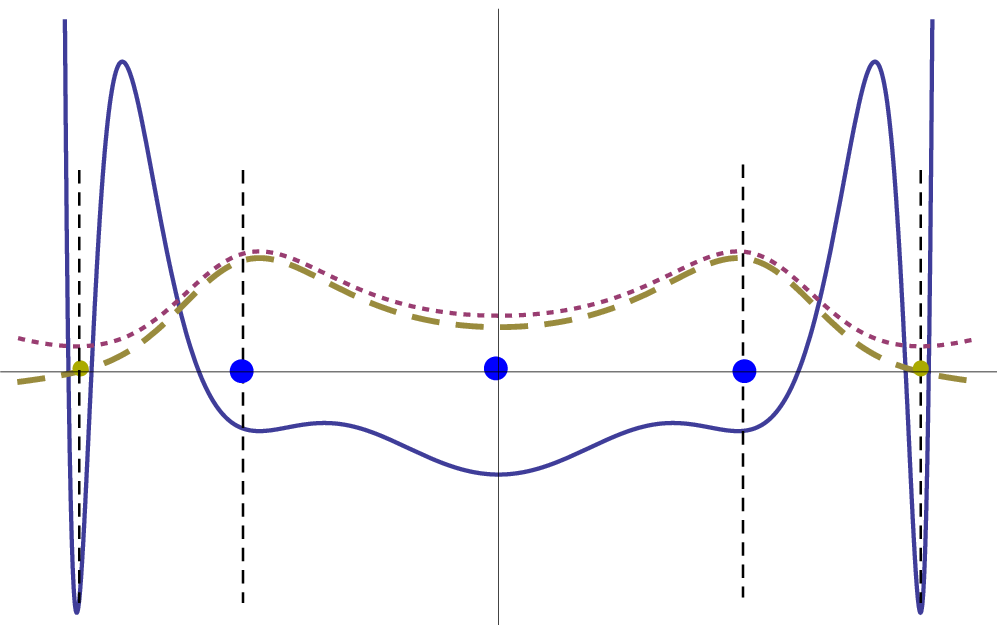}
\label{fig:6b}
}
\label{fig:6}
\caption{The matter potential $V(\s)$, in blue, for $d=5$  $\mu_+$-models with $\la<\frac{5}{27}$ and $L_{IR}^2>L^2/f^a_+(\la;\mu_+)$: on Fig.(a) $0<s_{UV}<4$, while on Fig.(b) $s_{UV}>4$. The $c(\s)$-function is the traced beige curve, while the $a(\s)$-function - the red dotted one.}
\end{figure}

$\bullet$ \textit{Comment about $s_{UV}>d-1$ case}. As we have mentioned in sect.2.2.2 (see also sect.4.4 of ref. \cite{lovedw}), the effect of this new restriction on the superpotential parameter $Bx_0$ (replacing the old one $s_{UV}<d-1$) is that the operator $\Phi_{\s}^{UV}(x_i)$ has now \emph{negative dimension}  $\Delta_{\Phi}<0$. Therefore the corresponding $CFT(UV)$ is  \emph{non-unitary}\footnote{another possibility is the so called $\Delta_-$-quantization \cite{witku} corresponding to spontaneous breaking of the conformal symmetry. Such DW's however do not describe RG flows in the dual $QFT_{d-1}$.}. The main difference between these two cases is that the corresponding \emph{old} $V_{(a)}(\s)$  and  \emph{new} $V_{(b)}(\s)$ matter potentials, shown on Figs. (\ref{fig:6a}) and (\ref{fig:6b}), have \emph{different number of extrema}, although the inverted ``double-well'' form of the superpotentials $W_{(a)}$ and $W_{(b)}$ is  preserved. The presence of two more ``maxima"  in the potential $V_{(b)}$  reflects the fact that together with  $m^2_{IR}$ and $m^2_{top}$, also the $m^2_{UV}(b)$ is now positive. Hence the new maxima must appear in between the ``neighbour'' minima $\s_{UV}$ and $\s_{top}$ (in the left) and $\s_{IR}$ (in the right). Such GB domain walls are interpolating now between  two minima of the matter potential $V_{(b)}$, while in the case of $V_{(a)}$  they are relating one  maximum with the neighbouring minima.

\subsection{RG Flows from Quasi-Topological Gravity DW's}

The flat DW's solutions of the Quasi-Topological Gravity  with inverted Higgs-like Superpotential in $d\geq5$ dimensions, constructed in ref. \cite{lovedw}, serve as a main tool in the analysis of the Holographic RG flows in the corresponding dual $QFT_{d-1}$. Again the basic ingredients in the description of the massless and massive $QFT$'s phases are the correlation length and the reduced free energy:
\begin{equation}
\xi(\s)\propto e^{A(\s)} = e^{A_\infty} \, |\s |^{- 1/s_{{\mathrm{IR}}}} \; \mid \! \s^2 - x_0 \! \mid^{- 1/s_{{\mathrm{UV}}}} \; \prod_{j = 1}^{8}  \mid \! \s^2 - \s_j^2 \! \mid^{- 1 / s_{top}^j} \; ,   \label{Solqt} 
\end{equation}
Similarly to the GB Gravity case, we have denoted the positions of the ``topological vacua" $\s_j^2 \equiv u_j + x_0 $, with $x_0=\s_{UV}^2$, independently of whether they are real or/and complex numbers. The parameters    
$ u_1 = - u_2 = u_+ \; ; \;\; u_3 = - u _4 = \tilde{u}_+ \; ; \;\; u_5 = - u_6 = u \quad u_7 = - u_8 = \tilde{u}_- $ 
are defined as follows : 
\begin{eqnarray}
u_\pm = \sqrt{\frac{(d-2)}{B}} \sqrt{\frac{1}{L_{top}^{\pm}}-\frac{1}{L_{UV}}} 
 \; ; \;\; 
 \tilde{u}_\pm = i\sqrt{\frac{(d-2)}{B}} \sqrt{\frac{1}{L_{top}^{\pm}}+\frac{1}{L_{UV}}} \label{udef}
\end{eqnarray}
The  explicit values of the  corresponding ``critical exponents" $s_k$ (\ref{criti}) are given by:
\begin{eqnarray}
&&s_{{\mathrm{UV}}}= 16 B x_0 L_{{\mathrm{UV}}} \left( 1 - \frac{(L_{top}^+)^2} { L_{{\mathrm{UV}}}^2} \right) \left( 1 - \frac{(L_{top}^-)^2 }{ L_{{\mathrm{UV}}}^2}\right),\nonumber\\ 
&&s_{{\mathrm{IR}}}= - 8 B x_0 L_{{\mathrm{IR}}} \left( 1 - \frac{(L_{top}^+)^2}{  L_{{\mathrm{IR}}}^2} \right) \left( 1 - \frac{(L_{top}^-)^2 }{ L_{{\mathrm{IR}}}^2}\right) \nonumber\\
&&s_{top}^p = - 64 (d-2)^{-1}  B^2 x_0^{-2} (L_{top}^+)^2 (\s^2_p - x_0)^2 \, \s^2_p \, \left( 1 - \frac{(L_{top}^-)^2}{(L_{top}^+)^2} \right) ; \; p = 1, \dots , 4. \nonumber\\
&& s_{top}^q = - 64 (d-2)^{-1}  B^2 x_0^{-2} (L_{top}^-)^2 (\s^2_q - x_0)^2 \, \s^2_q  \, \left( 1 - \frac{(L_{top}^+)^2}{ (L_{top}^-)^2} \right) ; \; q = 5, \dots , 8. \label{stop}  
\end{eqnarray}
According to their definition (see sect.2.2.2), they are related  to the scaling dimensions $\Delta_k$ of certain dual conformal fields $\Phi_{\s_k}(x_i)$ from the conjectured  CFT$_{d-1}$'s  ``attached'' to each one of the vacua of the considered Quasi-Topological Gravity-matter model. The values of the corresponding $OPE$'s structure constants (\ref{str}) for these operators are given by: 
\begin{eqnarray}
C_{\Phi\Phi\Phi}^{UV}&=&\frac{24(d-2)\Gamma((d-1)/2)}{D\pi^{(d-1)/2}}\sqrt{x_0}\C^{UV}, \quad C_{\Phi\Phi\Phi}^{IR}=0=C_{\Phi\Phi\Phi}^{top}\nonumber\\
\C^{UV}&=&1-2\la\left(\frac{L_+^{0top}}{L_{UV}}\right)^2-3\mu \left(\frac{L_+^{0top}}{L_{UV}}\right)^4\label{wstr}
\end{eqnarray}
They determine the singular part of the ``holographic'' $OPE$'s  (\ref{ope}) and are essential for the identification of the dual $QFT$'s as appropriate perturbed $CFT$'s (similarly to the GB case considered in Sect.7.3.1). By construction, the irrelevant operators $\Phi_{IR}^{top}$  for all the $CFT_{top}^{(j)}$  have vanishing $C_{\Phi\Phi\Phi}^{top}$, independently of the form of the superpotential, due to the fact that always  $\C(\s_{top}^j)=0$. The vanishing of the  IR structure constant however is a specific property of the $Z_2$ symmetric form of the quartic superpotential under investigation. It turns out that  the new superpotentials $\tilde{W}=W+K\s^3$ obtained by adding of a cubic term, give rise to dual $QFT$ models having both  $C_{\Phi\Phi\Phi}^{UV}\neq 0$ and $C_{\Phi\Phi\Phi}^{IR}\neq 0$ different from zero.

Observe that the scale factor $e^{2A(\s)}\sim\xi^2(\s)$ given by eq. (\ref{Solqt}), for all the considered $d\geq 5$ QT Gravity models, represent two specific DWs chains: $AdS_d(IR)/AdS_d(UV)/n.s.$ or the $AdS_d(IR)/AdS_d(UV)/AdS_d(top)$ one. Let us remind  that the holographic description of such ``two-phases structure'' of the dual $QFT_{d-1}$'s, depends not only on the analytic properties of $\xi(\s)$, but also  on the values and the sign of the $a(l)$ and $c(l)$-central functions. The holographic $a/c$-Theorem selects as \emph{physically admissible} those of the phases, for which both  $a(\s)$ and $c(\s)$ are positive for all the values $\s \in(\s_{IR}, \s_{UV})$ and $\s \in (\s_{UV}, \s_+^{top})$ or $\s \in (s_{UV}, \infty)$ and the corresponding \emph {critical values} $f_{UV/IR}$'s satisfy the p.e.f. conditions. The results of Sects.3-6 suggest that the extensive list of different QT Gravity  models  can be organized in the following three families of models of \emph{identical phase structures}:  (i) of minimal scale $f_{IR}<f^a_+$ or $f_{IR}<f^{top}_+$, (ii) of maximal scale $f_{UV}>f^{c'}_-$ and (iii) no minimal or maximal $L_{\pm}^{\eta}$ ($\eta=top, a ,c'$) scales at all.

\subsubsection{Minimal scale phase structures}

A common feature of all the $QFT_{d-1}$'s with \emph{minimal scale's} phase structure, duals to the $\mu_{\pm}$  QTG models for different ranges \footnote{notice that the particular form of these $\la$-intervals is $d$-dependent, as shown in Sect.5 and App. \ref{apexA}} of values of $\la$ is that they satisfy one of the following two type of restrictions: 

(a) $f_{UV}<f_{IR}<f^{a}_+(\la;\mu_{\pm})$ defining the GB-like \emph{massless} $(\s_{IR},\s_{UV})$ and \emph {$a$-massive} $(\s_{UV},\s^{a}_{cr})$ phases; 

(b) $f_{UV}<f_{IR}<f^{top}_+(\la;\mu_{\pm})$ leading to a \emph{new ``top-massless''}  $(\s_{UV}, \s_+^{top})$ phase, together with the standard  $(\s_{IR}, \s_{UV})$ massless one. 

It is important to mention that the p.e.f. requirements discussed in Sect.6.3. in the case of $d=5$ QT Gravity  models are in fact imposing \emph{stronger minimal scales} restrictions (\ref{peftheo}) and (\ref{pefminus}):
 $$f_{UV}<f_{IR}<f_{\pm}^{ep,k}<f^{a}_+<f^{top}_+,$$ 
where $f_{\pm}^{ep,k}=L^2/(L_{\pm}^{ep,k})^2$ with $k=1,2,3$  denote one of the the minimal p.e.f. scales (\ref{roots}), say $f_-^{ep,3}$, depending on the specific interval (\ref{peftheo}) of values of $\la$ under consideration. They determine  the \emph{physically consistent massless} phases, while in the case of massive phases they are imposing restrictions on the corresponding  $f_{UV}$'s values.  

\vspace{0.5cm}
7.4.1. \textit{Phase structure of $QFT_4$'s dual to $d=5$ QT Gravity.} The explicit analytic form of  the correlation length (\ref{Solqt}) for the $QFT_4$'s  dual to the $\mu_{\pm}$ QTG models for all the values of $\la\in (-\infty, 1/3)$ demonstrates  that the $(x_0,0)$-phase  has all the features of a true massless phase. If we further restrict the critical values $f_{UV}=f(x_0)$ and $f_{IR}=f(0)$  to belong to one of p.e.f. allowed intervals (\ref{peftheo}) or (\ref{pefminus}), then it describes  an \emph  {unitary consistent} massless RG flow in the dual $QFT_4$. 

Depending on the specific properties of the $a$-central function, we can realize two qualitatively different behaviours of the dual $QFT_4$. We first consider the models  of type (a): 
\begin{eqnarray} 
 \bullet \quad \mu_+(d=5),\quad\la\in(\frac{5}{27},\frac{1}{3});\quad\quad\quad\quad \bullet \quad\mu_-(d=5),\quad\la\in (-\infty,\frac{1}{3})\nonumber
\end{eqnarray} 
which are characterized by the existence of one zero $\s^{a}_{cr} \in (\s_{UV},\s_+^{top})$ of the $a$-central function, representing the end of the holographic description. As in the GB case, it  introduces a specific new mass scale $M_a=1/L_+^a$, which motivate us to consider  the physically consistent part of this phase $(\s_{UV},\s^{a}_{cr})$  as an ``$a$-massive phase''. The Holographic RG flows of the consider type (a) models within the interval $\la\in(5/27, 1/3)$ are practically identical to the ones  of the GB case, which is graphically represented on Figs.(\ref{fig:7a})-(\ref{fig:7d}) for a particular GB value $\la=1/4$ of the $\mu_+(d=5)$- model for $0<s_{UV}<4$.

\begin{figure}[ht]
\centering
\subfigure[]{
\includegraphics[scale=0.6]{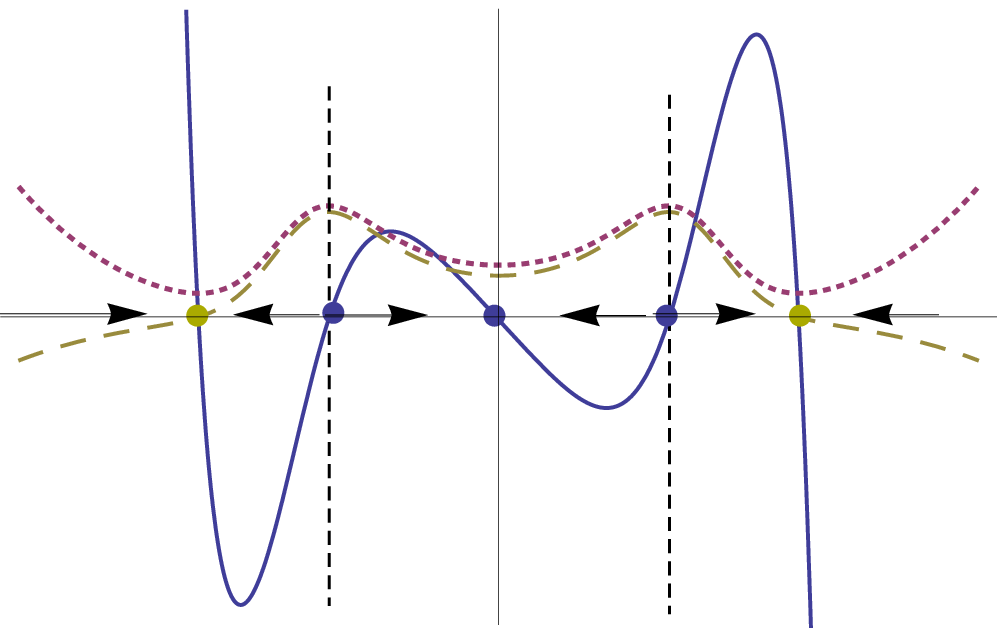}
\label{fig:5a}
}
\subfigure[]{
\includegraphics[scale=0.6]{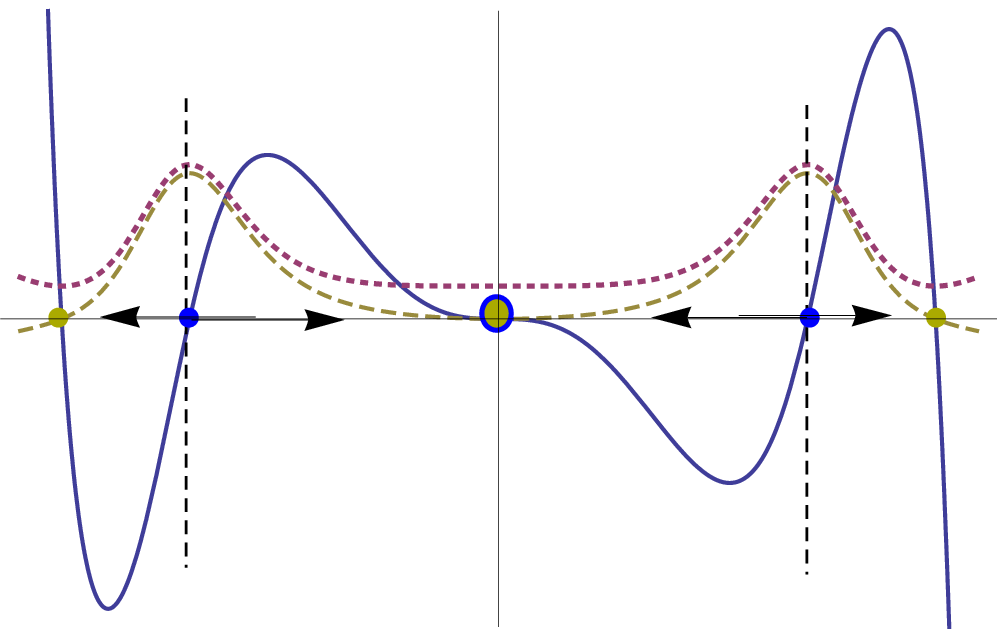}
\label{fig:5b}o
}
\label{fig:4}
\caption{RG flows in $d=5$ $\mu_+$ models with $\la<\frac{5}{27}$. The continues blue curve represents the $-\beta(\s)$-function, the $c(\s)$-function is the traced beige curve and the $a(\s)$-function is the red dotted one. The IR-scales restrictions: on Fig.(a) $L_{IR}^2>L^2/f^{top}_{+}(\la;\mu_+)$  and  on Fig. (b) $L_{IR}^2=L^2/f^{top}_{+}$, representing a marginal degenerated critical point.}
\end{figure}

While the holographic properties as well as the unitarity and causality of the massless phase $(\s_{UV}, \s_{IR})$ are well established, the complete understanding of this ``$a$-massive'' phase  and of its physical consistency requires further investigations of the corresponding  off-critical $<\Phi(x_i)\Phi(0)>$ correlation functions. Similarly, the identification of the second order phase transition at $\s_{UV}=x_0$ as being of massless-to-$a$-massive type needs more then the Holographic RG arguments we have used. Namely, the calculations of the off-critical mass spectrum by considering the small fluctuations around the considered DW background are required.

We next consider the models of \emph{type (b)}: $\mu_+(d=5,\la)$ with $\la\in(-\infty,\frac{5}{27})$, whose  physically consistent massless phase $(x_0,0)$ coincides with the one of the type (a) models described above. The main difference is that now $a(\s)$ is positive over  the entire interval $(\s_{UV},\s_{top}^+)$ that represents a new ``top-massless'' phase as indicated by the behaviour of the $\xi$: namely, $\xi(\s_{UV})=\infty$ and $\xi(\s_{top}^+)=0$. The corresponding massless-to-top-massless second order phases transition is graphically demonstrated on Fig.(\ref{fig:5a}) for the case of $L_{IR}>L_+^{top}$. The plots presented on Fig.(\ref{fig:5b}) describe the RG flows in the degenerate case $L_{IR}=L_+^{top}$, where the IR vacua $\s_{IR}=\s_{top}$ (the beige round central dot )  coincides with one of the topological vacua, giving rise to a particular second order phase transition between two ``top-massless'' phases.

Notice that independently of the positivity of  both central charges $a$ and $c$ during the ``physical-to-top- massless'' RG flow,  the ``top-massless'' phase \emph{is not satisfying the p.e.f conditions} (\ref{peftheo}). Therefore it does not represent consistent unitary and causal $QFT_4$. Although the proof of the causal inconsistency or of the eventual bulk instabilities of the corresponding $AdS_d(UV)/AdS_d(top)$ DW's is still missing, the field-theoretical  p.e.f.'s type of arguments are sufficient to classify the $d=5$ models of type (b) as {physically inconsistent}.

The above discussion leads us to the following conclusion concerning the dual $QFT_4$'s off-critical properties: 

\textit{Assuming the validity of the Holographic Renormalization Group for the cubic QT Gravity-matter models\footnote{which are expected to be ghost free and stable w.r.t. small fluctuations of the metrics and of the matter field} and imposing the  a/c-Theorem and p.e.f. requirements, no standard massive phases of $\xi(\s \rightarrow \infty)=1/M$ can be realized. Hence the corresponding physically consistent dual $QFT_4$'s contain only one well defined massless phase $(\s_{UV}, \s_{IR})$.}

\vspace {0.5 cm}

7.4.2. \textit{Comments on the phase structure of $QFT$'s dual to $d \geq 7$ models}.  The properties of the chains of DW's solutions and a part of the corresponding  $a/c$-Theorem requirements, corresponding to the $d \geq 7$ models satisfying the \emph{minimal scale restrictions}, turns out to be almost identical to the $d=5$ case considered above. In what follows we shall give the complete list of these $d \geq 7$ models with a  brief summary of their main features, emphasizing  the few important differences with the $d=5$ models.

$\bullet$ \textit{Massless and $a$-Massive phases.} The $d\geq 7$ Quasi-Topological Gravity models, whose dual $QFT_{d-1}$'s are of type (a), $i.e.$  representing massless and $a$-massive phases are given by:  
\begin{eqnarray}
\bullet & \mu_+(d\geq7),&   -\infty<\la<\frac{d(d-4)}{3(d-2)^2},\quad\quad\quad \bullet\quad\mu_-(d=7,8), \la\in (0,\frac{1}{3}),\nonumber\\
\bullet & \mu_-(d\geq 9),&  0<\la<\frac{(d-8)(d-4)}{3(d-6)^2}\label{mamass}
\end{eqnarray}
Again the explicit expression of their correlation lengths (\ref{Solqt})  confirms this phase structure. Notice that when one or two of the topological scales $L_{\pm}^{top}$ become imaginary, as in the case of $\mu_-(d=5)$ model with $\la\in (-\infty,0)$ and also for the $\mu_+(d\geq7)$ model with $\la\in ( -\infty,0)$, the replacement of $L_{\pm}^{top}$ with $i|L_{\pm}^{top}|$ transforms the ``topological vacua positions" $\s_j^2=(\s_{j+2}^2)^*$ (with $j=1,2,5,6$) and the  critical exponents $s_{top}^p$ and $s_{top}^q$ into complex numbers. Again as in the GB case (see App. \ref{apexB}) the contribution $G_{top}^{QT}(\s)$ of these ``topological vacua'' to the $\xi$ has the following \emph{real form}:
\begin{eqnarray}
G_{QT}^{top}&=&\prod_{j=1,5}^{2,6}\left[(\s^2-(\s_j^2)^R)^2+((\s_j^2)^I)^2\right]^{-\delta_j^R}exp\left(2\delta_j^I\arctan\left(\frac{(\s_j^2)^I}{\s^2-(\s_j^2)^R}\right)\right)\nonumber\\
\delta_j^R&=&\frac{s_j^R}{(s_j^R)^2+(s_j^I)^2},\quad\quad\quad    \delta_j^I= \frac{s_j^I}{(s_j^R)^2+(s_j^I)^2}\label{Glove}
\end{eqnarray}
The separation of the real and imaginary parts of the $s_{top}^{q/p}$ and $\s_j^2$ in the case  when both topological scales are imaginary can be realized as in the GB case, described in App. \ref{apexB}.

$\bullet$ \textit{Massless and top-massless phases.} The list of the $\mu_{\pm}$ models having one minimal topological scale $L_+^{top}$ and such that the both central functions $c(\s)$ and $a(\s)$ are  positive within the intervals $(\s_{IR},\s_{UV})$ and $(\s_{UV},\s_{top}^+$), includes:
\begin{eqnarray}
 \bullet\quad \mu_+(d\geq7), \frac{d(d-4)}{3(d-2)^2}<\la<\frac{1}{3}, \quad\bullet \quad \mu_-(d\geq 9),\quad  \frac{(d-8)(d-4)}{3(d-6)^2}<\la<\frac{1}{3}\label{mss}
\end{eqnarray}
Their holographic RG description is similar to the corresponding $d=5$ models.

Although the explicit holographic form of the p.e.f  requirements for  $QFT_{d-1}$ dual to $d \geq 7$ cubic  QT Gravity is yet unknown, we expect that the massless RG flows for $\s \in (s_{UV},s_{IR})$ do represent unitary $QFT_{d-1}$'s, while  the corresponding  ``top-massless phases'' are \emph{physically inconsistent}. This leads us to make a conjecture that only the holographic type (a) models can give rise to unitary and causal $QFT_{d-1}$ models.

$\bullet$ \textit{ $d\geq 7$ massless-to-massive phase transitions.} As we have shown in Sect.5, for negative values of $\la$ and $\mu$, both central functions $a(l)$ and $c(l)$ are automatically \emph{positive} for all the $d\geq 7$ QT gravity models. Therefore we have no topological vacua ($i.e.$ both $L_{\pm}^{top}$ are imaginary) and also no real zeros of the $a(\s)$ exist for all the values of $\s \in R$.  Under these conditions the corresponding  scale factor, representing a chain of two  DW's  $AdS(IR)/AdS(UV)/n.s.$, gives rise to one massless phase $(\s_{IR},\s_{UV})$ and one \emph{standard} massive phase $(\s_{UV},\infty)$ quite similar to the GB case described in Sect.7.2. As usual they are characterized by  $\xi(\s_{UV})=\infty$  and $\xi(\s_{IR})=0$ (the former one) and by the specific value $M_{QT}=1/\xi(\infty)$ of the mass gap (the later one), obtained by taking appropriate limits of the eqs. (\ref{Solqt}) and (\ref{Glove}). Let us emphasize that  the $QFT_{d-1}$'s dual to the $d \geq7$ QT Gravity for  negative $\la$ and $\mu$, exhibit \emph{ standard (GB-like) massless-to-massive phase transition}, which are absent in the corresponding $d=5$ models.

\subsubsection{Maximal scale massless phase}

An important new feature of  the $QFT_{d-1}$'s (with $d\geq 7$) dual to the  cubic Quasi-Topological Gravity  is that for a specific range of values of the Lovelock's couplings $\la$ and $\mu$, namely
\begin{eqnarray}
\bullet \mu_+(d\geq7),\quad \frac{1}{4}<\la<\frac{1}{3},\quad\quad\quad \bullet \mu_-(d\geq 7),\quad  0<\la<\frac{1}{3}\label{maxscale}
\end{eqnarray}
the a/c-Theorem requirements can be satisfied in two distinct (and disconnected) regions: (1) of minimal fundamental scales $L_+^{a/top}$ such that $L_{UV}>L_{IR}>L_+^{a/top}$ (studied in Sect.7.4.1); (2) of maximal fundamental  $L_-^{c'}$ scale\footnote{or $L_{top/h}^-$ in  case the requirements of the Modified a/c-Theorem are imposed} such that  $L_{IR}<L_{UV}<L_-^{c'}<L_-^{top}<L_+^{top}$.

\begin{figure}[ht]
    \centering
    \includegraphics[scale=0.6]{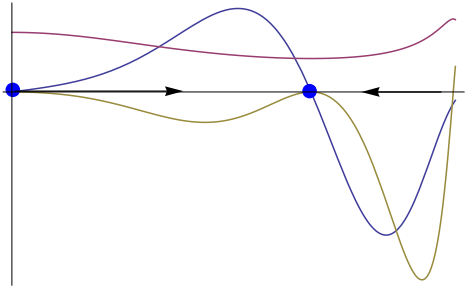}
    \begin{quotation}
    \caption[ed]{\small RG flows in the maximal scale region: the $-\beta(\s)$-function is the blue curve; the $c(\s)$-function is the red one and the $\frac{dc(\s)}{dl}$-function is drawn in beige; the arrows indicate the directions of the massless RG flow.}
     \label{fig:beta}
    \end{quotation}
\end{figure}

In the case of the considered quartic Superpotential $W(\s) = - B [ ( \s^2 - x_0 )^2 + D ]$ with the $B,D$ and $x_0$ all positive, when the above new maximal scale restriction is imposed, we have no real solutions of the  topological vacua equations  (\ref{sigma top eq}), as one can easily verify  from  their explicit solutions  
(\ref{udef}). Therefore we can not realize the  maximal scale requirements for  positive signs of the parameters of superpotential. As it was suggested in ref. \cite{lovedw}, certain flat DW solutions of the Quasi-Topological Gravity, satisfying the maximal scale conditions, can be constructed by  keeping the quartic form of $W(\s)$ and the condition $W<0$, but considering the case of \emph{negative} $B$ and $D$ and $x_0>0$, $i.e.$ by taking 
$\tilde{W}(\s)=  |B| [( \s^2 - x_0 )^2  -|D|]<0$. 
This Superpotential    represents the \emph{standard Higgs potential}, that satisfies the $\tilde{W}<0$ condition for the values of $\s$ restricted to the interval $\s^2\leq x_0+\sqrt{|D|}$ with $x_0\leq \sqrt{|D|}$. The solutions  for the scale factor of such maximal scale DWs and related to them correlations lengths $\xi(\s)$ are again given by eqs. (\ref{Solqt}), (\ref{udef}) and (\ref{stop}), but with $B$ and $D$ replaced by $-|B|$ and $-|D|$. As a consequence the critical points $\tilde{\s}_{IR}^2=x_0$, $i.e.$ the minima of the new  $\tilde{W}$  are now of IR type , while its maximum  $\s_{UV}=0$ is becoming the new UV -critical point, since  $\tilde{s}_{UV}(-|B|,-|D|)=s_{IR}(B,D)$ is  \emph{positive}. The topological $\s_{top}^-$ vacua is also of UV type. 

The above analysis of the critical points of the dual $QFT$ (respecting the maximal scale restrictions), completed by the behaviour of $\xi(\s)$, namely that $\xi(0)=\infty$ ,$\xi(\pm\sqrt{x_0})=0$ and $\xi(\pm\s_{top}^-)=\infty$, lead to the following conclusion: \emph{the phase structure of such maximal scale dual $QFT_{d-1}$'s consists of two identical massless phases $(-\sqrt{x_0},0)$ and $(0,\sqrt{x_0})$.} The plots of the corresponding $\beta(\s)$-, $c(\s)$- and $dc/dl$-functions, shown on Fig.(\ref{fig:beta}), confirm the existence of such maximal scale massless RG flows,  satisfying all the conditions of the $a/c$- Theorem for certain higher dimensional $QFT$'s dual of the $d\geq 7$ QT Gravity models.  

It is important to mention that contrary to the case of the minimal scale models that are characterized by $C_{\Phi\Phi\Phi}^{UV}\neq0$ , all the maximal scale models based on the above Higss-like superpotential \emph{have vanishing UV-structure constant} (and non-zero IR-ones) which makes unclear  their identification as certain perturbed $CFT_{d-1}$ models. This problem is however \emph{absent} in the case of more complicated maximal scale $QFT$ models dual to $QT$ Gravity with an appropriately \emph{modified superpotential} obtained from $\tilde{W}$ by adding to it a cubic term.


\section{Conclusions}

\setcounter{equation}{0}

\textit{8.1. Motivations and goals.} The Holographic RG group methods \cite{VVB,rg} are known to have many important applications in the investigation of the RG flows and the phase structure in the strong coupling limits of certain $SU(N_c)$  ${\cal N}=4$ supersymmetric $CFT_{d-1}$'s with an appropriate massive term for the chiral scalar super-fields added. The main tool in such off-critical version of the $AdS_5/CFT_4$ correspondence are the DW's solutions of the corresponding dual five-dimensional gauged ${\cal N}=8$ Einstein super-gravity. One of the main characteristics of the ${\cal N}=4$ susy $CFT_4$'s is that they must have equal central charges $c=a$. Since the less supersymmetric ${\cal N}=0,1,2$ $CFT_4$'s by definition have distinct central charges $c\neq a$, their holographic  critical and off-critical description are known to require  certain ``higher derivatives'' extensions of the Einstein (super)Gravity, as for example the GB and the cubic or quartic Quasi-topological Gravities \cite {My_qtop,c_th,mann}. Again for the construction of the corresponding ``holographic''  $\beta(\s)$- function the explicit forms of the DW's of such extended Gravities are needed. The  problem we have addressed and solved in the present paper concerns the holographic description of the massless and the massive phases and of the phase transitions occurring in a family of $c\neq a$ (zero temperature) $QFT_{d-1}$'s, by using  the recently constructed GB and cubic QT Gravities DW's \cite{lovedw}. As we have already mentioned, the main advantage of the superpotential method \cite{6,lovedw,nmg,3} is that it allows to easily implement all the unitarity and positive energy fluxes conditions on the basic UV- and IR- $CFT$ data ($L_k:s_k, c_k, a_k, t^k_4$) in order to derive a set of consistency conditions on the matter superpotential  parameters and on the  gravitational couplings $\la$ and $\mu$. The RG evolution of this conformal data, giving rise to the massless and massive  RG flows in the corresponding \emph{dual unitary} $QFT_{d-1}$'s, is described by specific chains of DW's. 

Let us emphasize once more that the purpose of the investigations presented in the this paper is to realize a comparative test of the way the off-critical properties of the dual $c\neq a$ holographic $QFT_{d-1}$'s  depend on the particular choice of the extended ``higher curvature''  Gravities and on the specific form of the scalar matter superpotential as well. Our attention is always concentrated on the question  of whether the specific requirements of the unitarity and causality consistency of corresponding massive and massless phases can be satisfied within the frameworks of the considered examples of extended EH gravity-matter models. The motivations to consider the particular case of  Holographic RG flows in the $QFT$'s dual to \emph{the GB and the cubic Quasi-Topological Gravity} \cite{My_qtop} are:
 
 $\bullet$ their remarkable feature to lead to second order equations for both the flat DWs and for the  corresponding linear fluctuations around such DW's background;
 
 $\bullet$ they  allow to reproduce the most general  $a\neq c$ and $t_4\neq0$ non-supersymmetric  $CFT$'s  \cite{2,MPS,bala};
 
 $\bullet$ for $d=5$ their dual $QFT_4$'s turns out to have quite a reasonable off-critical phase structure and describe phase transitions expected to take place in certain approximations of the $QCD_4$, as well as in the description of some features of the quark-gluon plasma hydrodynamics \cite{hydro,hidro,2,MPS}; 

 $\bullet$ they are expected to have an application to the consistent holographic definition and also for the derivation of some important off-critical characteristics of the entanglement entropy in the dual ``boundary'' \emph{non-conformal} $QFT_{d-1}$'s (in different geometries and b.c.'s), as well as for its calculation in such DW's backgrounds \cite{entjapa,entrop,Myers-new}.

\vspace{0.5 cm}
\textit{8.2. Summary of the main results}. Starting from the explicit constructions of the GB and QTG domain walls and the well known \emph{critical} unitarity and energy fluxes positivity conditions, we have addressed the following two  problems: 

 ${\cal P}1$: to establish the \emph{off-critical} consistency requirements that the corresponding $a(l)$, $c(l)$ and $\beta(l)$ -functions have to satisfy;

 ${\cal P}2$: to describe the phase structure, the holographic RG flows and the nature of the phase transitions in the dual \emph{non-conformal} $QFT_{d-1}$'s, which are compatible with the corresponding $a/c$-Theorem restrictions.

$\bullet$ \textit{Holographic $a/c$-Theorems.} Taking as a basic ingredient the unitarity and p.e.f. consistency of the $CFT_{d-1}(UV/IR)$-data, we have extended  them to the massless phases of the conjectured dual $QFT_{d-1}$'s, giving rise to the \emph{holographic versions} \cite{rg,My_thol,MPS} of the ``$a/c$-Theorems'' of Zamolodchikov and Cardy \cite{x,cardyth}. The main questions to be answered are: (1) about the behaviour and  of the eventual monotonic properties of the $c(l)$-central function and (2) about the conditions  the parameters of the considered models should satisfy in order to have both central functions always positive and respecting the p.e.f. requirements. Our proofs  have established the specific  minimal or/and maximal scales  restrictions on the UV and IR ``physical scales'' $L_{UV/IR}$, which  guarantee the validity of the different forms of the $a/c$-Theorem. Two qualitatively different behaviours have to be distinguished:

\textit{Minimal scale}: All the $QFT$'s models duals to the $d=5$ QT Gravity and to the $d\geq 5$ GB Gravity exhibit massless RG flows with \emph{both  $a$ and $c$-central functions positive and monotonically decreasing}, when the following minimal scale requirements $L_{UV}>L_{IR}>L_{min}$ are fulfilled. The specific exact (model dependent) values of these $L_{min}$ are derived in Sect.6 from the conditions of \emph{positivity of the energy fluxes}. For example, in all the $QFT_{d-1}$'s duals to the GB models of positive $\la$ we get $L^2_{min}=2L^2\la(d^2-5d+10)/(d-3)(d-4)$.   

\textit{Maximal scale}: All the $QFT$'s  duals to $d\geq 7$ QTG models admit two distinct type of massless RG flows: together with the above minimal scale version of the $a/c$-Theorem we have a new one with specific \emph{maximal scale} restrictions $L_{IR}<L_{UV}<L_{max}$, ensuring that again $c(l)$ is decreasing. However now we can also realize two quite different behaviours of the $c(l)$-function, namely: monotonically \emph{increasing} $dc/dl>0$  in the case when $L^*_{min}<L_{IR}<L_{UV}<L^*_{max}$ or \emph{non-monotonic} $c(l)$  for $L_{IR}<L_{max}<L_{UV}$. The values of the corresponding maximal/minimal scales $L^{*}_{min}$ and $L^*_{max}$ are derived in sect.5 and App. \ref{apexA}.

The realization of all the important off-critical $QFT$'s data \cite{MPS,My_thol,2} as functions of the running coupling $\s(l)$ and more precisely  in terms of the superpotential $W(\s)$, allows us to derive the extensions of the $a$ and $c$ -Theorems including  the energy fluxes parameters $t_2(\s)$ and $t_4(\s)$ as well. Eventual  applications of these results to the non-conformal $sQGP$-hydrodynamics for the calculation of the \emph{bulk viscosity $\zeta$}  and also for deriving formulas  $\left(\eta/s\right)_{UV}\geq \left(\eta/s\right)_{IR}$ relating  the UV and IR values of the the ratio of shear viscosity to entropy density, that takes place in the massless phase of the $QFT_4$  dual to the GB model of positive $\la$, are discussed in Sect. 6.

$\bullet$ \textit{Holographic RG flows and QFT's phase transitions}. Given the explicit form (\ref{rg}) of the holographic $\beta_W(\s)$-function for our particular choice of quartic inverted Higgs superpotential $W(\s)$. The solutions of the corresponding  Wilson RG equations (\ref{fs}) are known to provide  all the information  necessary for the description of the main characteristics of the $QFT$'s phases and the nature of the phase transitions \cite{cardy}. The knowledge of the analytic forms of the  running coupling constant $\s(l)$, of the correlation length $\xi(\s)$ and of the reduced free energy $F(\s)$  for all the values of the coupling $\s$, allows us to identify the distinct intervals $(\s_k,\s_{k+1})\in R$ between the critical points $\s_k$'s as representing  massive or massless phases. The detailed analysis of the off-critical behaviour of the considered models, performed in sect.7, leads us to the following conclusions about the ``holographic phase structure''  of the corresponding dual unitary $QFT_{d-1}$:

(1) \textit{$QFT$'s duals to GB for negative $\la$.} Massless unitary RG flow in $(\s_{IR},\s_{UV})$  and Massive phase $(\s_{UV},\infty)$ with mass gap  given by eq. (\ref{gbmass}); second order phase transition.

(2) \textit{$QFT$'s duals to GB for positive $\la$.} One massless phase $(\s_{IR}, \s_{UV})$ and  a new kind of a-massive phase $(\s_{UV}, \s^{cr}_{a})$  with $a(\s^{cr}_a)=0$, the physical consistency of which is under question.

(3) \textit{$QFT_4$ duals to $d=5$ QTG.} Only one well defined massless phase $(\s_{IR},\s_{UV})$. For the $\mu_+(d=5)$ model with $\la\in(\frac{5}{27},\frac{1}{3})$  and  for the $\mu_-(d=5)$ model with $\la\in (-\infty,\frac{1}{3})$ we have one massless and one a-massive phase as in the GB case (2). For the $\mu_+(d=5)$ model with $\la\in(-\infty,\frac{5}{27})$  together with the standard massless RG flow within $(\s_{IR},\s_{UV})$, we can have  a new ``top-massless'' phase $(\s_{UV},\s_{top}^+)$, which however is violating the p.e.f. requirements.

(4) \textit{$QFT_{d-1}$ duals to $d\geq 7$ QTG -minimal scale.} For $\la$ and $\mu$ both negative the massive-to-massless phase transition is as in GB case (1); the class of models listed in (\ref{mamass}) has the same phase structure as in the GB case (2) above; the phase structure of the models (\ref{mss})  includes the  physically inconsistent ``top-massless'' phase as in the $d=5$ case (3) above.

(5) \textit{$QFT_{d-1}$ duals to $d\geq 7$ QTG -maximal scale.} The $\mu_+$ models with $\frac{1}{4}<\la<\frac{1}{3}$ and the $ \mu_-$ ones with $0<\la<\frac{1}{3}$ exhibit only one consistent massless phase, satisfying the $a/c$- Theorem and its modifications. They are described in Sects.5. and realized in Sect.7.4. 

(6) \textit{Infinite order phase transitions.} We can have also two marginal degeneration limits $L_{UV}=L_{IR}$ and $L_{UV}=L_{top}$ (in the case when we have at least have  one topological vacuum) that lead to essential singularities in the free energy as in eq. (\ref{margb}). They are known to describe  infinite order BKT-type phase  transitions (or Miransky scaling \cite{Mira,lost,ven-kir,kutas}), that are important in the description of $QCD_4$ of large number of flavors $N_f>>1$ within the conformal window $4\leq N_f/N_c<11/2$ \footnote{the $QFT_4$'s duals to the marginally degenerated $d=5$ GB DW's are expected to represent ${\cal N}=1$ susy $SU(N_c)$ $QCD_4$ with $N_f$ flavors within the corresponding ``superconformal  window'' $3/2\leq N_f/N_c\leq 3$ \cite{seiberg}}.

\vspace{0.5 cm}
\textit{8.3. Open problems.} We have to mention that our investigation of the critical phenomena in certain unitary $QFT$'s duals to the GB and QT gravity-mater models  left  unanswered few important questions. The  description of the different massive and massless phases by using of RG methods, based on the analytic properties and the asymptotic behaviour of the corresponding reduced free energy,  needs to be completed by further analysis of the following three problems: 

$\bullet$ the stability and the bulk causality of the specially selected chains of DWs, describing the phase structure of the considered dual $QFT$'s; 
 
$\bullet$ the holographic calculation of the off-critical mass-spectrum and of the 2-point correlations functions of the relevant operator $\Phi_{\s}(x_i)$, driving the RG flows; 

$\bullet$ the off-critical  extension  of the $CFT$'s conditions for the positivity of the energy fluxes.

The solution of the first two problems requires further studies of the linear fluctuations (of both the metrics and of the matter field $\s$ ) around the special GB and cubic QTG DW's backgrounds, we have used in the description of the off-critical behaviour of the corresponding dual $QFT$'s. Although the physical vacua are by construction \emph{perturbatively} stable \cite{My_qtop,My_thol,MPS,espanha}, the knowledge of the spectrum of the small fluctuations around such DWs is further needed in order to establish their proper stability, as well as for revealing the eventual causality violations that might occur in the bulk, as in the examples studied in refs.\cite{MPS,espanha,BM}. An important fact to be mentioned here is that  the corresponding DW's fluctuation equations\footnote{differently from the case of the fluctuations equations around the cubic QTG black-hole solutions that are of fourth order \cite {My_qtop}.} obtained from the cubic  Quasi-Topological gravity-matter action (\ref{qtop}) are known to be  \emph{second} order differential equations \cite{My_qtop,My_thol}, rather similar to the ones of the EH gravity-matter case.

The third problem concerns further off-critical restrictions that eventually have to be imposed on the DW's  and on the gravitational couplings $\la$ and $\mu$ in order to ensure the unitarity of the dual $QFT$'s. Sect.6.4 contains a preliminary discussion of how one can define and calculate the energy fluxes in the perturbed $CFT$'s, representing the considered non-conformal $QFT$'s, by applying the well known conformal perturbations methods, quite similar to the ones used in ref. \cite{kleba} for the  calculation of the free energy.

\vspace{0.5 cm}

\textit{8.4. Few problems for further research.} The results we have obtained by using specially selected DW's of the GB and QT Gravities with the holographic $a/c$-Theorems and the energy fluxes positivity conditions properly implemented, provide  examples of dual unitary $QFT$'s having quite reasonable massless phases. However the only $QFT_4$'s that represent unitary (and eventually stable) massive phase are those ones duals to the GB models of negative $\la$  with $t_4=0$. They are known to correspond to ${\cal N}=1$ supersymmetric $QFT$'s. Therefore the problem of the holographic description of \emph{non-supersymmetric $c\neq a$ $QFT_4$'s with $t_4 \neq 0$, that have well defined massive phase} still remains open. Leaving aside the possible relation of such models with certain large $N_c$ limits of  the $QCD_4$, we shall mention  here few generalizations of the models and of the methods used in the present paper that might lead to  more realistic $t_4 \neq 0$ massive $QFT_4$:

$\bullet$ to take as a starting point different ``higher derivatives'' $d=5$ Gravities of Lovelock type, as for example the recently constructed \emph{quartic} Quasi-Topological gravity \cite{mann}, the extended Born-Infeld (BI) models \cite{correano} or certain extensions of the bi-metric massive gravity \cite{mpaulos}. In all these cases both the superpotential method and the proofs of the corresponding $a/c$-Theorems are expected to have rather straightforward generalizations;

$\bullet$ to keep unchanged the gravitational cubic QTG part of the action (\ref{qtop}), but to consider other more realistic forms of the matter superpotential. One can take as examples superpotentials, involving many scalar fields of sigma model type, or else certain well known super-gravity induced superpotentials, like in the case of EH super-gravity models \cite{rg}. Again an important technical problem to be resolved is the construction of the corresponding DW's by an appropriate extension of the superpotential method, that also allows to derive the holographic $\beta$-function and the form of the central functions \cite{lovemany}.

It is worthwhile to also mention another promising direction of related research, concerning the investigation of the marginally degenerated DW's in the GB and QTG models of at least three physical vacua \cite{lovedw}  and their applications to the holographic description of certain infinite order BKT (or Miransky ) phase transitions in the dual $QFT$'s. Such behaviour is expected to take place in the $QCD_4$ in the Veneziano limit \cite{venezi} of large number of flavors $N_f$ and colors $N_c$, but of finite ratio within the conformal (non-supersymmetric) window $x_c\leq N_f/N_c\leq 11/2$  \cite{Mira,ven-kir,lost,kutas} with $x_c\approx4$.

\vspace{1 cm}
\textbf{Acknowledgements.} We are grateful to A.L.A.Lima  for critical reading of the manuscript and for his suggestions for improvements.

\newpage

\appendix

\noindent{\Large {\bf Appendices}}

\vspace{0.5cm}

\section{The proof of $d\ge7$ $a/c$- Theorems }\label{apexA}
\setcounter{equation}{0}

\subsection{$\mu_+$ model}\label{app_A.1}

The Standard $\mu_+$ model $a/c$-Theorem valid for $d \ge 7$ reads: 

\emph{The conditions that $a(l)$- and $c(l)$-central functions are both positive and monotonically decreasing during the RG flow between two consecutive critical points $\s_{UV}$ and $\s_{IR}$ are given by}:
\begin{eqnarray}
&&(1)\bullet \quad -\infty<\la<\frac{1}{4}: \quad  f_{IR}<f^a_{+}(\la;\mu_+),\\
&&(2)\bullet \quad \frac{1}{4}<\la<\frac{d(d-4)}{3(d-2)^2}:\quad f_{IR}<f^a_{+}(\la;\mu_+) \textrm{ or } f_{UV}>f^{c'}_{-}(\la;\mu_+),\\
&&(3)\bullet \quad \frac{d(d-4)}{3(d-2)^2}<\la<\frac{1}{3}:\quad f_{IR}<f^{top}_{+}(\la;\mu_+) \textrm{ or } f_{UV}>f^{c'}_{-}(\la;\mu_+).
\end{eqnarray}
\emph{Proof.} Similarly to the $d=5$ case, the sign of $\mu_+$ is crucial in the investigation of the intervals of $\la$ where $f_{\pm}^{\eta}$ are real positive or complex. Taking into account their explicit forms (\ref{fac}) we realize that:
\begin{eqnarray}
  -\infty<\la<\frac{1}{4},\quad\quad \mu_+(\la)\geq 0,\quad\quad f_+^{\eta}(\la)>0,\quad f_-^{\eta}(\la)<0 ,\quad\quad \eta=top,a,c' \label{sevenmuposit}
\end{eqnarray}
and as a consequence the following ordering (see also Fig.(\ref{fig:3a})): 
\begin{eqnarray}
   0<f^a_{+}<f^{top}_+<f^{c'}_+,\label{ordermuposit}
\end{eqnarray}
takes place. For positive values of $\mu$ the requirements:
\begin{eqnarray}
    (f+|f_-^{\eta}|)(f-f_+^{\eta})<0,\quad \eta=top,a,c'\quad \textrm{or}\quad 0<f<f_+^{\eta},\quad\textrm{$i.e.$}\quad 0<f<f_+^{a}\label{muposrest}
\end{eqnarray} 
guarantee that the conditions of the Standard a/c-Theorem: $a>0$, $c>0$ and $dc/dl<0$  are fulfilled, which proves the case (1) of the above Theorem. 

As we have demonstrated in Sect.2.1., in the remaining two cases (a2) and (a3) where $\mu_+$ is \emph{negative} and both $f_{\pm}^{top}>0$ are positive we have the following restrictions ensuring $c(f)>0$:
\begin{eqnarray}
&&\bullet \quad \frac{1}{4}<\la<\frac{8}{27}: \quad h < 0,\quad c>0 \quad\textrm{for}\quad 0<f<f_+^{top}\quad \textrm{and} \quad f>f^h>f_-^{top},\nonumber\\
&&\bullet \quad \frac{8}{27}<\la<\frac{1}{3}: \quad h\ge 0,\quad c>0 \quad\textrm{for}\quad 0<f<f_+^{top}\quad \textrm{and} \quad f>f_-^{top}\label{cposit}
\end{eqnarray}
In order to satisfy all the requirements $(f-f_-^{\eta})(f-f_+^{\eta})\ge 0$ of the a/c-Theorem, we next observe  that according to the arguments presented in Sect.3.2. above, both $f_{\pm}^{c'}>0$ are positive within the interval $\la \in (1/4,1/3)$. Therefore the condition $dc/dl<0$ is satisfied for all the $f$'s such that:
\begin{eqnarray}
\bullet \quad 0<f<f_+^{c'} \quad \textrm{and} \quad \bullet f>f_-^{c'},\label{clposit}
\end{eqnarray}
Given the explicit form (\ref{fac}) of $f_{\pm}^a(\la,\mu_+)$, the description of their properties and of the corresponding conditions for positivity of $a(\s)-$central function requires a bit more involved case-by-case analysis. It is relatively easy to demonstrate that the solutions of the reality condition:
\begin{eqnarray}
\frac{9\la^2(d-6)(d-2)}{(d-4)^2} +2 -9\la \ge -2(1-3\la)^{3/2},\label{Dmuplus}
\end{eqnarray}
are given by:
\begin{eqnarray}
\bullet  \quad 0<f_+^a<f_-^a \quad\textrm{for}\quad \la \in(\frac{1}{4},\la_a^+) ,\quad \bullet \quad f_{\pm}^a \quad \textrm{complex for}\quad \la \in(\la_a^+,\frac{1}{3}),\quad\quad \la_a^+=\frac{d(d-4)}{3(d-2)^2}\nonumber
\end{eqnarray} 
 As we have mentioned in Sect.2.1. in the region $\la \in(1/4,8/27)$, where  $h < 0$, we have to introduce a \emph{new maximal scale} $f^h= - \la / 2 \m_+ + \sqrt{ \left( \la^2 + 4 \m_+ \right)/4 \m_+^2}>f_-^{top}$, instead of $f_-^{top}$. Therefore the description of the conditions of validity of a/c-Theorem for $\mu_+$-model, depends on whether $\la_+^a<8/27$ or $\la_+^a \ge 8/27$. In the proof $d=7$ a/c-Theorem, due to the fact that $\la_+^a=7/25<8/27$, we have to distinguish the following three cases:
\begin{eqnarray}
\bullet \ \frac{1}{4}<\la<\frac{7}{25}: && \ 0<f_{+}^a<f_{+}^{top}<f_{+}^{c'}<f_{-}^a<f_{-}^{top}<f^h<f_{-}^{c'}\quad \textrm{or}\nonumber\\
&&\ 0<f_{+}^a<f_{+}^{top}<f_{-}^a<f_{+}^{c'}<f_{-}^{top}<f^h<f_{-}^{c'};\nonumber\\
\bullet  \frac{7}{25}<\la<\frac{8}{27}: && \ 0<f_{+}^{top}<f_{+}^{c'}<f_{-}^{top}<f^h<f_{-}^{c'},\nonumber\\
\bullet \ \frac{8}{27}<\la<\frac{1}{3}: && \ 0<f_{+}^{top}<f_{+}^{c'}<f_{-}^{top}<f_{-}^{c'}.\label{seven}
\end{eqnarray}
The particular ordering of all the $f^{\eta}_{\pm}(\la,d=7)$ with $\eta=a,c',h,top$ is established graphically as plotted on Fig.(\ref{fig:3b}). Since $\la_+^a=8/27$ for $d=8$, we have to consider separately only the following two regions : (1) $\frac{1}{4}<\la<\frac{8}{27}$ with the same properties of all the $f^{\eta}$'s as in the first region in (\ref{seven}) above, and (2) $\frac{8}{27}<\la<\frac{1}{3}$ literally repeating the third region of $d=7$ case. Finally, for $d\ge 9$ we always have that $\frac{8}{27}<\la_+^a<\frac{1}{3}$. Therefore, similarly to the $d=7$ case, the following three cases should be analysed separately: 
\begin{eqnarray}
\bullet \ \frac{1}{4}<\la<\frac{8}{27}:\quad\quad\quad && \ 0<f_{+}^a<f_{+}^{top}<f_{+}^{c'}<f_{-}^a<f_{-}^{top}<f^h<f_{-}^{c'}\quad \textrm{or}\nonumber\\
&&\ 0<f_{+}^a<f_{+}^{top}<f_{-}^a<f_{+}^{c'}<f_{-}^{top}<f^h<f_{-}^{c'};\nonumber\\
\bullet \ \frac{8}{27}<\la<\frac{d(d-4)}{3(d-2)^2}: && \ 0<f_{+}^a<f_{+}^{top}<f_{+}^{c'}<f_{-}^a<f_{-}^{top}<f_{-}^{c'}\quad \textrm{or}\nonumber\\
&&\ 0<f_{+}^a<f_{+}^{top}<f_{-}^a<f_{+}^{c'}<f_{-}^{top}<f_{-}^{c'};\nonumber\\
\bullet \ \frac{d(d-4)}{3(d-2)^2}<\la<\frac{1}{3}:\quad && \ 0<f_{+}^{top}<f_{+}^{c'}<f_{-}^{top}<f_{-}^{c'}.\label{nine}
\end{eqnarray}
Notice that (a) the only difference between the first two cases in (\ref{nine}) is in the presence of the new intermediate scale $f^{h}$ when the interval $1/4<\la<8/27$ is considered, due to the fact that $h<0$ is negative there; (b) the properties of the $f^{\eta}$'s that are positive in the last region \emph{are common} for all the $d\ge 7$. It is important to mention that the two different orderings $f_{+}^{c'}<f_{-}^a$ and $f_{-}^a<f_{+}^{c'}$ present in all the cases when $\la \in (1/4,\la_+^a)$ reflect the fact that it exists one critical value $\la_{cr}^{ac'}$,
where the two scales $L_+^{c'}(\la_{cr})=L_-^{a}(\la_{cr})$ do coincide. As one can see from fig.({\ref{fig:3b}) it is exactly the point when the curves $f_{+}^{c'}(\la)$ and $f_{-}^a(\la)$ cross each other.

It is now straightforward to conclude that the conditions : $(f-f_-^{\eta})(f-f_+^{\eta})\ge 0$ for validity of the Standard $a/c$-Theorem for the $\mu_+$-model in the region $\la \in (1/4,1/3)$ are given by : 

(1a) $0<f_{vac}< f^a_{+}(\la;\mu_+)$ and (1b) $f>f^{c'}_{-}(\la;\mu_+)$ for $\la \in(\frac{1}{4},\frac{d(d-4)}{3(d-2)^2})$ due to the fact that $f_+^a$ is defining the \emph{minimal scale} and $f^{c'}_{-}$ - the \emph{maximal} one in this region, while for $\la \in (\frac{d(d-4)}{3(d-2)^2},1/3)$ we have: 

(2a) $0< f <f^{top}_{+}(\la;\mu_+)$ of \emph{different minimal} scale and (2b) $f>f^{c'}_{-}(\la;\mu_+)$, $i.e.$ with the same \emph{maximal} scale. 

Notice that  the of presence of the new  \emph{intermediate} scale defined by $f^h$ in the regions of negative $h<0$ that leads to certain differences between the cases $d=7, d=8$ and $d\ge 9$ as described by eqs. (\ref{seven}) and (\ref{nine}) above \emph{is not changing}  the requirements of the \emph{Standard Theorem}. They have  in fact the same form for  all the models with $d \ge 7$. 
\begin{figure}[ht]
\centering
\subfigure[]{
\includegraphics[scale=0.7]{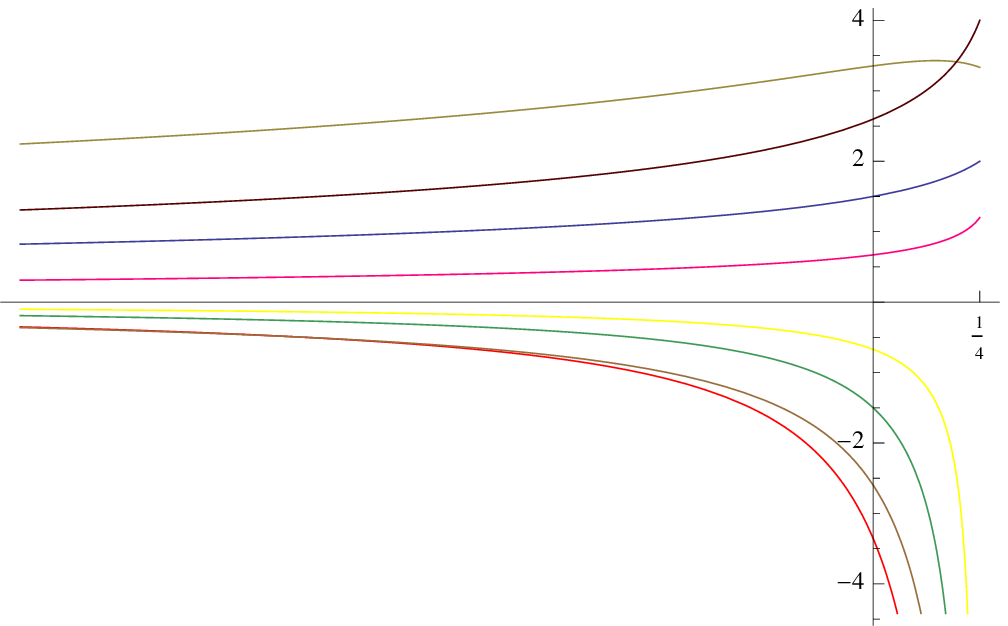}
\label{fig:3a}
}
\subfigure[]{
\includegraphics[scale=0.7]{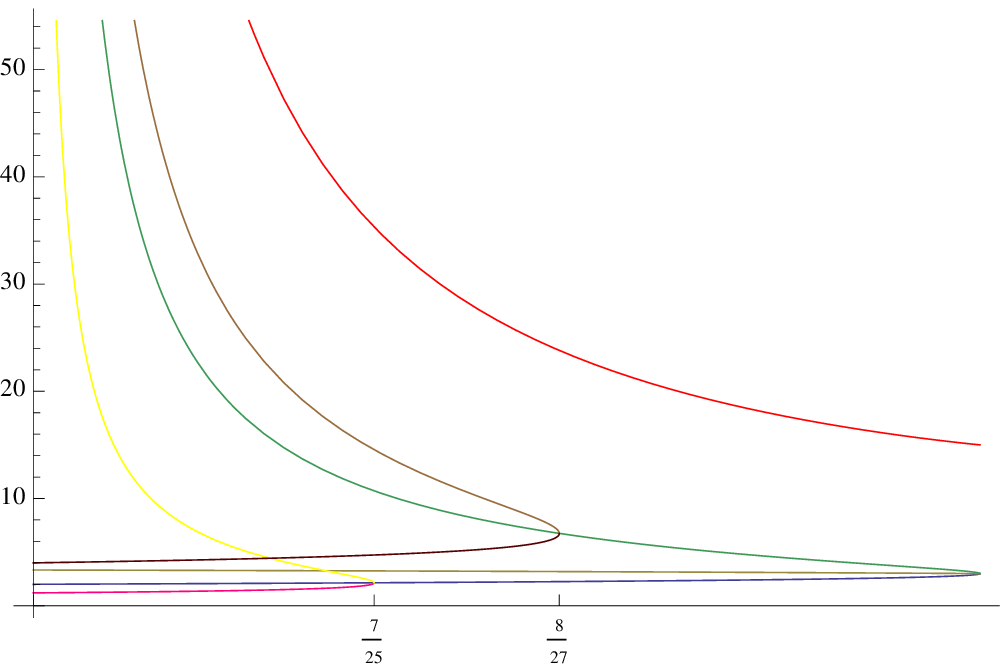}
\label{fig:3b}
}
\label{fig:3}
\caption{The $f^{\eta}_{\pm}(\la;\mu_+)$ curves in $d=7$ $\mu_+$ model: $f^{top}_+$ is depicted in blue; $f^a_+$ in pink; $f^{c'}_+$ in beige; $f^h_+$ in dark brown; $f^{top}_-$ in green; $f^{a}_-$ in yellow; $f^{c'}_-$ in red and $f^h_-$ in bright brown. Fig.(a) represents the $\la<\frac{1}{4}$ interval and Fig.(b) the $\frac{1}{4}<\la<\frac{1}{3}$ one; both with reference point ($1/4,0$). The special point $\la=\frac{7}{25}$ marks the end of the curves $f^a_{\pm}$.}
\end{figure}

\subsection{$\mu_-$ model}\label{app_A.2}

\textit{The conditions under which the Standard a/c-Theorem for the $\mu_-$-model is satisfied  are given by}:  
\begin{eqnarray}
\bullet \ &&d=7,8: \quad 0<\la<\frac{1}{3}: \quad\quad\quad  f<f^{top}_+(\la;\mu_-) \quad\textrm{or}\quad f>f^{c'}_-(\la;\mu_-),\label{mumenosoito}\\
\bullet \ &&d \ge 9 :  \quad 0<\la<\frac{(d-8)(d-4)}{3(d-6)^2}: \quad f<f^{a}_+(\la;\mu_-) \quad \textrm{or} \quad f>f^{c'}_-(\la,\mu_-)\label{mumenosteo}\\
\bullet \ &&d \ge 9 :  \quad \frac{(d-8)(d-4)}{3(d-6)^2}<\la<1/3: \quad f<f^{top}_+(\la;\mu_-) \quad \textrm{or}\quad f >f^{c'}_-(\la;\mu_-)\label{mumenosteoc}
\end{eqnarray}
\textit{where f is denoting $f_{IR}$ or $f_{UV}$ in the case of minimal/maximal scale correspondingly}.

\textit{Proof.} Since both $f^{\eta}_{\pm}>0$ (for $\eta=c',top$) are known to be positive for $\la \in (0,1/3]$,
it remains the analysis of the properties of $f^a_{\pm}$'s. Taking into account their form (\ref{fac}), similarly to the the $\mu_+$-model case (see eq. (\ref{Dmuplus}) is straightforward  to realize  that the corresponding  reality condition: 
\begin{eqnarray}
\frac{9\la^2(d-6)(d-2)}{(d-4)^2} +2 -9\la \ge 2(1-3\la)^{3/2},\label{Dmumenos}
\end{eqnarray}
is satisfied only for $\la \le \la^a_-=\frac{(d-8)(d-4)}{3(d-6)^2}$. Within the interval $\la \in (\la^a_-,1/3]$ the $f^a_{\pm}$'s are complex. In the case $d=7$ we have $\la^a_-=-1$ and therefore $f^a_{\pm}$ are indeed complex  in the interval $(0,1/3)$ we are investigating. Similarly, for $d=8$ the $\la^a_-(d=8)=0$ is vanishing and again we have complex $f^a_{\pm}$'s. The ordering of the remaining real and positive $f_{\eta}$'s is plotted on Fig.({\ref{fig:de}) (for $d=7$) and without lost of generality for $d=7,8$ it is given by:
\begin{eqnarray}
0<f^{top}_+<f^{c'}_+<f^{top}_-<f^{c'}_-,\label{mu_1}
\end{eqnarray}
Remembering that the conditions $(f-f_-^{\eta})(f-f_+^{\eta})\ge 0$ for validity of the standard Theorem are identical to the ones of the $\mu_+$ model for negative $\mu_+<0$, we derive the restrictions on $f_{UV/IR}$ for $d=7,8$ as stated by eq. (\ref{mumenosoito}) in the beginning of this subsection.

Due to the fact that for $d \ge 9$  the corresponding values of $\la^a_-<1/3$ are always \emph{positive}, we now have to consider separately the case $\la \in (0,\la^a_-)$, where both $f^a_{\pm}(\la,\mu_-)>0$ are positive, from the remaining part of the interval,$i.e.$ $\la \in(\la^a_-,1/3)$ where $f^a_{\pm}(\la,\mu_-)$ still remain complex. The corresponding (real and positive) $f^{\eta}(\la)$ curves for $d=9$ are shown on Fig.(\ref{fig:df}). Hence, quite  similarly to the $\mu_+$-model, we  can conclude that  the following orderings 
\begin{eqnarray}
&&\bullet \quad \la \in(0,\la_a^-)\quad \quad 0<f_+^a<f_+^{top}<f_+^{c'}<f_-^a<f_-^{top}<f_-^{c'};\nonumber\\ 
&&\bullet \quad  \la \in(\la_a^- ,\frac{1}{3})\quad\quad 0<f^{top}_+<f^{c'}_+<f^{top}_-<f^{c'}_-\nonumber
\end{eqnarray} 
take place. It is then evident that the conditions of validity of the Standard Theorem (same as above) are indeed the ones given by eqs. (\ref{mumenosteo}) and (\ref{mumenosteoc}). 
\begin{figure}[ht]
\centering
\subfigure[]{
\includegraphics[scale=0.7]{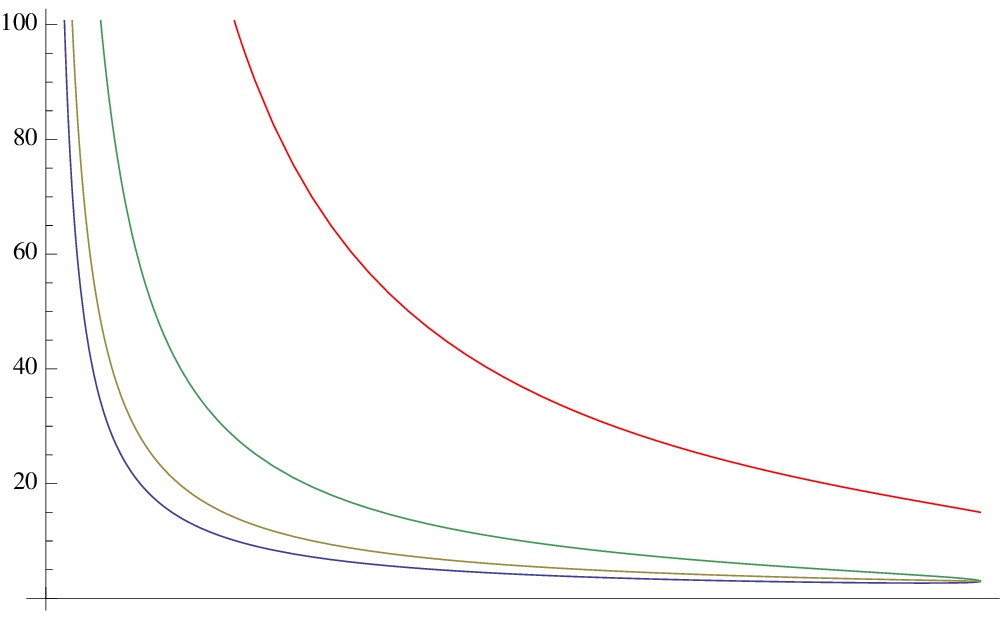}
\label{fig:de}
}
\subfigure[]{
\includegraphics[scale=0.7]{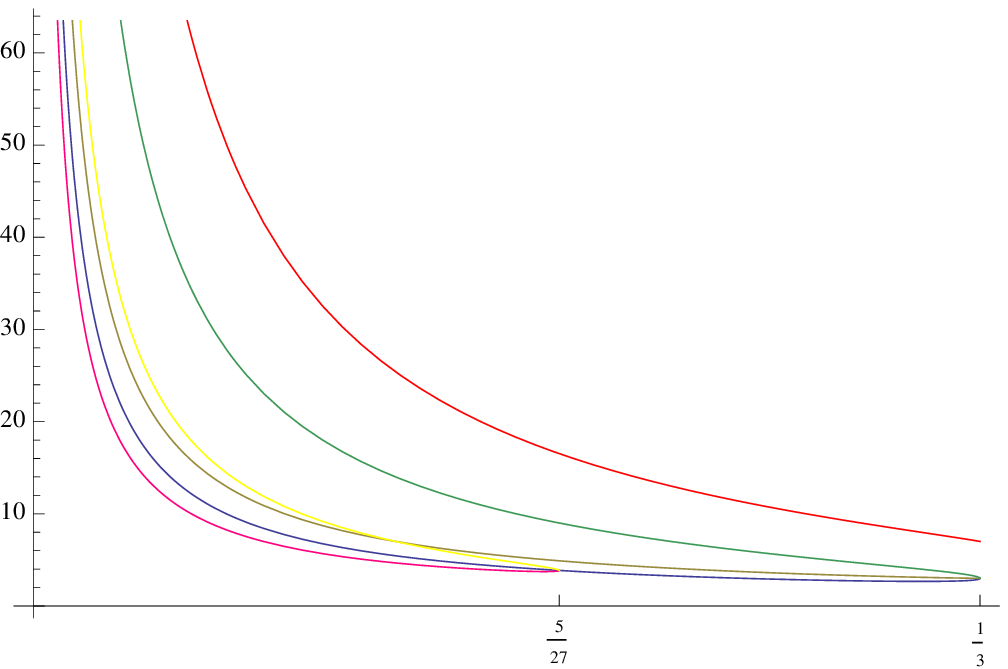}
\label{fig:df}
}
\label{fig:d}
\caption{The  $f^{\eta}_{\pm}(\l;\mu_-)$ curves in $d=7$ for $0<\l<\frac{1}{3}$ are represented on Fig.(a): the  curve $f^{top}_+$ is depicted in blue; $f^{c'}_+$ in beige; $f^{top}_-$ in green and the $f^{c'}_-$ in red. The $f^{\eta}_{\pm}(\l;\mu_-)$ curves  for  $d=9$ $\mu_-$ model are plotted on Fig.(b): the $f^{top}_+$ is in blue; $f^a_+$ is in pink;  $f^{c'}_+$ in beige; $f^{top}_-$ in green; $f^{a}_-$ in yellow and $f^{c'}_-$ in red.}
\end{figure}


\section{GB domain walls for negative $\la$}\label{apexB}
\setcounter{equation}{0}

The scale factor $e^{2A(\s)}$ of GB DW's for the considered Higgs-like superpotential $W$  can be easily obtained by integrating the RG equation (\ref{rg}), which gives rise to eq. (\ref{gbsol}) representing its general solution. In the case of \emph{negative values} of $\la$, due to the absence of topological vacua, the topological scale $L_{top}=L\sqrt{2\la}$ that is used in the parametrisation of the corresponding ``topological vacua'' positions $\s_j^2$ and of the  ``critical exponents'' $s_{top}^j$  has to be  replaced by $i|L_{top}|=iL\sqrt{2|\la|}$. Although both the $s_{top}^j$ and $\s_j^2$ become complex numbers, we are going to show that the scale factor (\ref{gbsol}) remains indeed a \emph{well defined real function of $\s$}. We next assume that the $\la>0$ vacua stability requirement  $L_{UV}>L_{top}$, $i.e.$ $x=L_{top}/L_{UV}<1$, take place in the considered $\la<0$ case as well, in order to keep the considered $\la<0$ GB DW's as a simple analytic continuation of the GB DW's for positive $\la>0$, given by eq. (\ref{Gnegtop}). 

\vspace{0.5 cm}
\textit{B.1. The $\s_j^2$ real and imaginary parts.} We start by rewriting 
$$\s_1^2= (\s_3^2)^*=(\s_1^2)^R+i(\s_1^2)^I,$$ 
defined by eq. (\ref{slaneg}) in sect.7.3.1, as follows:
\begin{eqnarray}
\s_1^2=x_0+e^{-i\frac{\pi}{4}}\sqrt{\frac{D}{x}}\sqrt{1-ix},\nonumber
\end{eqnarray}
Since $x<1$, we can expand it in a convergent power series:
\begin{eqnarray}
\sqrt{1-ix}&\equiv&\mathcal{P}(x)+i\mathcal{I}(x)=\frac{1}{\Gamma(-1/2)}\sum_{n=0}^{\infty}\frac{\Gamma(n-1/2)}{n!}i^n x^n\nonumber\\
&=&\frac{1}{\Gamma(-1/2)}\sum_{n=0}^{\infty}\frac{\Gamma(2n-1/2)}{(2n)!}(-1)^n x^{2n}+i\frac{1}{\Gamma(-1/2)}\sum_{n=0}^{\infty}\frac{\Gamma(2n+1/2)}{(2n+1)!}(-1)^n x^{2n+1}.\nonumber
\end{eqnarray}
The series representing the corresponding real $\mathcal{P}(x)$ and imaginary $\mathcal{I}(x)$ parts can be easily resumed to give the following  well known functions:
\begin{eqnarray}
\mathcal{P}(x)=(1+x^2)^{1/4}\cos\left(\frac{\arctan(x)}{2}\right),
\quad\quad\quad\mathcal{I}(x)=-\frac{x}{\sqrt{2}\sqrt{1+\sqrt{1+x^2}}}.\nonumber
\end{eqnarray}
With the above results at hand, we finally get the explicit form of the real and imaginary parts for all the $\s_j^2$:
\begin{eqnarray}
(\s_1^2)^R(x)&=&x_0+\sqrt{\frac{D}{2x}}\left(\mathcal{P}(x)+\mathcal{I}(x)\right)=(\s_3^2)^R(x),\quad (\s_1^2)^I(x)=\sqrt{\frac{D}{2\xi}}\left(\mathcal{I}(x)-\mathcal{P}(x)\right)=-(\s_3^2)^I\nonumber\\
(\s_2^2)^R(x)&=&x_0-\sqrt{\frac{D}{2x}}\left(\mathcal{P}(x)+\mathcal{I}(x)\right)=(\s_4^2)^R(x),\nonumber\\
(\s_2^2)^I(x)&=&-\sqrt{\frac{D}{2\xi}}\left(\mathcal{I}(x)-\mathcal{P}(x)\right)=-(\s_4^2)^I(x). \label{sigmatop}
\end{eqnarray}

\vspace{0.5 cm}
\textit{B.2.The $s_{top}^{j}$ real and imaginary parts.} Taking into account the explicit form of the corresponding GB ``topological critical exponents'' $s_{top}^j$ (see ref. \cite{lovedw}): 
\begin{eqnarray}
s_{top}^{j}&=&\frac{64B^2L_{top}^2}{(d-2)}(\s_j^2)(\s_j^2-x_0)^2=(s_{top}^{j+2})^*=(s_{top}^{j})^R+i(s_{top}^{j})^I, \ j=1,2 \label{stopneg}
\end{eqnarray}
and  eqs. (\ref{sigmatop}) for the ``topological vacua positions''  $\s_j^2$, we can separate their real and imaginary parts as follows: 
\begin{eqnarray}
(s^j_{top})^R&\equiv&\frac{64B^2(L_{top}^2)}{(d-2)}\left\{(\s^2_j)^R\left[((\s^2_j)^R-x_0)^2-(\s^4_j)^I\right]-2(\s^4_j)^I((\s^2_j)^R-x_0)\right\},\nonumber\\
(s^j_{top})^I&\equiv&\frac{64B^2(L_{top}^2)}{(d-2)}\left\{(\s^2_j)^I\left[((\s^2_j)^R-x_0)^2-(\s^4_j)^I\right]+2(\s^2_j)^R(\s^2_j)^I((\s^2_j)^R-x_0)\right\}\label{topexpon}
\end{eqnarray}

\textit{B.3.The scale factor real form.} In order to derive the real form of the GB scaling factor we are looking for, we first rewrite the following product of two complex conjugate factors 
\begin{eqnarray}
U_j(\s)=(\s^2-\s^2_j)^{-\frac{1}{s_j^{top}}}(\s^2-\s^2_{j+2})^{-\frac{1}{s_{j+2}^{top}}}, \ j=1,2,\nonumber
\end{eqnarray}
in terms of certain \emph{real functions}. By simple complex numbers manipulations we realize that:
\begin{eqnarray}
U_j(\s)&=&(\s^2-(\s^2_j)^R+i(\s^2_j)^I)^{\frac{-s_j^R+is_j^I}{(s_j^R)^2+(s_j^I)^2}} (\s^2-(\s^2_j)^R-i(\s^2_j)^I)^{\frac{-s_j^R-is_j^I}{(s_j^R)^2+(s_j^I)^2}},\nonumber\\
&=&\left[(\s^2-(\s^2_j)^R)^2+(\s^4_j)^I\right]^{\frac{-s_j^R}{(s_j^R)^2+(s_j^I)^2}}\left(\frac{1-i\frac{(\s^2_j)^I}{\s^2-(\s^2_j)^R}}{1+i\frac{(\s^2_j)^I}{\s^2-(\s^2_j)^R}}\right)^{i\frac{s_j^I}{(s_j^R)^2+(s_j^I)^2}}\nonumber\\
&=&\left[(\s^2-(\s^2_j)^R)^2+(\s^4_j)^I\right]^{\frac{-s_j^R}{(s_j^R)^2+(s_j^I)^2}}e^{2\frac{s_j^I}{(s_j^R)^2+(s_j^I)^2}\arctan\left(\frac{(\s^2_j)^I}{\s^2-(\s^2_j)^R}\right)}\label{uprod}
\end{eqnarray}

Finally, it remains to substitute the above product formula (\ref{uprod}) into the original generic form (\ref{gbsol}) of the scale factor:
\begin{eqnarray}
e^{A(\sigma)}&=& e^{A_{\infty}}|\s|^{-\frac{1}{s_{IR}}} |\s^2-x_0|^{-\frac{1}{s_{UV}}} G^{-}_{top}\nonumber\\
G^{-}_{top}&=&\prod_{j=1}^2\left[(\s^2-(\s_j^2)^R)^2+((\s_j^2)^I)^2\right]^{-\frac{s_j^R}{(s_j^R)^2+(s_j^I)^2}}exp\left(\frac{2s_j^I}{(s_j^R)^2+(s_j^I)^2}\arctan\left(\frac{(\s_j^2)^I}{\s^2-(\s_j^2)^R}\right)\right),\nonumber
\end{eqnarray}
which reproduces the form of the non-singular function $G_{top}^{-}(\s)$  (for $\la$  negative), given  by eq. (\ref{Gnegtop}), we have announced in Sect.7.3.1 above.


\end{document}